\documentclass{amsart}
\usepackage[utf8]{inputenc}
\usepackage[cm]{fullpage}
\usepackage{amssymb,amsfonts,amsthm,amsbsy,amscd}
\usepackage{graphicx}
\usepackage{courier}
\usepackage{caption}
\usepackage{subcaption}
\usepackage{float}
\usepackage{yfonts}
\usepackage{mathrsfs}
\usepackage{mathtools}
\usepackage{color}
\usepackage[table]{xcolor}
\usepackage{multirow}
\usepackage{tikz} \usetikzlibrary{positioning}
\usepackage{makecell}
\usepackage{verbatim}
\usepackage{enumitem}
\usepackage{hyperref}
\usepackage[T1]{fontenc}
\usepackage{scalerel,stackengine}
\usepackage[style=numeric,backend=biber]{biblatex}
\usepackage{quiver}

\newtheorem{Lemma}{Lemma}[section]
\newtheorem{Theorem}{Theorem}
\newtheorem*{theorem*}{Theorem}
\newtheorem{Proposition}[Lemma]{Proposition}

\newtheorem{Remark}[Lemma]{Remark}
\newtheorem{Definition}[Lemma]{Definition}

\newtheorem{Criteria}[Lemma]{Criteria}
\newtheorem{Criterion}[Lemma]{Criterion}

\graphicspath{ {./pics/} } 

\usepackage{todonotes}

\newcommand{\uvec}[1]{\boldsymbol{\hat{\textbf{#1}}}}

	\definecolor{Gray}{gray}{0.9}
	
	\def\Ray{\mathsf{R}}
	\def\Pra{\mathsf{P}}
	\def\Rot{\mathsf{S}}
	\def\ShaOne{\mathsf{k}_1}
	\def\ShaTwo{\mathsf{k}_2}
	\def\ShaThree{\mathsf{k}_3}
	\def\uVecOne{\uvec{e}^{\scriptscriptstyle 1}}
	\def\uVecTwo{\uvec{e}^{\scriptscriptstyle 2}}
	\def\uVecThree{\uvec{e}^{\scriptscriptstyle 3}}
	\def\uVecJ{\uvec{e}^{\scriptscriptstyle j}}
	\def\normVec{\boldsymbol{\nu}}
	\def\truncInd{\boldsymbol{\tau}}
	\def\vorticity{\boldsymbol{\omega}}
	\def\parameters{\boldsymbol{\mathfrak{P}}}
	\def\imOp{\mathcal{I}^{\truncInd}}
	\def\compParam{\boldsymbol{\chi}}

	\usepackage{subfiles} 
	
	\title{Energetic consistency and heat transport in Fourier-Galerkin truncations of free slip 3D rotating convection}
	
	\author[F1.RW]{Jens D. M. Rademacher$^1$}
	\author[F2.RW]{Roland Welter$^1$}
	
	\address{$^1$Universität Hamburg\\
		Fachbereich Mathematik \\
		Bundesstraße 55 \\
		20146 Hamburg, DE}
	\email{roland.welter@uni-hamburg.de}
	
	
	\date{}
	
\begin{document}

		\begin{abstract}
			This paper examines the effects of energetic consistency in Fourier truncated models of the 3D Boussinesq-Coriolis (BC) equations as a case-study towards improving the realism of convective processes in climate models.  As a benchmark we consider the Nusselt number, defined as the average vertical heat transport of a convective flow.  A set of formulae are derived which give the ODE projection of the BC model onto any finite selection of modes.  It is proven that projected ODE models obey energy relations consistent with the PDE if and only if a mode selection Criterion regarding the vertical resolution is satisfied.  It is also proven that the energy relations imply the existence of a compact attractor for these ODE's, which then implies bounds on the Nusselt number.  By contrast, it is proven that a broad class of energetically inconsistent models admit solutions with unbounded, exponential growth, precluding the existence of a compact attractor and giving an infinite Nusselt number.  On the other hand, certain energetically inconsistent models can admit compact attractors as shown via a simple model.  The above formulas are implemented in MATLAB, enabling a user to study any desired Fourier truncated model by selecting a desired finite set of Fourier modes.  All code is made available on GitHub.  Several numerical studies of the Nusselt number are conducted to assess the convergence of the Nusselt number with respect to increasing spatial resolution for consistent models and measure the distorting effects of inconsistency for more general solutions.  Our results indicate that large errors can arise due to energy inconsistency, and we advocate the use of consistent models.
		\end{abstract}
		
		\maketitle
		
		\tableofcontents
		
		\section{Introduction}
		
		In the study of climate, general circulation models aim to accurately represent phenomena in the atmosphere and ocean using equations from fluid dynamics. In order to be computationally tractable these models must make approximations and, in so doing, can violate energy conservation and other physical principles \cite{eden2019energy}. In this paper we derive energetically consistent Fourier-Galerkin truncated rotating convection models and analyse their properties in contrast to energetically inconsistent truncations, in particular regarding vertical heat transport.  Building on past work \cite{Welter2025Rotating}, this sequel paper extends to fully three dimensional fluid flows.    
		
		The fundamental continuum convection model considered in this paper is the three-dimensional, non-dimensionalized Boussinesq Oberbeck equations with a Coriolis force
		\begin{equation}
			\label{BoussinesqCoriolis}  \begin{split}
				\frac{1}{\Pra} \Big [ \partial_{t} \textbf{u} + \textbf{u} \cdot \nabla \textbf{u} \Big ] + \nabla p & = \Delta \textbf{u} + \frac{\pi}{\ShaThree} \Ray T \uVecThree - \Rot \hspace{.5 mm} \uVecThree \times \textbf{u} , \\
				\nabla \cdot \textbf{u} & = 0 , \\
				\partial_t T + \textbf{u}\cdot \nabla T & =  \Delta T.
		\end{split} \end{equation}
		Here $\textbf{u} = (u_1,u_2,u_3)$, $T$, $p$ are the non-dimensionalized velocity, temperature and pressure of a fluid in a box $\Omega = [0,\frac{2\pi}{\ShaOne}]\times[0,\frac{2\pi}{\ShaTwo}] \times [0,\frac{\pi}{\ShaThree}]$, where $\ShaOne, \ShaTwo > 0$ and $\ShaThree := (\ShaOne \ShaTwo)^{-1}$. The space variable is $\textbf{x} = (x_1,x_2,x_3) \in \Omega$ with $(x_1,x_2)$ the horizontal and $x_3$ the vertical direction, and $\uVecThree = (0,0,1)$ denoting the vertical unit vector.  We consider horizontally periodic boundary conditions, and uniform heating from below and cooling from above, so that the non-dimensionalized temperature $T$ satisfies the boundary conditions
		\begin{equation} \label{BC_Temp} T(x_1,x_2,0,t) = 1 \hspace{1 cm} \text{ , } \hspace{1 cm} T(x_1,x_2,\frac{\pi}{\ShaThree},t) = 0. \end{equation}
		Furthermore, the fluid is assumed to satisfy the impenetrability and free slip conditions at the top and bottom boundaries
		\begin{equation} \label{BC_Velocity} u_3 = 0 \hspace{1 cm} \text{ , }  \hspace{1 cm} \partial_{x_3} u_1 = \partial_{x_3} u_2 = 0  \hspace{2 cm} \text{ for } \hspace{.5 cm} x_3 = 0,\frac{\pi}{\ShaThree}, \end{equation}
		which means that the fluid parcels can freely slip along the boundary with no drag.  While lacking some physically realistic features such as moisture, topography and no-slip boundaries, the model \eqref{BoussinesqCoriolis} was chosen as suitably sophisticated to capture core features of turbulent atmospheric convection, yet simple enough to make some analytic statements.  
		
		The non-dimensional parameters $\Pra, \Ray, \Rot$ in \eqref{BoussinesqCoriolis} are the Prandtl, Rayleigh and Coriolis numbers, respectively.  Note that we use Coriolis number $\Rot$ (proportional to the angular speed of the rotating frame) rather than the Rossby number (used for studying very rapid rotation), since the case $\Rot = 0$ is also a relevant parameter value for general geophysical flows.  If the fluid is supposed to represent air in order to model atmospheric convection, and furthermore the domain is taken to be the troposphere (see for instance \cite{Welter2025Rotating}) then the relevant parameter values are approximately as follows:
		\begin{equation} \label{ParameterVals} \Pra \approx 1 \hspace{1 cm} \text{ , } \hspace{1 cm}  \Ray \approx 10^{16} \hspace{1 cm} \text{ , } \hspace{1 cm} \Rot \approx \left \{ \begin{array}{ll}
				0  & \text{ near the equator},  \\
				10^{15} & \text{ near the poles} .
			\end{array} \right . \end{equation}
		
		As mentioned, the main issue at hand here regards energetic consistency, and more generally balance law consistency.  Sufficiently regular solutions $\textbf{u}, T$ of \eqref{BoussinesqCoriolis} satisfy the following balance equations, where $\vorticity := \nabla \times \textbf{u}$:
		\begin{align} \label{Balance_KinEnergy_Tform} \frac{d}{dt} \langle \frac{1}{2}|\textbf{u}|^2 \rangle & = - \Pra \langle |\nabla \textbf{u} |^2 \rangle + \frac{\pi}{\ShaThree} \Pra \Ray \langle u_3 T \rangle \text{ , } \\ \label{Balance_PotEnergy_Tform} \frac{d}{dt} \langle (1-\frac{\ShaThree x_3}{\pi}) T \rangle & = - \frac{\ShaThree}{\pi} \langle u_3 T \rangle + \langle (1-\frac{\ShaThree x_3}{\pi}) \partial_{x_3}^2 T \rangle \text{ , } \\ \label{Balance_Vort}
			\frac{d}{dt} \langle \vorticity \rangle & = \Pra \langle \partial_{x_3}^2 \vorticity \rangle + \langle ( \vorticity \cdot \nabla ) \textbf{u} \rangle + \Pra \Rot \langle  \partial_{x_3} \textbf{u} \rangle . \end{align}
		These balance relations are physically meaningful, since they express the time evolution of the kinetic energy, potential energy and total vorticity for the PDE \eqref{BoussinesqCoriolis}, respectively, and specify exact rates of dissipation and energy conversion.  Except in rather special cases, one must introduce a discretization to approximate solutions of \eqref{BoussinesqCoriolis}, and in so doing one can violate these laws.  We say that a discretization is energetically consistent if \eqref{Balance_KinEnergy_Tform}, \eqref{Balance_PotEnergy_Tform} hold also for the solutions of the discretized equations.  In particular, we consider spectral spatial discretizations of \eqref{BoussinesqCoriolis}, which can be analyzed very cleanly using theoretical techniques.  It has been recognized in several works that such balance laws can fail to hold in spectral ODE models unless the modes are selected in a particular way \cite{HermizGuzdarFinn_1995,ThiffeaultHorton_1996,GluhovskyTongAgee_2002,OlsonDoering2022}, and that this failure can lead to non-physical effects such as unbounded trajectories \cite{HowardKrishnamurti_1986}.  While these works consider low dimensional or restricted truncations of the 2D problem, here we make precise Criteria regarding when these balances hold for arbitrary truncations of 3D model \eqref{BoussinesqCoriolis}. 
		
		Of interest for climate applications is the average vertical heat transport, as measured by the Nusselt number
		\begin{equation} \label{NusseltDef} \mathsf{Nu} := 1 + \frac{1}{4\pi^3} \overline{\langle u_3 T \rangle},  \end{equation}
		where $\langle \cdot \rangle$ denotes the volume integral, and an overbar denotes an infinite time average, i.e., 
		\[ \langle f \rangle := \int_{\Omega} f(\textbf{x}) d\textbf{x} \hspace{.5 cm} \text{ , } \hspace{.5 cm} \bar{f} := \lim_{t \to \infty} \frac{1}{t} \int_0^t f(s) ds \text{ . } \]
		The Nusselt number depends in a complicated way on the parameters, $\Ray,\Pra,\Rot,\ShaOne,\ShaTwo$, and in general also on the initial states $(\textbf{u}_0,T_0)$.  Relatively few results are known analytically, and these seem to focus on scaling laws for maximal heat transport.  For instance it has been proven that, holding all other parameters constant, one has $\max_{(\textbf{u}_0,T_0)} \mathsf{Nu} \lesssim 1+\Ray^{\frac{5}{12}}$ in the 2D case \cite{WhiteheadDoering_2011} for general Prandtl and the 3D infinite Prandtl case \cite{WhiteheadDoering_2012}, $\max_{(\textbf{u}_0,T_0)} \mathsf{Nu} \lesssim 1+\Ray^{\frac{1}{2}}$ in the 3D no slip case for general Prandtl \cite{ConstantinDoering_1996}, among many other such results.  In numerical studies, the Nusselt number also depends on the discretisation for which an important aspect is energetic consistency, and more generally the preservation of physically realistic balances.  While there are abundant numerical studies regarding the Nusselt number \cite{wen_goluskin_doering_2022, WenDingChiniKerswell2022, iyer2020classical, StevensClercxLohse_2010}, the question of energetic consistency, and its influence on the approximate Nusselt number, seems to be relatively unaddressed.  
		
		In order to systematically examines the effects of energetic consistency, we first determine an orthonormal basis for Fourier-Galerkin truncations in the present setting (Proposition \ref{prop:DivFreeVectorBasis}) which we enumerate in terms of index vectors, essentially the choice of wave-vectors in the model.  With this basis we derive a set of general formulae \eqref{GeneralBoussinesqODE_Vel} - \eqref{def_SignCoefs} by which one can explicitly generate the ODE problem found by projecting \eqref{BoussinesqCoriolis} onto any desired finite selection of modes.  We confirm that any such truncated ODE possesses global-in-time solutions, which is required for defining the Nusselt number, and we show that for sufficiently small Rayleigh number the conducting state is always globally attractive (Lemma \ref{lem:ODE_Existence}).  However, potential energy balance does not hold in general, and the Nusselt number can in general be infinite.  We give a precise characterization of models for potential energy balance holds (Proposition \ref{prop:BalancePreservation}), and show that the potential energy balance implies the existence of a compact attractor for these ODE's, and hence a bound on the Nusselt number (Lemma \ref{lem:TruncatedNusseltBound}). We additionally identify hierarchies for which the vorticity balance from the PDE holds (Proposition \ref{prop:VortPreservation}).

		\begin{figure}[H]
			\begin{center}
				\includegraphics[height=70mm]{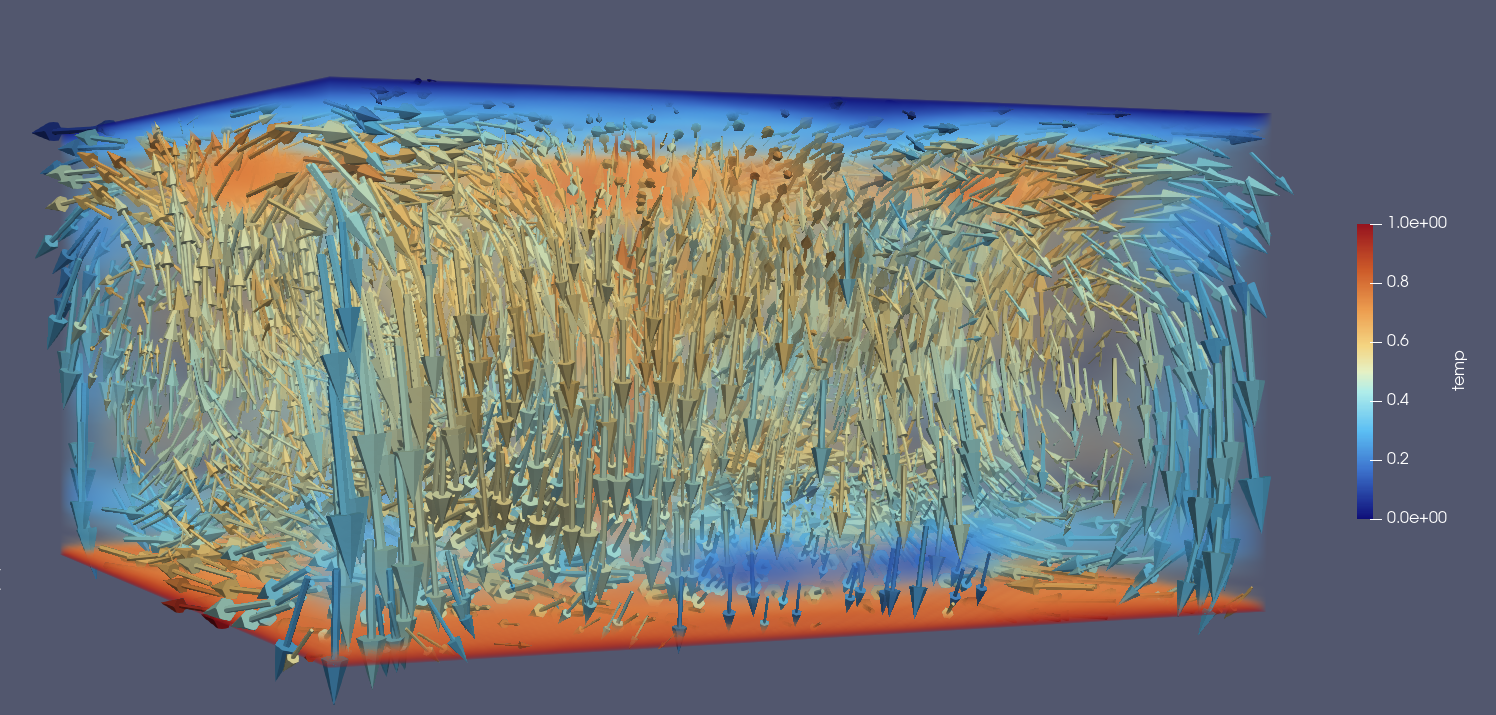}
				\caption{For $\Pra =10$, $\ShaOne = \ShaTwo = 1$, a depiction of a fluid flow from the $\truncInd = (1,5)$ model with $\Ray = 2000, \Rot = 0$, with the color indicating the temperature and the arrows indicating the velocity.  Full video available at \url{https://www.youtube.com/watch?v=qCHgRDqIOGM}.}
				\label{fig:FluidVisualization}
			\end{center}
		\end{figure}  
		
		Several energetically consistent hierarchies of truncated ODE models are then constructed, where one ascends a given hierarchy by adjoining more Fourier modes, obtaining higher spatial resolution by considering a higher dimensional ODE.  These hierarchies are implemented in MATLAB, with code made available on GitHub at \url{https://github.com/rkwelter/RRBC_SpectralCodeRepo}.  As a basic illustration, Figure \ref{fig:FluidVisualization} displays a fluid flow computed from one of these models.  These energetically consistent models are used to compute approximations of $\mathsf{Nu}$ for relatively modest values of the Rayleigh number and Coriolis numbers.  Indeed while these models are numerically stable even for very large parameter values, we find that low dimensional models fail to accurately represent the PDE as the parameter values become larger (Figures \ref{fig:HeatTransport_Linf} - \ref{fig:HeatTransport_HKC}).  Although this implementation can easily handle moderate spatial resolution (i.e. ODE's of dimension up to around 1000), it is not suitable for very high resolution numerical computations since the ODE is encoded explicitly through a MATLAB file that becomes very large for increasing dimension.  This can likely be optimised, but this is not our goal here.  We view the truncations as finite dimensional reduction that are amenable to  analysis for criteria of energetic consistency and the qualitative impact of energetic inconsistency, while still ranging to sufficient complexity for turbulent dynamics. 
		
		In contrast, truncations that violate the potential energy balance, such as the standard ``Fourier-box''-models (containing all wave vectors with wave length above a given value), have less controlled dynamics and can indeed generate unphysical infinite Nusselt number. It has been noted in \cite{HowardKrishnamurti_1986,ThiffeaultHorton_1996,OlsonDoering2022} for a 6-dimensional truncation that `runaway-modes' can occur, i.e. Fourier-modes which grow exponentially and unboundedly. These have also been found in various other fluid models on the PDE level, in particular with kinetic energy backscatter parameterisation \cite{Holst2024,Prugger02112022}, which immediately implies their occurrence in Fourier-Galerkin truncations. Such solutions might provide one potential explanation for the frequently reported problems with numerical blow-up in practice for climate models. 
		
		In the present context, on the one hand, we prove that the appearance of runaway modes is a widespread property of energetically inconsistent truncations (Theorem \ref{thm:runaway}), referred to as type-one inconsistent, and stems from one-dimensional invariant subspaces with linear dynamics. This in particular occurs in the box-truncation for all medium to small scale modes, and for any Rayleigh number so large that the mode in question is linearly unstable. Since the 2D Boussinesq equations are known to feature a compact global attractor, this highlights that there are many arbitrarily high dimensional reductions that do not respect this basic structure of the PDE dynamics, and are thus not necessarily good approximations of the long-term dynamics of the PDE. On the other hand, we prove that for any box-like truncation, the ODE possesses a compact global attractor if the Rayleigh number is below a threshold that is proportional to the box size (Lemma \ref{lem:inconsRay}). Hence, the runaway can be suppressed by including sufficiently many additional modes. Concerning other inconsistency, we refer to inconsistent truncations that do not admit such one dimensional runaway modes as type-two inconsistent, and find that these  can behave differently. For such a 5-dimensional truncation we prove the existence of a compact global attractor for any Rayleigh number (Lemma \ref{lem:type2sample}). Hence, also the Nusselt number is bounded.  Finally, we study the effects of energetic inconsistency also numerically.  Although we prove that runaway modes occur on invariant subspaces, it is unclear whether this unbounded growth occurs also for more generic initial conditions.  We find for several Fourier-box models, solutions with random initial conditions seem to exhibit bounded behavior, but the effects of inconsistency can be seen via wildly large amplitudes and fluctuations.  For low dimensional inconsistent truncations the resulting Nusselt numbers differ by orders of magnitude from comparable energetically consistent models (Figures \ref{fig:FourierBox_L2Norms}, \ref{fig:FourierBox_HeatTransport}).  These effects are mitigated for higher dimensional truncations, but seem to remain considerably large.  It is perhaps more insidious that the apparent errors remain large but are not so dramatically obvious in higher resolution models.

		The paper is organised as follows. In the remainder of the introduction we fix some notation and cast \eqref{BoussinesqCoriolis} as an evolution equation. In \S\ref{sec:GeneralSpectralModels} we derive general Fourier-Galerkin truncations and note some general properties, while in \S\ref{s:energyconsistent} we characterize energetically consistent truncations and derive bounds, in particular for the Nusselt number.  Section \ref{sec:EnergyInconsis} analyses effects of energetic inconsistency and corresponding bounds for the Rayleigh number. Finally, \S\ref{s:numerics} concerns numerical implementation and simulation results. The paper ends with a conclusion in \S\ref{s:conclusion}.

		\subsection{Notation}
		
		In this paper, scalars are written in normal font, vectors $\textbf{v}$ in bold, matrices $\mathcal{M}$ and operators in calligraphic, parameters $\mathsf{P}$ in serif and sets $\mathscr{S}$ in script.  Generally subscripts will be used to denote the component of a vector or matrix, and superscripts will be used to indicate indexed quantities.  There will be a couple of exceptions to this rule, namely that some quantities will be indexed by the symbols $\textbf{u}, \theta$ as subscripts to indicate a different definition for the two different variables, and the subscript zero (e.g. $\textbf{u}_0$) will be used to indicate initial conditions.  This should not cause confusion, since these could not indicate a component of a vector or matrix.
		
		$C^{k}$, $W^{k,p}$, $H^k = W^{k,2}$ and $L^p$ will denote the space of $k$ times differentiable functions, the Sobolev spaces, the Hilbert Sobolev spaces and the Lebesgue spaces respectively.  Unless otherwise specified these will be used to denote functions defined on $\Omega$, and when referring to functions defined on another domain $\Omega'$, the notation $C^k(\Omega')$ and so on will be used.  For all functions defined on $\Omega$ assume periodicity in $x_1$ and $x_2$, and let $C_0^{k}$ be the corresponding subspace of functions which are zero on the vertical boundaries, and $W_0^{k,p}$, $H_0^k$, $L_0^p$ be the closure of $C^k_0$ in $W^{k,p}$, $H^k$, $L^p$.  Similarly let $\textbf{C}^{k}_{\sigma}$ be the space of incompressible vector fields $\textbf{u} = (u_1,u_2,u_3)^T$ with components in $C^{k}$ and satisfying the boundary conditions \eqref{BC_Velocity}, and let $\textbf{W}^{k,p}_{\sigma}$, $\textbf{H}^{k}_{\sigma}$, $\textbf{L}^{k}_{\sigma}$ denote the closure of $\textbf{C}^{\infty}_{\sigma}$ in $W^{k,p}$, $H^k$ and $L^p$, respectively.  Since one can translate to a moving frame if necessary, without loss of generality one can take these vector valued function spaces such that $u_1$, $u_2$ have zero mean.  Finally, for a Banach space $X$, let $C^{k}( [0,T]; X)$ denote the space of mappings $f:[0,T]\mapsto X$ such that $t \mapsto \|f(t)\|_X$ belongs to $C^k([0,T])$, and so on for the spaces $W^{k,p}( [0,T]; X)$, $H^k( [0,T]; X)$ and $L^p( [0,T]; X)$.
		\[ \mathbb{Z}_\theta:=\mathbb{Z}_{\geq 0}^2 \times \mathbb{Z}_{>0} \hspace{.5 cm}  \text{ , } \hspace{.5 cm} \mathbb{Z}_\textbf{u}:=\mathbb{Z}_{\geq 0}^3\setminus \{\textbf{0}\} . \]
		
		For a vector $\textbf{v} \in \mathbb{R}^d$ the notation $|\textbf{v}|_{p} = (\sum_i v_i^p)^{1/p}$ will be used and $|\textbf{v}| = |\textbf{v}|_{2}$.  The notation $\textbf{x}_h = (x_1,x_2)$ will be used for the horizontal spatial variables.  Let $\mathcal{K} = \mathsf{diag}[\ShaOne,\ShaTwo,\ShaThree]$ be the diagonal matrix with $\ShaOne,\ShaTwo,\ShaThree$ on the diagonal.  In the context of a wave vector $\textbf{m} \in \mathbb{Z}^3_{\geq 0}$, we will use $\iota^{m_k}, \eta^{\textbf{m}}, V$ to denote normalizing constants defined by 
		\begin{equation} \label{Normalizers} \iota^{m_k} := \left \{ \begin{array}{ll} 0 & \text{ if } m_k = 0 \\ 1 & \text{ if } m_k > 0 \end{array} \right .  \hspace{.5 cm} \text{ , } \hspace{.5 cm}  \eta^{\textbf{m}} =  \prod_{k\leq 3} (\sqrt{2})^{\iota^{m_k}} \hspace{.5 cm} \text{ , } \hspace{.5 cm} V := \big ( 4\pi^3 \big )^{1/2} . \end{equation}
		Since $m_k\geq0$, $\iota^{m_k} = \mathrm{sgn}(m_k)$. We will also make use of indicator functions as follows:
		\[ \delta^{i,j} := \left \{ \begin{array}{cc} 1 & \text{ if } i=j , \\ 0 & \text{ if } i\neq j , \end{array} \right .  \hspace{1 cm} \delta^{\textbf{v},\tilde{\textbf{v}}} := \Pi_{i=1}^d  \delta^{v_i,\tilde{v}_i} , \hspace{1 cm} \chi^{m,m'} := \left \{ \begin{array}{cc} 0 & \text{ if } m = 0 \text{ or } m' = 0 , \\ 1 & \text{ otherwise. } \end{array} \right . \]
		$\parameters = (\Ray,\Rot,\Pra, \ShaOne,\ShaTwo)$ will denote the vector of parameters for \eqref{BoussinesqCoriolis}, and $\parameters$ will be said to be admissible if $\Ray, \Rot \geq 0$, $\Pra, \ShaOne, \ShaTwo > 0$.
		
		\subsection{Reformulation as an evolution equation}
		
		It is desirable to write the system \eqref{BoussinesqCoriolis} as an evolution equation on a function space.  To this end, define $\theta$ to be the temperature deviation from the pure conduction profile:
		\[ \theta := \frac{\pi}{\ShaThree} (T - 1) + x_3 . \]
		Furthermore, let $\mathcal{P}$ denote the Leray projector, which projects a vector field \textbf{v} onto its divergence free part.  By changing variables to $\theta$ in \eqref{BoussinesqCoriolis} and applying the Leray projector $\mathcal{P}: (L^2)^3 \mapsto \textbf{L}^2_{\sigma}$ to the momentum equation, one can write \eqref{BoussinesqCoriolis} as an evolution equation:
		\begin{equation} \label{EvolutionPDE} \begin{split}
				\partial_{t} \textbf{u} & = \Pra \Delta \textbf{u} + \Pra \Ray \mathcal{P} \big [ \theta \uVecThree \big ] - \Pra \Rot \mathcal{P} \big [ \uVecThree \times \textbf{u} \big ] - \mathcal{P} \big [ \textbf{u} \cdot \nabla \textbf{u} \big ] \\
				\partial_t \theta & =  \Delta \theta + u_3 - \textbf{u}\cdot \nabla \theta .
		\end{split} \end{equation}
		The velocity field $\textbf{u}$ should have finite kinetic energy, hence a natural choice of function space for the velocity is $\textbf{L}^2_{\sigma}$, and the function $\theta$ is equal to zero at both boundaries, hence a natural choice of function space is $L^2_0$.
		
		
		\section{General spectral projections of the Boussinesq-Coriolis system}
		
		\label{sec:GeneralSpectralModels}
		
		\subsection{Model construction}
		
		First, we lay out the projection process by which one obtains an ODE reduction of \eqref{EvolutionPDE} given an arbitrary finite selection of Fourier modes.  For maximal generality this process is described abstractly using a condensed notation, and in subsequent sections we provide concrete examples for clarity.  We start by looking for a general expansion of the form
		\begin{equation} \label{GeneralFourierExpansion} \begin{split} \textbf{u}(\textbf{x},t) = \sum_{\textbf{n}} u^{\textbf{n}}(t) \textbf{v}^{\textbf{n}}(\textbf{x}) \hspace{.5 cm} \text{ , } \hspace{.5 cm} \theta(\textbf{x},t) = \sum_{\textbf{n}} \theta^{\textbf{n}}(t) f^{\textbf{n}}(\textbf{x}) . \end{split} \end{equation}
		where $u^{\textbf{n}}(t),\theta^{\textbf{n}}(t)$ are Fourier coefficients corresponding to some sinusoidal functions $\textbf{v}^{\textbf{n}}(\textbf{x}),f^{\textbf{n}}(\textbf{x})$ yet to be defined.  One could naively choose the functions $\textbf{v}^{\textbf{n}}(\textbf{x}),f^{\textbf{n}}(\textbf{x})$ to be complex exponentials typical for Fourier expansions, but eventually the nonlinear analysis becomes much messier due to the complicated implicit relationships between the Fourier coefficients required to enforce the real-valued condition, boundary conditions and divergence free conditions.  Since we avoid complex exponentials, let us briefly motivate the choice of the definitions of $\textbf{v}^{\textbf{n}}(\textbf{x}),f^{\textbf{n}}(\textbf{x})$ by recalling some basic facts about Fourier expansions in one dimension.  In this case a real-valued, periodic function $f(x)$ defined on $[0,2\pi]$ can be expanded in terms of Fourier coefficients $\hat{f}^{m,p}$ via
		\[ f(x) = \frac{1}{(2\pi)^{1/2}} \hat{f}^{0,0} + \sum_{m \in \mathbb{Z}_{>0}} \hat{f}^{m,0} \frac{\cos(m x)}{(\pi)^{1/2}} + \hat{f}^{m,1} \frac{\sin(m x_1)}{(\pi)^{1/2}} . \]
		Here we see that using sinusoids requires that the Fourier coefficients be indexed by two quantities, namely a wave number $m$ and a phase $p$, describing whether the coefficient corresponds to cosine or sine.  Furthermore, the number of phases depends on the wave number, e.g. for $m = 0$ there is only one phase, corresponding to $\cos(0) = 1$, whereas $\sin(0) = 0$ is not included in the expansion.  For this reason, we define the phase index set 
		\[ \mathscr{P}^{m} := \left \{ \begin{array}{cc}
			\{ 0 \} & \text{ if } m = 0 \text{ , } \\
			\{0,1\} & \text{ if } m > 0 \text{ . } 
		\end{array} \right .   \]
		Since one has $\frac{d}{dx} \cos (x) = -\sin(x)$, $\frac{d}{dx} \sin (x) = \cos(x)$ it is natural to take the phase indices mod 2.  Finally, for a Fourier expansion of a vector field one must additionally specify which component of the vector is being expanded, which is done here by using a "component" index $c$.  For a $d$-dimensional vector field, one expects that $c$ can take $d$ values, but the components of a divergence-free vector field cannot be chosen independently, so $c$ can take at most $d-1$ values. 
		
		With this in mind, the index $\textbf{n} = (m_1,m_2,m_3,p_1,p_2,c)$ will consist of a wave vector $\textbf{m}=(m_1,m_2,m_3)$, a phase vector $\textbf{p}= (p_1,p_2)$ and a component index $c$.  We will slightly abuse the notation and write $\textbf{n} = (\textbf{m},\textbf{p},c)$ at times.  For a wave number $m$ and phase $p$, we use the following shorthand for the sinusoidal functions:
		\begin{equation} \label{SinFuncs} s^{m,0}(x) := \cos(m x ) \hspace{.5 cm} \text{ , } \hspace{.5 cm}  s^{m,1}(x) := \sin(mx ) \text{ . } \end{equation}
		As mentioned, all calculations involving the phase indices $p_1,p_2$ are understood to be mod $2$, which enables one to concisely write the derivatives for both phases via 
		\begin{equation} \label{SinusoidDerivative} \frac{d}{dx} s^{m,p} = (-1)^{p+1} m s^{m,p+1} \text{ . } \end{equation}
		The space $L^2_0$ consists of scalar functions, hence one can just use the usual Fourier expansion satisfying the Dirichlet boundary conditions for $\theta$:
		\begin{equation} \label{def:ThetaFourierBasis} f^{(\textbf{m},\textbf{p},c)}(\textbf{x}) := \frac{\eta^{\textbf{m}}}{V} s^{m_1,p_1}(\ShaOne x_1) s^{m_2,p_2}(\ShaTwo x_2) s^{m_3,1} (\ShaThree x_3)  . \end{equation} 
		Since $\theta$ is a scalar, $f^{(\textbf{m},\textbf{p},c)}$ is defined only for $c = 1$, but adding $c$ will provide a uniform notation with the velocity expansion.  As described above, a sinusoid can equal the zero function when one of the wave numbers $m_j$ is equal to zero.  We must exclude zero functions from our basis, hence we define the following set $\mathscr{N}_{\theta}$ of admissible indices:
		\[ \mathscr{P}^{\textbf{m}}_{\theta} := \mathscr{P}^{m_1} \times \mathscr{P}^{m_2} \hspace{.5 cm} \text{ , } \hspace{.5 cm} \mathscr{N}_{\theta} := \{ \textbf{n} : \textbf{m} \in \mathbb{Z}_{\theta} \text{ , } \textbf{p} \in \mathscr{P}^{\textbf{m}}_{\theta} \text{ , }  c = 1 \} . \]
		Henceforth, whenever referring to $f^{\textbf{n}}(\textbf{x})$ or the corresponding Fourier coefficient it is always assumed that $\textbf{n} \in \mathscr{N}_{\theta}$.
		
		Turning now to the space $\textbf{L}^2_{\sigma}$ for the velocity field, first note that due to the Galilean invariance of the Navier-Stokes equations one can assume each component of the velocity field is mean free, hence one need not consider $\textbf{m} = \textbf{0} $.  We let $\boldsymbol{\gamma} := (1,1,-1)$, and for given $\parameters$, $\textbf{m} \in \mathbb{Z}_{\textbf{u}}^3$, define the matrices $\mathcal{G}^{\textbf{n}}$ and the normalizing vectors $\normVec^{\textbf{n}}$ via
		\begin{align} \label{VectorFieldDef_MatsVecs} \mathcal{G}^{(\textbf{m},\textbf{p},1)} := \mathsf{diag} \big [ \boldsymbol{\gamma} \times \mathcal{K}\textbf{m} \big ] \hspace{.5 cm} \text{ , } \hspace{.5 cm} \mathcal{G}^{(\textbf{m},\textbf{p},2)} := \mathsf{diag} \big [ \mathcal{K} \textbf{m} \times \big ( \boldsymbol{\gamma} \times \mathcal{K}\textbf{m} \big ) \big ] \hspace{.5 cm} \text{ , } \hspace{.5 cm} \nu_i^{\textbf{n}} := \prod \nolimits_{j \leq 3} \big ( \iota^{m_j} \big )^{p_j + \delta^{i,j}}  \text{ , } \end{align}
		where we use the dummy phase variable $p_3 := 0$ and the convection $0^0 = 1$ for $\nu_i^{\textbf{n}}$.  We note $\nu_i^{\textbf{n}}\in \{0,1\}$ and it is zero if $m_j=0$ and $p_j+\delta^{i,j}>0$ for some $j=1,2,3$.
		Define also the sign functions
		\[ \varsigma_{1,1}^{\textbf{p}} := (-1)^{p_1} \hspace{.5 cm} \text{ , } \hspace{.5 cm} \varsigma_{2,2}^{\textbf{p}} := (-1)^{p_2} \hspace{.5 cm} \text{ , } \hspace{.5 cm} \varsigma_{3,3}^{\textbf{p}} := 1 \hspace{.5 cm} \text{ , } \hspace{.5 cm} \varsigma_{i,j}^{\textbf{p}} := - \varsigma_{i,i}^{\textbf{p}} \hspace{.25 cm} \text{ for } j \neq i \text{ . } \]
		One can then define the following collection of vector fields $\textbf{v}^{\textbf{n}}(\textbf{x}) $:
		\begin{equation} \label{VectorFieldDef_FreeSlip}  v_i^{\textbf{n}}(\textbf{x}) := \frac{\eta^{\textbf{m}}}{|\mathcal{G}^{\textbf{n}}\normVec^{\textbf{n}}|V} \mathcal{G}^{\textbf{n}}_{i,i} \varsigma_{i,i}^{\textbf{p}} \prod \nolimits_{j \leq 3}   s^{m_j,p_j + \delta^{i,j}}(\mathsf{k}_{j} x_j) \text{ . }  \end{equation}
		The explicit forms of these vector fields are lengthier to write, but are provided in \ref{app:VectorFieldDef_Explicit} for clarity.  By comparing \eqref{VectorFieldDef_MatsVecs}, \eqref{VectorFieldDef_FreeSlip} one sees that $\nu_i^{\textbf{n}}$ indicates whether the sinusoid in $v_i^{\textbf{n}}$ is the zero function, i.e. $v_i^{\textbf{n}}(\textbf{x}) = 0$ if $\nu_i^{\textbf{n}} = 0$.  In order to avoid the case when all components of $\textbf{v}^{\textbf{n}}(\textbf{x})$ are zero functions, as well as ensure linear independence, we define
		\begin{equation} \label{PhaseIndexSets} \mathscr{P}^{\textbf{m}}_{\textbf{u}} := \left \{ \begin{array}{cc}
				\{0,1\} \times \{ 1 \} & \text{ if } m_1 > 0 \text{ , } m_2 , m_3 = 0 \text{ , }   \\
				\{ 1 \} \times \{0,1\}  & \text{ if } m_2 > 0 \text{ , } m_1 , m_3 = 0 \text{ , }   \\
				\{(0,1),(1,0) \} & \text{ if } m_3 > 0 \text{ , } m_1 , m_2 = 0 \text{ , }   \\
				\{0,1\}^2 & \text{ otherwise, } 
			\end{array} \right . \hspace{.5 cm}  \mathscr{C}^{\textbf{m}}_{\textbf{u}} := \left \{ \begin{array}{cc}
				\{1,2\} & \text{ if } \textbf{m} \in \mathbb{Z}^3_{> 0} \text{ , } \\
				\{1\}  & \text{ otherwise, }
			\end{array} \right .  \end{equation}
		and we can then define the set $\mathscr{N}_{\textbf{u}}$ of admissible indices
		\[ \mathscr{N}_{\textbf{u}} := \{ \textbf{n} : \textbf{m} \in \mathbb{Z}_{\textbf{u}} \text{ , } \textbf{p} \in \mathscr{P}^{\textbf{m}}_{\textbf{u}} \text{ , } c \in \mathscr{C}_{\textbf{u}}^{\textbf{m}} \} .  \]
		Henceforth, whenever referring to $\textbf{v}^{\textbf{n}}(\textbf{x})$ or the corresponding Fourier coefficient it is always assumed that $\textbf{n} \in \mathscr{N}_{\textbf{u}}$.  
		
		In the following proposition, it is verified that this collection of functions $f^{\textbf{n}}$ and vector fields $\textbf{v}^{\textbf{n}}$ is precisely the desired complete Fourier basis satisfying the boundary conditions and divergence free condition, and several inner products involving the basis elements are computed.  The proof is given in Appendix \ref{app:FourierExpDeriv}. 
		
		\begin{Proposition}
			\label{prop:DivFreeVectorBasis}
			For each ${\normalfont \textbf{n} \in \mathscr{N}_{\textbf{u}}}$, the vector fields ${\normalfont \textbf{v}^{\textbf{n}} }$ are divergence-free and satisfy the boundary conditions \eqref{BC_Velocity}.  The sets ${\normalfont \{ f^{\textbf{n}} \}_{\textbf{n} \in \mathscr{N}_{\theta}}}$, ${\normalfont \{ \textbf{v}^{\textbf{n}} \}_{\textbf{n} \in \mathscr{N}_{\textbf{u}}}}$ form an orthonormal basis for the spaces $L^2_{0}, \textbf{L}^2_{\sigma}$ respectively.  These basis elements satisfy the following inner product relations:
			\begin{equation} \label{BasisRelations} \begin{split}
					{\normalfont \langle f^{\textbf{n}} f^{\tilde{\textbf{n}}} \rangle = \delta^{\textbf{n},\tilde{\textbf{n}}} } \hspace{.25 cm} & \text{ , } \hspace{.25 cm} {\normalfont \langle \textbf{v}^{\textbf{n}} \cdot \textbf{v}^{\tilde{\textbf{n}}} \rangle = \delta^{\textbf{n},\tilde{\textbf{n}}} } \hspace{.25 cm} \text{ , } \hspace{.25 cm} {\normalfont \langle \textbf{v}^{\textbf{n}} \cdot (f^{\tilde{\textbf{n}}} \uVecThree ) \rangle } = {\normalfont \frac{\mathcal{G}_{3,3}^{\textbf{n}}\nu_3^{\textbf{n}}}{|\mathcal{G}^{\textbf{n}} \boldsymbol{\nu}^{\textbf{n}}|}  \delta^{\textbf{m},\tilde{\textbf{m}}} \delta^{\textbf{p},\tilde{\textbf{p}}} \text{ , } } \\ 
					{\normalfont \langle \big ( \uVecThree \times \textbf{v}^{\tilde{\textbf{n}}} \big ) \cdot \textbf{v}^{\textbf{n}} \rangle } & = {\normalfont (-1)^{|\textbf{p}|_1} \frac{ \delta^{\textbf{m},\tilde{\textbf{m}}} \delta^{p_1+1,\tilde{p}_1} \delta^{p_2+1,\tilde{p}_2}}{|\mathcal{G}^{\tilde{\textbf{n}}} \boldsymbol{\nu}^{\tilde{\textbf{n}}}||\mathcal{G}^{\textbf{n}} \boldsymbol{\nu}^{\textbf{n}}| } \big ( \mathcal{G}_{2,2}^{\tilde{\textbf{n}}} \mathcal{G}_{1,1}^{\textbf{n}} \nu_1^{\textbf{n}} - \mathcal{G}_{1,1}^{\tilde{\textbf{n}}} \mathcal{G}_{2,2}^{\textbf{n}} \nu_2^{\textbf{n}} \big ) \text{ . } }
			\end{split}  \end{equation}
		\end{Proposition}
		
		The above basis can be used to construct reduced order models via Galerkin truncation of the full PDE model.  In general, this can be done by considering expansions of the form
		\begin{equation} \label{FiniteFourierExpansion} \textbf{u}^{\truncInd}(\textbf{x},t) = \sum_{\textbf{n} \in \mathscr{N}_{\textbf{u}}^{\truncInd}} u^{\truncInd,\textbf{n}}(t) \textbf{v}^{\textbf{n}}(\textbf{x}) \hspace{.5 cm} \text{ , } \hspace{.5 cm} \theta^{\truncInd}(\textbf{x},t) = \sum_{\textbf{n} \in \mathscr{N}_{\theta}^{\truncInd}} \theta^{\truncInd,\textbf{n}}(t) f^{\textbf{n}}(\textbf{x}) . \end{equation}
		in which the truncated index sets $\mathscr{N}^{\truncInd}_{\textbf{u}} \subset \mathscr{N}_{\textbf{u}}$, $\mathscr{N}^{\truncInd}_{\theta} \subset \mathscr{N}_{\theta}$ are finite, and the truncation index $\truncInd$ describes which specific truncation is made.  For now, we leave the choice of truncation unspecified and focus on the derivation of a reduced model given arbitrary index sets $\mathscr{N}^{\truncInd}_{\textbf{u}}$, $\mathscr{N}^{\truncInd}_{\theta}$. 
		%
		%
		Define the projection operators $\mathcal{P}_{\textbf{u}}^{\truncInd},\mathcal{P}_{\theta}^{\truncInd}$ associated to these index sets via
		\[ \mathcal{P}_{\textbf{u}}^{\truncInd} \big [ \textbf{v} \big ] = \sum_{\textbf{n} \in \mathscr{N}_{\textbf{u}}^{\truncInd}} \langle \textbf{v} \cdot \textbf{v}^{\textbf{n}} \rangle \textbf{v}^{\textbf{n}}(\textbf{x}) \hspace{.5 cm} \text{ , } \hspace{.5 cm} \mathcal{P}_{\theta}^{\truncInd} \big [ g \big ] = \sum_{\textbf{n} \in \mathscr{N}_{\theta}^{\truncInd}} \langle g f^{\textbf{n}} \rangle f^{\textbf{n}}(\textbf{x}) . \] 
		We informally note that as the index sets $\mathscr{N}^{\truncInd}_{\textbf{u}},\mathscr{N}^{\truncInd}_{\theta}$ tend toward the full admissible sets $\mathscr{N}_{\textbf{u}},\mathscr{N}_{\theta}$ the projection operators $\mathcal{P}_{\textbf{u}}^{\truncInd},\mathcal{P}_{\theta}^{\truncInd}$ converge to the Leray projector and the identity operator on $L^2_0$, respectively.  
		
		With these preparations we can formulate the truncated model obtained by projecting of the full problem \eqref{EvolutionPDE} onto this basis, as follows
		\begin{equation} \label{TruncatedEvolutionPDE} \begin{split}
				\partial_{t} \textbf{u}^{\truncInd} & = \Pra \Delta \textbf{u}^{\truncInd} + \Pra \Ray \mathcal{P}^{\truncInd}_{\textbf{u}} \big [ \theta^{\truncInd} \uVecThree \big ] - \Pra \Rot \mathcal{P}^{\truncInd}_{\textbf{u}} \big [ \uVecThree \times \textbf{u}^{\truncInd} \big ] - \mathcal{P}^{\truncInd}_{\textbf{u}} \big [ \textbf{u}^{\truncInd} \cdot \nabla \textbf{u}^{\truncInd} \big ] , \\
				\partial_t \theta^{\truncInd} & =  \Delta \theta^{\truncInd} + \mathcal{P}^{\truncInd}_{\theta} \big [ u_3^{\truncInd} \big ] - \mathcal{P}^{\truncInd}_{\theta} \big [ \textbf{u}^{\truncInd} \cdot \nabla \theta^{\truncInd} \big ].
		\end{split} \end{equation}
		Note that solutions of this truncated PDE do not solve the original PDE \eqref{EvolutionPDE}, hence the superscript notation $\textbf{u}^{\truncInd},\theta^{\truncInd}$, but rather one aims for an approximation of \eqref{EvolutionPDE} via \eqref{TruncatedEvolutionPDE}.  This truncated PDE can also be viewed as a system of ODE's for the finite set of Fourier coefficients, where the equation for $u^{\textbf{n},\truncInd}$ can be recovered by computing the inner product of the first equation with $\textbf{v}^{\textbf{n}}$, and the equation for $\theta^{\textbf{n},\truncInd}$ can be recovered by computing the inner product of the second with $f^{\textbf{n}}$.  For each $\textbf{n} \in \mathscr{N}_{\textbf{u}}^{\truncInd}$ one thereby obtains an equation of the following form:
		\begin{equation} \label{GeneralBoussinesqODE_Vel} \frac{d}{dt} u^{\truncInd,\textbf{n}} = - \Pra |\mathcal{K}\textbf{m}|^2 u^{\truncInd,\textbf{n}} + \Pra \Ray \frac{\mathcal{G}_{3,3}^{\textbf{n}}\nu_3^{\textbf{n}}}{|\mathcal{G}^{\textbf{n}} \boldsymbol{\nu}^{\textbf{n}}|} \theta^{\truncInd,(\textbf{m},\textbf{p},1)} + (-1)^{|\textbf{p}|_1} \Pra \Rot \sum_{\substack{\tilde{\textbf{m}} = \textbf{m} \\ \tilde{\textbf{p}} = \textbf{p} + (1,1) \\ \tilde{c} \in \mathscr{C}^{\textbf{m}}_{\textbf{u}}}} \frac{\mathcal{G}_{1,1}^{\tilde{\textbf{n}}} \mathcal{G}_{2,2}^{\textbf{n}} \nu_2^{\textbf{n}} - \mathcal{G}_{2,2}^{\tilde{\textbf{n}}} \mathcal{G}_{1,1}^{\textbf{n}} \nu_1^{\textbf{n}} }{|\mathcal{G}^{\tilde{\textbf{n}}} \boldsymbol{\nu}^{\tilde{\textbf{n}}}||\mathcal{G}^{\textbf{n}} \boldsymbol{\nu}^{\textbf{n}}| } u^{\truncInd,\tilde{\textbf{n}}} - N_{\textbf{u}}^{\truncInd,\textbf{n}} \text{ , } \end{equation} 
		and for each $\textbf{n} \in \mathscr{N}_{\theta}^{\truncInd}$ one obtains an equation of the following form: 
		\begin{equation} \label{GeneralBoussinesqODE_Temp} \frac{d}{dt} \theta^{\truncInd,\textbf{n}} = - |\mathcal{K}\textbf{m}|^2 \theta^{\truncInd,\textbf{n}} + \sum_{\substack{\tilde{\textbf{m}} = \textbf{m} \\ \tilde{\textbf{p}} = \textbf{p} \\ \tilde{c} \in \mathscr{C}^{\textbf{m}}_{\textbf{u}}}} \frac{\mathcal{G}_{3,3}^{\tilde{\textbf{n}}}\nu_3^{\tilde{\textbf{n}}}}{|\mathcal{G}^{\tilde{\textbf{n}}} \boldsymbol{\nu}^{\tilde{\textbf{n}}} |}  u^{\truncInd,\tilde{\textbf{n}}} - N_{\theta}^{\truncInd,\textbf{n}} \text{ . } \end{equation}
		The linear terms in these equations are obtained directly from \eqref{BasisRelations}.  Abstractly, the nonlinear terms are given by 
		\begin{equation} \label{AbstractNonlinear} N_{\textbf{u}}^{\truncInd,\textbf{n}} = \big \langle \textbf{v}^{\textbf{n}} \cdot \big [ \big ( \textbf{u}^{\truncInd} \cdot \nabla \big ) \textbf{u}^{\truncInd} \big ] \big \rangle \hspace{.5 cm} \text{ , } \hspace{.5 cm} N_{\theta}^{\truncInd,\textbf{n}} = \big \langle f^{\textbf{n}} \cdot \big [ \textbf{u}^{\truncInd} \cdot \nabla \theta^{\truncInd} \big ] \big \rangle . \end{equation}
		However, to write \eqref{TruncatedEvolutionPDE} as an ODE system one must insert the expansions for $\textbf{u}^{\truncInd}$, $\theta^{\truncInd}$ to determine these sums explicitly in terms of the Fourier variables $u^{\truncInd,\textbf{n}},\theta^{\truncInd,\textbf{n}}$, namely one must find the time-independent constants $I_{\textbf{u}}^{\boldsymbol{\alpha}}, I_{\theta}^{\boldsymbol{\alpha}}$ such that
		\[ N_{\textbf{u}}^{\truncInd,\textbf{n}} = \sum_{ \textbf{n}' \in \mathscr{N}_{\textbf{u}}^{\truncInd} } \sum_{\textbf{n}'' \in \mathscr{N}_{\textbf{u}}^{\truncInd}} I^{\boldsymbol{\alpha}}_{\textbf{u}} u^{\truncInd,\textbf{n}'} u^{\truncInd,\textbf{n}''} \hspace{.5 cm} \text{ , } \hspace{.5 cm} N_{\theta}^{\truncInd,\textbf{n}} = \sum_{ \textbf{n}'\in \mathscr{N}_{\textbf{u}}^{\truncInd} } \sum_{\textbf{n}''\in \mathscr{N}_{\theta}^{\truncInd}}   I^{\boldsymbol{\alpha}}_{\theta} u^{\truncInd,\textbf{n}'} \theta^{\truncInd,\textbf{n}''} \text{ , } \]
		where the multi-index $\boldsymbol{\alpha} := (\textbf{n},\textbf{n}',\textbf{n}'')$ is used to indicate that $I_{\textbf{u}}^{\boldsymbol{\alpha}}, I_{\theta}^{\boldsymbol{\alpha}}$ depend on all of the indices involved in the triad interaction.  These constants are explicitly derived in Appendix \ref{app:NonlinearDerivation} by computing inner products of the basis elements inserted into \eqref{AbstractNonlinear}.  This derivation is somewhat involved, but much of the coming analysis only requires general properties of $I_{\textbf{u}}^{\boldsymbol{\alpha}}, I_{\theta}^{\boldsymbol{\alpha}}$ rather than their precise form, so the reader could skip this derivation on a first read.  However, for a complete definition of the ODE model we give the explicit formulas here.  First, we introduce some notation to concisely describe the triad interactions, where again the dummy phase variable $p_3 = 0$ is used:
		\begin{equation} \label{TriadNotation}
			\boldsymbol{\mu}^{k} := (m_k,m_k',m_k'') \hspace{.5 cm} \text{ for } \hspace{.5 cm} k = 1,2,3 \hspace{.5 cm} \text{ , } \hspace{.5 cm} \boldsymbol{\phi}^{k} := (p_k,p_k',p_k'') \hspace{.5 cm} \text{ for } \hspace{.5 cm} k = 1,2,3 \text{ . }  
		\end{equation}
		Due to the orthogonality properties of the sinusoids, it turns out that the constants $I_{\textbf{u}}^{\boldsymbol{\alpha}}, I_{\theta}^{\boldsymbol{\alpha}}$ are zero unless the wave number and phase triads $\boldsymbol{\mu}^k, \boldsymbol{\phi}^k$ satisfy certain compatibility conditions.  First, such non-trivial wave number triads must each satisfy convolution type conditions:
		\begin{equation} \label{WaveCompatibilityCond} \text{ For each } \hspace{.5 cm} k = 1,2,3 \hspace{.5 cm} \text{ either  }  \hspace{.5 cm} m_k = m_k' + m_k'' \hspace{.5 cm} \text{ or  }  \hspace{.5 cm} m_k = |m_k' - m_k''| \text{ . } \end{equation} 
		While the phase triads in general belong to $\{0,1\}^3$, it turns out that only phase triads satisfying the following compatibility conditions have non-zero interaction:
		\begin{equation} \label{PhaseCompatibilityCond}
			\text{ For each } \hspace{.5 cm} k = 1,2 \hspace{.5 cm} \text{ one must have  }  \hspace{.5 cm} p_k = p_k' + p_k'' \text{ . }
		\end{equation}
		Since only phases triads belonging to certain subspace of $\{0,1\}^3$ interact nonlinearly, we adopt the following notation
		\begin{equation} \label{AdmissiblePhaseTriads}
			\boldsymbol{\xi}^1 := (0,0,0) \hspace{.5 cm} \text{ , } \hspace{.5 cm} \boldsymbol{\xi}^2 := (0,1,1) \hspace{.5 cm} \text{ , } \hspace{.5 cm} \boldsymbol{\xi}^3 := (1,0,1) \hspace{.5 cm} \text{ , } \hspace{.5 cm} \boldsymbol{\xi}^4 := (1,1,0) \text{ . }
		\end{equation}
		Note \eqref{PhaseCompatibilityCond} can also be expressed as $\boldsymbol{\phi}^k \in \{ \boldsymbol{\xi}^j \}_{j=1}^4$.  We introduce next the following phase maps, which are defined to take a phase triad $\boldsymbol{\phi}^k$ and give the phases appearing for the $i,j^{th}$ component of the dot product in \eqref{AbstractNonlinear}:
		\begin{equation} \label{PhaseMaps} \pi^{i,j,k}(\boldsymbol{\phi}) := \boldsymbol{\phi} + (\delta^{j,k},\delta^{i,k},\delta^{i,k}+\delta^{j,k} ) \text{ . } \end{equation}
		If conditions \eqref{WaveCompatibilityCond}, \eqref{PhaseCompatibilityCond} are satisfied the constants $I_{\textbf{u}}^{\boldsymbol{\alpha}}, I_{\theta}^{\boldsymbol{\alpha}}$ are given 
		by 
		\begin{equation} \label{def:NonlinearCoefs} \begin{split} I^{\boldsymbol{\alpha}}_{\textbf{u}} = \sum_{i,j \leq 3} \frac{ 8 \mathcal{G}_{j,j}^{\textbf{n}} \mathcal{G}_{i,i}^{\textbf{n}'}\mathcal{G}_{j,j}^{\textbf{n}''} \varsigma^{\textbf{p}}_{j,j} \varsigma^{\textbf{p}'}_{i,i} \varsigma^{\textbf{p}''}_{j,j} \varsigma^{\textbf{p}''}_{i,j} \mathsf{k}_i m_i'' I_{i,j}^{\boldsymbol{\alpha}} }{ ( \eta^{\textbf{m}}\eta^{\textbf{m}'}\eta^{\textbf{m}''} ) |\mathcal{G}^{\textbf{n}}\boldsymbol{\nu}^{\textbf{n}}||\mathcal{G}^{\textbf{n}'}\boldsymbol{\nu}^{\textbf{n}'}||\mathcal{G}^{\textbf{n}''}\boldsymbol{\nu}^{\textbf{n}''}| V} \hspace{.5 cm} & \text{ , } \hspace{.5 cm} I^{\boldsymbol{\alpha}}_{\theta} = \sum_{i \leq 3} \frac{ 8\mathcal{G}_{i,i}^{\textbf{n}'} \varsigma^{\textbf{p}'}_{i,i} \varsigma^{\textbf{p}''}_{i,3} \mathsf{k}_i m_i'' I_{i,3}^{\boldsymbol{\alpha}} }{( \eta^{\textbf{m}}\eta^{\textbf{m}'}\eta^{\textbf{m}''} ) |\mathcal{G}^{\textbf{n}'}\boldsymbol{\nu}^{\textbf{n}'}| V} \text{ , } \end{split}  \end{equation}
		in which the terms $I^{\boldsymbol{\alpha}}_{i,j}$ are given in terms of amplitude and sign coefficients via:
		\begin{equation} \label{def:NonlinearCoefs2} I_{i,j}^{\boldsymbol{\alpha}} = \prod \nolimits_{k \leq 3} A^{(\boldsymbol{\mu}^{k},\pi^{i,j,k}(\boldsymbol{\phi}^k))} \sigma^{(\boldsymbol{\mu}^{k},\pi^{i,j,k}(\boldsymbol{\phi}^k))} \text{ . } \end{equation}
		Finally, the amplitude and sign coefficients are defined as follows:
		\begin{equation} \label{def_SignCoefs} \begin{split}
				A^{(\boldsymbol{\mu},\boldsymbol{\xi}^1)} := \frac{1}{2}(\eta^{\max(m,m',m'')})^2 \hspace{.5 cm} \text{ , } \hspace{.5 cm}  A^{(\boldsymbol{\mu},\boldsymbol{\xi}^2)} = \chi^{m',m''} & \hspace{.5 cm} \text{ , } \hspace{.5 cm} A^{(\boldsymbol{\mu},\boldsymbol{\xi}^3)} = \chi^{m,m''}\hspace{.5 cm} \text{ , } \hspace{.5 cm} A^{(\boldsymbol{\mu},\boldsymbol{\xi}^4)} = \chi^{m,m'} \text{ , } \\ 
				\sigma^{(\boldsymbol{\mu},\boldsymbol{\xi}^1)} := 1  \hspace{.5 cm} \text{ , } \hspace{.5 cm} \sigma^{(\boldsymbol{\mu},\boldsymbol{\xi}^2)} := S^{m,m',m''} & \hspace{.5 cm} \text{ , } \hspace{.5 cm} \sigma^{(\boldsymbol{\mu},\boldsymbol{\xi}^3)} := S^{m',m'',m} \hspace{.5 cm} \text{ , } \hspace{.5 cm} \sigma^{(\boldsymbol{\mu},\boldsymbol{\xi}^4)} := S^{m'',m,m'} \text{ , } \\
				\chi^{a,b} := \left \{ \begin{array}{cc}
					1 & \text{ if } a > 0 \text{ and } b > 0, \\
					0 & \text{ otherwise, } \end{array} \right . & \hspace{.5 cm } \text{ , } \hspace{.5 cm} S^{a_1,a_2,a_3} := \left \{ \begin{array}{cc}
					-1 & \text{ if } a_1 = a_2+a_3, \\
					1 & \text{ otherwise. } \end{array} \right . 
		\end{split} \end{equation}
		The above formulas \eqref{GeneralBoussinesqODE_Vel} - \eqref{def_SignCoefs} yield an algorithm to write down a system of ODE's for any choice of finite sets $\mathscr{N}_{\textbf{u}}^{\truncInd},\mathscr{N}_{\theta}^{\truncInd}$.  Indeed, this is included in the software provided; see \S\ref{s:code}.  For convenience, we let $\textbf{L}^{2,\truncInd}_{\sigma}$, $L^{2,\truncInd}_{0}$ be the spaces on which these ODE's are defined:
		\begin{equation} \label{TruncatedODESpaces} \textbf{L}^{2,\truncInd}_{\sigma}:=\mathcal{P}_{\textbf{u}}^{\truncInd} \textbf{L}^2_{\sigma} = \mathbb{R}^{|\mathscr{N}_{\textbf{u}}^{\truncInd}|} \hspace{.5 cm} \text{ , } \hspace{.5 cm} L^{2,\truncInd}_{0}:=\mathcal{P}_{\theta}^{\truncInd} L^2_{0}  = \mathbb{R}^{|\mathscr{N}_{\theta}^{\truncInd}|} \text{ . } \end{equation}
		Formulas \eqref{GeneralBoussinesqODE_Vel} - \eqref{def_SignCoefs} can be somewhat challenging to decipher, so for illustration we provide concrete examples of the construction procedure for very simple, low-dimensional truncated models in \ref{app:ConsProced}.  This procedure follows three steps: mode selection, evaluation of the linear coupling structure and evaluation of the nonlinear coupling structure.  For instance, with the following choice of index sets
		\[ \textbf{n}^1 = (1,0,1,0,0,1) \hspace{.25 cm} \text{ , } \hspace{.25 cm} \textbf{n}^2 = (0,0,2,0,0,1) \hspace{.25 cm} \text{ , } \hspace{.25 cm} \textbf{n}^3 = (1,0,1,1,1,1) \hspace{.25 cm} \text{ , } \hspace{.25 cm}  \mathscr{N}_{\textbf{u}}^{LS} = \{ \textbf{n}^1, \textbf{n}^3 \} \hspace{.25 cm} \text{ , } \hspace{.25 cm} \mathscr{N}_{\theta}^{LS} = \{ \textbf{n}^1, \textbf{n}^2 \} \text{ , }  \]
		one obtains the ODE model
		\begin{equation} \label{e:Lorenz63} \begin{split}
				\frac{d}{dt} u^{LS,\textbf{n}^1} & = - \Pra (\ShaOne^1 + \ShaThree^2) u^{LS,\textbf{n}^1} + \frac{\Pra \Rot \ShaThree}{\sqrt{\ShaOne^1 + \ShaThree^2}}  u^{LS,\textbf{n}^3} - \frac{\Pra \Ray \ShaOne}{\sqrt{\ShaOne^1 + \ShaThree^2}} \theta^{LS,\textbf{n}^1} \text{ , } \\
				\frac{d}{dt} u^{LS,\textbf{n}^3} & = - \Pra (\ShaOne^1 + \ShaThree^2) u^{LS,\textbf{n}^3} - \frac{\Pra \Rot \ShaThree}{\sqrt{\ShaOne^1 + \ShaThree^2}}  u^{LS,\textbf{n}^1} \text{ , } \\
				\frac{d}{dt} \theta^{LS,\textbf{n}^1} & = - (\ShaOne^1 + \ShaThree^2) \theta^{LS,\textbf{n}^1} - \frac{\ShaOne}{\sqrt{\ShaOne^1 + \ShaThree^2}} u^{LS,\textbf{n}^1} - \frac{\sqrt{2} \ShaOne \ShaThree}{\sqrt{\ShaOne^2 + \ShaThree^2}V} u^{LS,\textbf{n}^1}\theta^{LS,\textbf{n}^2} \text{ , } \\
				\frac{d}{dt} \theta^{LS,\textbf{n}^2} & = - 4\ShaThree^2 \theta^{LS,\textbf{n}^2} + \frac{\sqrt{2} \ShaOne \ShaThree}{\sqrt{\ShaOne^2 + \ShaThree^2}V} u^{LS,\textbf{n}^1}\theta^{LS,\textbf{n}^1} \text{ , }
			\end{split}   
		\end{equation}
		and by rescaling the variables and choosing $\ShaTwo = \frac{1}{\ShaOne}$ (i.e. $\ShaThree = 1$) one obtains the classical Lorenz-Stenflo model \cite{LStenflo_1996}, which contains the famous Lorenz '63 model as an invariant subspace for $\Rot=0$.

		\subsection{Results for general truncated ODE models}
		In general the ODE models described in the previous section feature at least a minimal set of desirable properties.  For any choice of index sets $\mathscr{N}_{\textbf{u}}^{\truncInd}$, $\mathscr{N}_{\theta}^{\truncInd}$ the corresponding ODE model involves only linear and quadratic terms, hence is locally Lipschitz and the Cauchy-Lipschitz theorem provides existence of smooth, short-time solutions.  Furthermore, by computing inner products of \eqref{TruncatedEvolutionPDE} with $\textbf{u}^{\truncInd},\theta^{\truncInd}$ respectively, one finds that these solutions satisfy the following:
		\begin{align}
			\frac{d}{dt} \langle \frac{1}{2}|\textbf{u}^{\truncInd}|^2 \rangle & = - \Pra \langle |\nabla \textbf{u}^{\truncInd} |^2 \rangle + \langle \mathcal{P}_{\textbf{u}}^{\truncInd} \big [ \Pra \Ray \theta^{\tau} \uVecThree - \Pra \Rot (\uVecThree \times \textbf{u}^{\tau}) - ( \textbf{u}^{\tau} \cdot \nabla ) \textbf{u}^{\tau} \big ] \cdot \textbf{u}^{\tau} \rangle , \notag  \\
			\frac{d}{dt} \langle \frac{1}{2} \big (\theta^{\truncInd} \big )^2 \rangle & = - \langle |\nabla \theta^{\truncInd} |^2 \rangle + \langle u_3^{\truncInd} \theta^{\truncInd} \rangle - \langle \mathcal{P}_{\theta}^{\truncInd} \big [ ( \textbf{u}^{\tau} \cdot \nabla ) \theta^{\tau} \big ] \cdot \theta^{\tau} \rangle . \notag
		\end{align}
		Note that the projectors $\mathcal{P}^{\truncInd}_{\textbf{u}},\mathcal{P}^{\truncInd}_{\theta}$ have a self-adjointness property, i.e. for any scalar functions $F,G \in L^2$ one has
		\begin{equation} \label{SelfAdjointProp} \langle \mathcal{P}^{\truncInd}_{\theta} [ F ] G \rangle = \Big \langle \sum_{\textbf{n} \in \mathscr{N}_{\theta}^{\truncInd}} \langle  F f^{\textbf{n}} \rangle f^{\textbf{n}} G \Big \rangle =  \Big \langle F \sum_{\textbf{n} \in \mathscr{N}_{\theta}^{\truncInd}} \langle f^{\textbf{n}} G \rangle f^{\textbf{n}} \Big \rangle = \langle F \mathcal{P}^{\truncInd}_{\theta} [ G ] \rangle , \end{equation}
		and a similar identity holds for $\mathcal{P}_{\textbf{u}}^{\truncInd}$.  Since also $\mathcal{P}^{\truncInd}_{\theta} [ \theta^{\truncInd} ] = \theta^{\truncInd}$ it follows that the projectors disappear when computing these "self" inner products, so for instance with the nonlinear terms one can then integrate by parts:
		\[ \langle \mathcal{P}_{\theta}^{\truncInd} \big [ \textbf{u}^{\truncInd} \cdot \nabla \theta^{\truncInd} \big ] \cdot \theta^{\truncInd} \rangle = \langle \big [ \textbf{u}^{\truncInd} \cdot \nabla \theta^{\truncInd} \big ] \cdot \theta^{\truncInd} \rangle = \langle \textbf{u}^{\truncInd} \cdot \nabla \frac{1}{2} (\theta^{\truncInd} )^2 \rangle = 0 . \]
		Thus one finds that all of the truncated models obey the following balance equations for the kinetic energy and temperature variance, which hold analogously for solutions of the PDE:
		\begin{align}
			\label{Balance_KinEnergy}
			\frac{d}{dt} \langle \frac{1}{2}|\textbf{u}^{\truncInd}|^2 \rangle & = - \Pra \langle |\nabla \textbf{u}^{\truncInd} |^2 \rangle + \Pra \Ray \langle u_3^{\truncInd} \theta^{\truncInd} \rangle ,  \\
			\label{Balance_TempVar}
			\frac{d}{dt} \langle \frac{1}{2} \big (\theta^{\truncInd} \big )^2 \rangle & = - \langle |\nabla \theta^{\truncInd} |^2 \rangle + \langle u_3^{\truncInd} \theta^{\truncInd} \rangle .
		\end{align}    
		The same reasoning can be applied to extend the short time solutions to global solutions, and indeed for sufficiently small Rayleigh number one even has a trivial compact global attractor.  This is proven in the following Lemma:
		\begin{Lemma}
			\label{lem:ODE_Existence}
			For any admissible parameters $\parameters$, index sets $\mathscr{N}_{\textbf{u}}^{\truncInd},\mathscr{N}_{\theta}^{\truncInd}$ and initial condition $\textbf{X}^{\truncInd}_0 = (\textbf{u}_{0}^{\truncInd}, \theta_{0}^{\truncInd})$ there exists a unique, globally-defined smooth solution $\textbf{X}^{\truncInd}(t) = (\textbf{u}^{\truncInd}(t), \theta^{\truncInd}(t))$ of the truncated ODE model in \eqref{GeneralBoussinesqODE_Vel} - \eqref{GeneralBoussinesqODE_Temp} with $\textbf{X}^{\truncInd}(0) = \textbf{X}^{\truncInd}_0$, and the following semi-group $\mathcal{S}^{\truncInd}(t)$ is well-defined:
			\begin{equation} \label{ODE_Semigroup} \mathcal{S}^{\truncInd}(t) \big [  \textbf{X}_0^{\truncInd} \big ] := \textbf{X}^{\truncInd}(t) \text{ . } \end{equation}
			Letting $\lambda_1^{\truncInd} = \min \{ |\mathcal{K}\textbf{m} | : \textbf{n} \in \mathscr{N}_{\textbf{u}}^{\truncInd} \cup \mathscr{N}_{\theta}^{\truncInd} \}$ the origin is globally exponentially attractive for $\Ray < (\lambda_1^{\truncInd})^4$ .
		\end{Lemma}
		\begin{proof}
			For $\rho \geq 0$ yet to be chosen, let $(\tilde{\textbf{u}}^{\truncInd}(t),\tilde{\theta}^{\truncInd}(t)) = e^{-\rho t}(\textbf{u}^{\truncInd}(t),\theta^{\truncInd}(t))$.  By repeating the arguments above, in particular by showing that the nonlinear term drops out, one finds the analogues of \eqref{Balance_KinEnergy}, \eqref{Balance_TempVar} for $\tilde{\textbf{u}}^{\truncInd},\tilde{\theta}^{\truncInd}$.  By adding these together, one finds
			\[ \frac{1}{2} \frac{d}{dt} \Big \langle \frac{|\tilde{\textbf{u}}^{\truncInd}|^2}{\Pra} + |\tilde{\theta}^{\truncInd}|^2 \Big \rangle = - \big ( 1 + 2\rho \big ) \langle |\nabla \tilde{\textbf{u}}^{\truncInd} |^2 \rangle + \big ( \Ray + 1 \big ) \langle \tilde{u}_3^{\truncInd} \tilde{\theta}^{\truncInd} \rangle - \big ( 1 + 2\rho \big ) \langle |\nabla \tilde{\theta}^{\truncInd} |^2 \rangle \text{ . } \]
			For any $\epsilon > 0$ one can apply Young's inequality to obtain 
			$\Ray\langle \tilde{u}_3^{\truncInd} \tilde{\theta}^{\truncInd} \rangle 
			\leq \frac{\Ray}{2\epsilon} \langle |\tilde{u}_3^{\truncInd}|^2 \rangle + \frac{\epsilon \Ray}{2} \langle | \tilde{\theta}^{\truncInd}|^2 \rangle$, $\langle \tilde{u}_3^{\truncInd} \tilde{\theta}^{\truncInd} \rangle \leq \frac{\epsilon}{2} \langle |\tilde{u}_3^{\truncInd}|^2 \rangle + \frac{1}{2\epsilon} \langle | \tilde{\theta}^{\truncInd}|^2 \rangle$ and combined with Poincar\'e's inequality we get
			\[ \frac{1}{2} \frac{d}{dt} \Big \langle \frac{|\tilde{\textbf{u}}^{\truncInd}|^2}{\Pra} + |\tilde{\theta}^{\truncInd}|^2 \Big \rangle \leq - r  \Big \langle \frac{|\tilde{\textbf{u}}^{\truncInd}|^2}{\Pra} + |\tilde{\theta}^{\truncInd}|^2 \Big \rangle \hspace{.5 cm}  \text{ , } \hspace{.5 cm} 
			r = \min \Big [ \Pra \Big ( (\lambda_1^{\truncInd})^2(1 + 2\rho ) - \frac{\Ray}{2\epsilon} - \frac{\epsilon}{2}  \Big ) , 
			(\lambda_1^{\truncInd})^2(1 + 2\rho) - \frac{1}{2\epsilon} - \frac{\epsilon \Ray}{2}  \Big ] \text{ . }  \]
			
			On the one hand choosing $\epsilon = 1$ and $\rho$ sufficiently large, we have $r > 0$, hence the $L^2$ norm is a strict Lyapunov function and the origin is globally attractive.  Thus all short time solutions $\tilde{\textbf{u}}^{\truncInd},\tilde{\theta}^{\truncInd}$ can be extended to global solutions, and hence $\textbf{u}^{\truncInd},\theta^{\truncInd}$ can be as well.  On the other hand, choosing $\rho = 0$ and optimizing in $\epsilon$, one sees that for $\Ray <  (\lambda_1^{\truncInd})^4$ the origin is the global attractor for the original system.
		\end{proof}
		Since the solutions for these models are defined for all time, one can define associated infinite time averages analogous to \eqref{NusseltDef} for the truncated ODE. For a given truncation we  thus define the Nusselt number associated to an initial condition $(\textbf{u}_0^{\truncInd},\theta_0^{\truncInd})$ at parameter $\parameters$ as follows:
		\begin{equation} \label{def:ODENusselt} \mathsf{Nu}^{\truncInd} = \mathsf{Nu}^{\truncInd}(\textbf{u}_0^{\truncInd},\theta_0^{\truncInd},\parameters) := 1 + \frac{1}{4\pi^4} \overline{\langle u^{\truncInd}_3 \theta^{\truncInd} \rangle} \text{ . } \end{equation}
		Note the extra factor of $\frac{1}{\pi}$ relative to \eqref{NusseltDef} comes from the definition of $\theta^{\truncInd}$ via $T^{\truncInd} = (1-\frac{\ShaThree x_3}{\pi}) + \frac{\theta^{\truncInd}}{\pi}$.  Note however that without further bounds on the growth of the solutions, the Nusselt number defined in this way could equal an infinite value, in stark contrast to the PDE.  Although the authors could not find a scaling law for the 3D free slip case at finite Prandtl number, it may be the case that such bounds existence.  Such infinite Nussult numbers are in clear violation of known bounds for other cases, and indeed a clear violation of the known bounds in the 2D case which is a sub-case of what we study here.  We will see in \S\ref{s:Nusselt} that energetically consistent truncations discussed next give finite values, while in \S\ref{sec:EnergyInconsis} we will see that many other truncations do not.
		
		\section{Energetically consistent hierarchies of ODE models}\label{s:energyconsistent}
		
		\subsection{Mode selection criteria}
		
		We turn now to the question of how to choose the sets $\mathscr{N}_{\textbf{u}}^{\truncInd},\mathscr{N}_{\theta}^{\truncInd}$ to best retain the behavior of the full PDE system.  In terms of the variable $\theta$ the potential energy balance \eqref{Balance_PotEnergy_Tform} becomes
		\begin{align}
			\label{Balance_PotEnergy}
			\frac{d}{dt} \langle (1-\frac{\ShaThree x_3}{\pi}) \theta \rangle & = \langle (1-\frac{\ShaThree x_3}{\pi}) \partial_{x_3}^2 \theta \rangle - \frac{\ShaThree}{\pi} \langle u_3 \theta \rangle .
		\end{align}
		As mentioned in the introduction, it has been recognized in past works that this balance can fail to hold in low dimensional models unless the truncation is chosen in specific way.  Consider now the the following generalized mode selection Criterion regarding the index sets s $\mathscr{N}_{\textbf{u}}^{\truncInd}, \mathscr{N}_{\theta}^{\truncInd}$:
		
		\begin{Criterion}
			(Energy balance):  Suppose a pair of indices $\textbf{n}' \in \mathscr{N}_{\textbf{u}}^{\truncInd}, \textbf{n}'' \in \mathscr{N}_{\theta}^{\truncInd}$ satisfy
			\begin{equation} \label{Eq_EnergyCrit} (m_1',m_2') = (m_1'',m_2'') \neq \textbf{0} \hspace{.5 cm} \text{ , } \hspace{.5 cm} \textbf{p}' = \textbf{p}''  \hspace{.5 cm} \text{ , } \hspace{.5 cm}  \text{ and either } \hspace{.5 cm} \ShaOne m_1' \neq \ShaTwo m_2' \hspace{.5 cm} \text{ or } \hspace{.5 cm}  c' = 2 \text{ , } \end{equation}
			in which case define
			\[ \textbf{n}^{\pm} = (0,0,|m_3' \pm m_3''|,0,0,1) \text{ . } \] 
			If $m_3' = m_3''$, then $\textbf{n}^{+} \in \mathscr{N}^{\truncInd}_{\theta}$.  If $m_3' \neq m_3''$, then either $\textbf{n}^{+} \in \mathscr{N}^{\truncInd}_{\theta}$, $\textbf{n}^{-} \in \mathscr{N}^{\truncInd}_{\theta}$ or $\textbf{n}^{+} \notin \mathscr{N}^{\truncInd}_{\theta}$, $\textbf{n}^{-} \notin \mathscr{N}^{\truncInd}_{\theta}$.
			\label{Crit:EnergyCrit}
		\end{Criterion}
		
		\noindent In the following proposition, it is shown that this Criterion is necessary and sufficient to prove the potential energy balance \eqref{Balance_PotEnergy} also holds for the corresponding truncated model, and that \eqref{Balance_KinEnergy}-\eqref{Balance_PotEnergy} then imply the existence of a global attractor.
		
		\begin{Proposition}
			\label{prop:BalancePreservation}
			Consider the ODE model \eqref{GeneralBoussinesqODE_Vel} - \eqref{GeneralBoussinesqODE_Temp} determined by a mode selection  $\mathscr{N}_{\textbf{u}}^{\truncInd},\mathscr{N}_{\theta}^{\truncInd}$. The potential energy balance \eqref{Balance_PotEnergy} holds for all solutions $\textbf{u}^{\truncInd},\theta^{\truncInd}$ (i.e. for all initial data  $\textbf{u}_0^{\truncInd},\theta_0^{\truncInd}$) if and only if $\mathscr{N}_{\textbf{u}}^{\truncInd},\mathscr{N}_{\theta}^{\truncInd}$ satisfy Criterion \ref{Crit:EnergyCrit}.  In this case all solutions are exponentially attracted to the following ellipsoidal ball:
			\begin{equation} \label{ExplicitAttractingBall}
				\sum_{\textbf{n}\in \mathscr{N}_{\textbf{u}}^{\truncInd}} \frac{(u^{\truncInd,\textbf{n}})^2}{\Pra \Ray}  + \sum_{\textbf{n}\in \mathscr{N}_{\theta}^{\truncInd}} \big ( \theta^{\truncInd,\textbf{n}} + \frac{2(2\pi)^{3/2}}{\ShaThree m_3} \big )^2  \leq \frac{16\pi^3 }{\min_{i\leq 3}(\mathsf{k}_i^2)} \Big ( \frac{|\mathscr{N}_{\theta}^{\truncInd} \cap \mathscr{N}^{*}_{\theta}|}{\min(\Pra,1)} + \frac{\pi^3}{6} \Big ) \text{ . } \end{equation}
			In particular, the semi-group $\mathcal{S}^{\truncInd}(t)$ admits a compact global attractor $\mathscr{A}^{\truncInd}$. 
		\end{Proposition}
		
		\begin{proof}
			
			Computing the inner product of the temperature equation in \eqref{TruncatedEvolutionPDE} with $(1-\frac{\ShaThree x_3}{\pi})$, one obtains
			\begin{equation} \label{EnerConsis_Eq2} \begin{split} \frac{d}{dt} \langle (1-\frac{\ShaThree x_3}{\pi}) \theta^{\truncInd} \rangle & = \langle (1-\frac{\ShaThree x_3}{\pi} ) \partial_{x_3}^2 \theta^{\truncInd} \rangle - \langle (1-\frac{\ShaThree x_3}{\pi} ) \mathcal{P}^{\truncInd}_{\theta} \big [ \textbf{u}^{\truncInd} \cdot \nabla \theta^{\truncInd} \big ] \rangle . \end{split} \end{equation}
			Let $\ell^{\truncInd}(x_3)$ denote the projection of the linear background state onto the modes included in the index set $\mathscr{N}_{\theta}^{\truncInd}$: 
			\begin{equation} \label{ConductStateFourier} \ell^{\truncInd}(x_3) := \mathcal{P}^{\truncInd}_{\theta} \big [ 1-\frac{\ShaThree x_3}{\pi} \big ] = \sum_{\substack{ (0,0,m_3,0,0,1) \in \mathscr{N}_{\theta}^{\truncInd} \\ m_3 > 0 }} \frac{2\sqrt{2\pi }}{m_3}  \frac{\sqrt{2} \sin(\ShaThree m_3 x_3) }{V} . \end{equation}
			This function of course depends only on the vertical variable, and in fact the set of vertically stratified modes will occur so often in our analysis that we define 
			\begin{equation} \label{VertStratNodes}
				\mathscr{M}^* = \{ \textbf{m} \in \mathbb{Z}^3_{\geq 0}: m_1 = m_2 = 0 \text{ , } m_3 > 0 \}  \text{ , } \hspace{.5 cm} \mathscr{N}^*_{\textbf{u}} = \{ \textbf{n} \in \mathscr{N}_{\textbf{u}} : \textbf{m} \in \mathscr{M}^* \} \text{ , } \hspace{.5 cm} \mathscr{N}^*_{\theta} = \{ \textbf{n} \in \mathscr{N}_{\theta} : \textbf{m} \in \mathscr{M}^* \} . 
			\end{equation}
			Focusing on the last term in \eqref{EnerConsis_Eq2}, one finds the following by using the self-adjoint property \eqref{SelfAdjointProp} and integrating by parts
			\[ \langle (1-\frac{\ShaThree x_3}{\pi} ) \mathcal{P}^{\truncInd}_{\theta} \big [ \textbf{u}^{\truncInd} \cdot \nabla \theta^{\truncInd} \big ] \rangle = \langle \ell^{\truncInd}  \textbf{u}^{\truncInd} \cdot \nabla \theta^{\truncInd} \rangle = - \langle \partial_{x_3} \ell^{\truncInd}  u_3^{\truncInd} \theta^{\truncInd}  \rangle . \]
			Comparing with the expression in \eqref{Balance_PotEnergy}, one sees that the ODE model obeys a consistent potential energy balance iff
			\begin{equation} \label{EnerConsis_Condition} \langle \partial_{x_3} \ell^{\truncInd} u_3^{\truncInd} \theta^{\truncInd}  \rangle = -\frac{\ShaThree}{\pi} \langle u_3^{\truncInd} \theta^{\truncInd} \rangle . \end{equation}
			Inserting the Fourier expansions into the left hand side of \eqref{EnerConsis_Condition} gives
			\begin{equation} \label{EnerConsis_LHS} \begin{split} \langle \partial_{x_3} \ell^{\truncInd} u_3^{\truncInd} \theta^{\truncInd}  \rangle & = \frac{2\ShaThree}{\pi} \sum_{ \textbf{n} \in \mathscr{N}_{\theta}^{\truncInd} \cap \mathscr{N}_{\theta}^*} \langle  \cos( \ShaThree m_3 x_3) u_3^{\truncInd} \theta^{\truncInd} \rangle \\ & = \frac{2\ShaThree}{\pi} \sum_{\textbf{n} \in \mathscr{N}_{\theta}^{\truncInd} \cap \mathscr{N}_{\theta}^*} \sum_{\textbf{n}' \in \mathscr{N}_{\textbf{u}}^{\truncInd} } \sum_{\textbf{n}'' \in \mathscr{N}_{\theta}^{\truncInd}} u^{\truncInd,\textbf{n}'} \theta^{\truncInd,\textbf{n}''} \langle  \cos(\ShaThree m_3 x_3) v^{\textbf{n}'}_3 f^{\textbf{n}''} \rangle . \end{split} \end{equation}
			The orthogonality properties of the sinusoidal functions imply that the only terms which survive in this sum satisfy $m_1' = m''_1$, $m_2' = m''_2$, $\textbf{p}' = \textbf{p}''$ and either $m_3 = m_3' + m''_3$,  $m_3 = m_3' - m''_3$ or  $m_3 = -m_3' + m''_3$.  On the other hand, inserting the Fourier expansions into the right hand side of \eqref{EnerConsis_Condition} one has
			\[ \langle u_3^{\truncInd} \theta^{\truncInd} \rangle = \sum_{\textbf{n}' \in \mathscr{N}_{\textbf{u}}^{\truncInd} } \sum_{\textbf{n}'' \in \mathscr{N}_{\theta}^{\truncInd} } u^{\truncInd,\textbf{n}'} \theta^{\truncInd,\textbf{n}''} \langle v^{\textbf{n}'}_3 f^{\textbf{n}''} \rangle , \] 
			so using \eqref{BasisRelations} one must have $\textbf{m}' = \textbf{m}''$ and $\textbf{p}' = \textbf{p}''$:
			\begin{equation} \label{EnerConsis_RHS} \begin{split}  \langle u_3^{\truncInd} \theta^{\truncInd} \rangle & =  \sum_{\textbf{n}'' \in \mathscr{N}_{\theta}^{\truncInd} } \sum_{(\textbf{m}'',\textbf{p}'',c) \in \mathscr{N}_{\textbf{u}}^{\truncInd} } u^{\truncInd,(\textbf{m}'',\textbf{p}'',c)} \theta^{\truncInd,\textbf{n}''} \frac{\mathcal{G}_{3,3}^{(\textbf{m}'',\textbf{p}'',c)}\nu_3^{\textbf{n}''}}{|\mathcal{G}^{(\textbf{m}'',\textbf{p}'',c)}\boldsymbol{\nu}^{\textbf{n}''}|} . \end{split} \end{equation}
			Assume now that Criterion \ref{Crit:EnergyCrit} is satisfied.  Then for each term $\textbf{n}',\textbf{n}''$ in the sum \eqref{EnerConsis_LHS} with $m_3' \neq m_3''$ both of the following two terms appear:
			\[ \langle \cos \big ( \ShaThree ( m_3' + m''_3 ) x_3 \big ) v_3^{\textbf{n}'} f^{\textbf{n}''} \rangle u^{\truncInd,\textbf{n}'} \theta^{\truncInd,\textbf{n}''} + \langle \cos \big ( \ShaThree | m_3' - m''_3 | x_3 \big ) v_3^{\textbf{n}'} f^{\textbf{n}''} \rangle u^{\truncInd,\textbf{n}'} \theta^{\truncInd,\textbf{n}''} \text{ , } \]
			One can easily check that
			\begin{equation} \label{EnerConsis_Cancel} \langle \cos \big ( \ShaThree ( m_3' + m''_3 ) x_3 \big ) v_3^{\textbf{n}'} f^{\textbf{n}''} \rangle = - \langle \cos \big ( \ShaThree | m_3' - m''_3 | x_3 \big ) v_3^{\textbf{n}'} f^{\textbf{n}''} \rangle \text{ , } \end{equation}
			thus all such terms cancel out.  Furthermore, for each term appearing in \eqref{EnerConsis_LHS} with $m_3' = m''_3$ the index $m_3 = 2m_3'$ is included in the sum, and one can similarly check that
			\begin{equation}\label{eq:canceldiag} -2 \langle  \cos( 2 \ShaThree m_3' x_3) v^{\textbf{n}'}_3 f^{\textbf{n}'} \rangle = \frac{\mathcal{G}_{3,3}^{\textbf{n}'}\nu_3^{\textbf{n}'}}{|\mathcal{G}^{\textbf{n}'}\boldsymbol{\nu}^{\textbf{n}'}|} . \end{equation}
			Thus \eqref{EnerConsis_LHS} and \eqref{EnerConsis_RHS} are equal and \eqref{EnerConsis_Condition} holds.  On the other hand if Criterion \ref{Crit:EnergyCrit} is not satisfied, then it is clear that at least one term which appears on one side of \eqref{EnerConsis_LHS}, \eqref{EnerConsis_RHS} does not appear on the other side, either because it is not included in the sum, or because a cancellation fails to occur. 
			
			For part (b) note that by combining the balance equations \eqref{Balance_KinEnergy} - \eqref{Balance_PotEnergy} one obtains the following:
			\begin{equation} \label{ODE_AttractingBall} \frac{1}{2} \frac{d}{dt} \Big \langle \frac{|\textbf{u}^{\truncInd}|^2}{\Pra \Ray}  + \big ( \theta^{\truncInd} \big )^2 + \frac{4\pi}{\ShaThree} (1-\frac{\ShaThree x_3}{\pi} )\theta^{\truncInd}   \Big \rangle = - \Big \langle \frac{|\nabla \textbf{u}^{\truncInd}|^2}{\Ray} + |\nabla \theta^{\truncInd} |^2 - \frac{2\pi}{\ShaThree} (1-\frac{\ShaThree x_3}{\pi} )\partial_{x_3}^2 \theta^{\truncInd} \Big \rangle .
			\end{equation}
			This alone implies the existence of an attracting ball $\mathscr{B}^{\truncInd}$ for the ODE system.  To see why, one can use \eqref{SelfAdjointProp} and by adding a constant term inside the time derivative to complete the square, one has the following for the left hand side:
			\[ \frac{1}{2} \frac{d}{dt} \Big \langle \frac{|\textbf{u}^{\truncInd}|^2}{\Pra \Ray}  + \big ( \theta^{\truncInd} \big )^2 + \frac{4\pi}{\ShaThree} (1-\frac{\ShaThree x_3}{\pi} )\theta^{\truncInd}   \Big \rangle = \frac{1}{2} \frac{dE}{dt} \hspace{.5 cm} \text{ , } \hspace{.5 cm} E := \Big \langle \frac{|\textbf{u}^{\truncInd}|^2}{\Pra \Ray}  + \big ( \theta^{\truncInd} + \frac{2\pi}{\ShaThree} \ell^{\truncInd} \big )^2 \Big \rangle . \]
			For the right hand side one can use \eqref{SelfAdjointProp}, integrate by parts and add and subtract a term to obtain the following:
			\[ \Big \langle - |\nabla \theta^{\truncInd} |^2 + \frac{2\pi}{\ShaThree} (1-\frac{\ShaThree x_3}{\pi} )\partial_{x_3}^2 \theta^{\truncInd} \Big \rangle =  \Big \langle \big ( \theta^{\truncInd} + \frac{\pi}{\ShaThree} \ell^{\truncInd} \big ) \Delta \theta^{\truncInd} + \theta^{\truncInd} \Delta \frac{\pi}{\ShaThree} \ell^{\truncInd} + \tilde{T} - \tilde{T} \Big \rangle \text{ , } \]
			where $\tilde{T} = \frac{\pi}{\ShaThree} \ell^{\truncInd} \Delta \frac{\pi}{\ShaThree} \ell^{\truncInd} $.  Hence by integrating by parts, one obtains 
			\[ \Big \langle - |\nabla \theta^{\truncInd} |^2 + \frac{2\pi}{\ShaThree} (1-\frac{\ShaThree x_3}{\pi} )\partial_{x_3}^2 \theta^{\truncInd} \Big \rangle =  \Big \langle - \big | \nabla \big ( \theta^{\truncInd} + \frac{\pi}{\ShaThree} \ell^{\truncInd} \big ) \big  |^2 + \frac{\pi^2}{\ShaThree^2}  \big | \nabla \ell^{\truncInd} \big |^2 \Big \rangle \text{ . } \]
			Using the identity $\frac{\pi^2}{\ShaThree^2} \langle |\nabla \ell^{\truncInd}|^2 \rangle = 8\pi^3 |\mathscr{N}_{\theta}^{\truncInd} \cap \mathscr{N}^{*}_{\theta}|$ obtained from the explicit Fourier expansion \eqref{ConductStateFourier}, \eqref{ODE_AttractingBall} can thus be rewritten as 
			\begin{equation} \label{ODE_AttractingBall2} \frac{1}{2} \frac{dE}{dt} = - \Big \langle \frac{|\nabla \textbf{u}^{\truncInd}|^2}{\Ray} + \big | \nabla \big ( \theta^{\truncInd} + \frac{\pi}{\ShaThree} \ell^{\truncInd} \big ) \big  |^2 \Big \rangle  + 8\pi^3 |\mathscr{N}_{\theta}^{\truncInd} \cap \mathscr{N}^{*}_{\theta}| \text{ . } \end{equation}
			On the left hand side one sees the time derivative of a weighted $\textbf{H}^{0}$ norm of $(\textbf{u}^{\truncInd},\theta^{\truncInd} + \frac{2\pi}{\ShaThree} \ell^{\truncInd} )$, and on the right one has a weighted $\textbf{H}^{1}$ norm of $(\textbf{u}^{\truncInd},\theta^{\truncInd} + \frac{\pi}{\ShaThree} \ell^{\truncInd} )$ subtracted from a positive constant.  The weighted $\textbf{H}^0$ norm expresses an ellipsoid in the finite dimensional phase space with principle semi-axes $\Pra \Ray$, $1$ along all $\textbf{u}^{\truncInd}, \theta^{\truncInd}$ directions respectively, whereas the $\textbf{H}^1$ weighted norm expresses an ellipsoid with principle semi-axes $\frac{\Ray}{|\mathcal{K}\textbf{m}|} $, $\frac{1}{|\mathcal{K}\textbf{m}|}$ along the $\textbf{u}^{\truncInd,\textbf{n}}, \theta^{\truncInd,\textbf{n}}$ directions.  Hence \eqref{ODE_AttractingBall2} shows that trajectories move along progressively smaller $\textbf{H}^{0}$ ellipsoids until they belong to an $\textbf{H}^{1}$ ellipsoid with principle semi-axes $\Pra \Ray$, $1$.  Concretely, one obtains the following estimate using Young's inequality, the explicit form of the Fourier expansion for $\ell^{\truncInd}$ and the Euler formula for the Basel problem: 
			\begin{align}
				\Big \langle \frac{|\textbf{u}^{\truncInd}|^2}{\Pra \Ray}  + \big ( \theta^{\truncInd} + \frac{2\pi}{\ShaThree} \ell^{\truncInd} \big )^2 \Big \rangle & \leq \Big \langle \frac{|\textbf{u}^{\truncInd}|^2}{\Pra \Ray}  + 2 \big ( \theta^{\truncInd} + \frac{\pi}{\ShaThree} \ell^{\truncInd} \big )^2 \Big \rangle + \Big \langle 2 \big ( \frac{\pi}{\ShaThree} \ell^{\truncInd} \big )^2 \Big \rangle \notag \\ & \leq \frac{2}{\min(\Pra,1) \min(\ShaOne^2,\ShaTwo^2,\ShaThree^2)} \Big \langle \frac{|\nabla \textbf{u}^{\truncInd}|^2}{\Ray}  + \big | \nabla ( \theta^{\truncInd} + \frac{\pi}{\ShaThree} \ell^{\truncInd} )\big |^2 \Big \rangle + \frac{8 \pi^6}{3\ShaThree^2} \text{ , } \notag 
			\end{align}
			Hence using this estimate in \eqref{ODE_AttractingBall2} one has 
			\begin{align} \frac{dE}{dt} \leq - \min(\Pra,1)\min_{i\leq 3}(\mathsf{k}_i^2) \Big ( E - \frac{16\pi^3 }{\min_{i\leq 3}(\mathsf{k}_i^2)} \big ( \frac{|\mathscr{N}_{\theta}^{\truncInd} \cap \mathscr{N}^{*}_{\theta}|}{\min(\Pra,1)} + \frac{\pi^3}{6} \big ) \Big ) \text{ . } \notag \end{align}
			Hence the ellipsoidal ball given by \eqref{ExplicitAttractingBall} is exponentially attracting.  To prove existence of an attractor one only needs an attracting ball, but we point out that using these ellipsoids gives a bound on $\theta^{\truncInd}$ which is independent of $\Ray$.  Denoting this attracting ball by $\mathscr{B}^{\truncInd}$, recall that the $\omega$-limit set of $\mathscr{B}^{\truncInd}$ with respect to $\mathcal{S}^{\truncInd}(t)$ is defined
			\[ \omega(\mathscr{B}^{\truncInd}) := \cap_{s > 0} \overline{\cup_{t > s} \mathcal{S}^{\truncInd}(t)\mathscr{B}^{\truncInd}} \text{ , } \]
			where here the overline denotes the closure.  Since $\mathscr{B}^{\truncInd}$ is forward invariant, this is an intersection of compact sets, hence compact.  It is easy to check that it is forward invariant, attracts the bounded sets of the phase space, and that it is the maximal set with these properties.  Thus $\mathscr{A}^{\truncInd} := \omega(\mathscr{B}^{\truncInd})$ is the global attractor.
			
			\vspace{-.75 cm}
			\[ \textcolor{white!100}{ . } \]
			
		\end{proof}
		
		Turning now to the vorticity balance \eqref{Balance_Vort}, consider the following mode selection Criteria:
		\begin{Criteria}
			\begin{enumerate}[label=(\roman*)]
				\item (Vortex stretching balance): Suppose a pair of indices $\textbf{n}',\textbf{n}'' \in \mathscr{N}_{\textbf{u}}^{\truncInd}$ satisfy
				\[ (m_1',m_2') = (m_1'',m_2'') \neq \textbf{0} \hspace{.5 cm} \text{ , } \hspace{.5 cm} m_3' + m_3'' \text{ odd } \hspace{.5 cm} \text{ , } \hspace{.5 cm} \textbf{p}' + \textbf{p}'' = \textbf{p} \text{ , } \] 
				for either $\textbf{p} = (0,1)$ or $(1,0)$, and either $\ShaOne m_1' \neq \ShaOne m_2'$ or $(c',c'') \neq (1,1)$.  In this case define
				\[ \textbf{n}^{\pm} = (0,0,|m_3' \pm m_3''|,\textbf{p},1) \text{ . } \]
				\begin{enumerate}[label=(\alph*)]
					\item If $(m_1',m_2') \in \mathbb{Z}_{> 0}^2$, $m_1' = 0$ and $p_1' \neq p_1''$, or $m_2' = 0$ and $p_2' \neq p_2''$, then $\textbf{n}^{+}, \textbf{n}^{-} \in \mathscr{N}^{\truncInd}_{\textbf{u}}$.
					
					\item If $m_1' = 0$ and $p_1' = p_1'' = 0$ or $m_2' = 0$ and $p_2' = p_2'' = 0$, then either $\textbf{n}^{+} \in \mathscr{N}^{\truncInd}_{\textbf{u}}$, $\textbf{n}^{-} \in \mathscr{N}^{\truncInd}_{\textbf{u}}$ or $\textbf{n}^{+} \notin \mathscr{N}^{\truncInd}_{\textbf{u}}$, $\textbf{n}^{-} \notin \mathscr{N}^{\truncInd}_{\textbf{u}}$ .
				\end{enumerate}
				
				\item (Coriolis balance): If $\Rot \neq 0$, then $(0,0,m_3,\textbf{p},1) \in \mathscr{N}_{\textbf{u}}^{\truncInd}$, $m_3$ odd $\Rightarrow$ $(0,0,m_3,p_1+1,p_2+1,1) \in \mathscr{N}_{\textbf{u}}^{\truncInd}$.
			\end{enumerate}
			\label{Crit:VortCrit}
		\end{Criteria}
		In the following Proposition, it is proven that these Criteria are necessary and sufficient to ensure that \eqref{Balance_Vort} also holds for the corresponding truncated model.  Similar to Prop. \ref{prop:BalancePreservation}, the essence of the proof involves checking when the curl operator commutes with the projection operator $\mathcal{P}_{\textbf{u}}^{\truncInd}$.  However, the analysis is more technical due to the curl operator, and so the proof is given in \ref{app:VortBal}.
		\begin{Proposition}
			\label{prop:VortPreservation}
			For the ODE model determined by $\mathscr{N}_{\textbf{u}}^{\truncInd},\mathscr{N}_{\theta}^{\truncInd}$ and any initial condition $(\textbf{u}_0^{\truncInd},\theta^{\truncInd}_0)$, the vorticity balance \eqref{Balance_Vort} holds for $\vorticity^{\truncInd} := \nabla \times \textbf{u}^{\truncInd}, \textbf{u}^{\truncInd}$ if and only if $\mathscr{N}_{\textbf{u}}^{\truncInd}$ satisfies Criteria \ref{Crit:VortCrit}.  
		\end{Proposition}
		
		\subsection{Bounds on the Nusselt number for energetically consistent models} \label{s:Nusselt}
		
		For an energetically consistent truncated model, one has the following equivalent expressions for the Nusselt number $\mathsf{Nu}^{\truncInd}$, which are found by taking the infinite time averages of the balances in \eqref{Balance_TempVar}, \eqref{Balance_PotEnergy}:
		\begin{equation} \label{NusseltNumExpressions} \mathsf{Nu}^{\truncInd} = 1 + \frac{1}{4\pi^4} \overline{\langle u_3^{\truncInd} \theta^{\truncInd} \rangle} = 1 + \frac{1}{4\pi^4} \overline{\langle | \nabla \theta^{\truncInd} |^2 \rangle} = 1 + \frac{1}{4\pi^3 \ShaThree} \overline{\langle (1-\frac{\ShaThree x_3}{\pi}) \partial_{x_3}^2 \theta^{\truncInd} \rangle} . \end{equation}
		These various expressions are useful in different contexts, for instance while the first has a clear physical interpretation as the average vertical heat transport, it is clear from the second that the Nusselt number must be greater than or equal to 1, i.e. a convective flow transports more heat than the pure conductive state.  Together with the explicit bound \eqref{ExplicitAttractingBall}, the third expression for the Nusselt number can be used to prove the following Lemma regarding the Nusselt number in a truncated model:
		\begin{Lemma}
			\label{lem:TruncatedNusseltBound}
			For given admissible $\parameters$ and $\mathscr{N}_{\textbf{u}}^{\truncInd}, \mathscr{N}_{\theta}^{\truncInd}$ satisfying Criterion \ref{Crit:EnergyCrit}, there exists a $C> 0$ independent of $\Ray,\Rot$, such that one has the following bound:
			\begin{equation} \label{TruncatedNusseltBound} \sup_{(\textbf{u}_0^{\truncInd},\theta_0^{\truncInd})} \mathsf{Nu}^{\truncInd} \leq C  (1 + |\mathscr{N}_{\theta}^{\truncInd} \cap \mathscr{N}_{\theta}^*|) \max_{\textbf{n}\in\mathscr{N}_{\theta}^{\truncInd}\cap \mathscr{N}_{\theta}^*}(m_3) \text{ . } \end{equation}
		\end{Lemma}
		Note in particular that this result says that the Nusselt numbers for a fixed truncated is bounded above by some constant with respect to $\Ray$, in strong contrast to the behavior for the PDE, where one expects that the Nusselt number behaves either according to the classical scaling $\mathsf{Nu} \sim C \Ray^{1/3}$ or possibly the ultimate scaling  $\mathsf{Nu} \sim C \Ray^{1/2}$ \cite{iyer2020classical}.  This implies that a fixed truncated model can only give a good approximation for $\mathsf{Nu}$ for some finite range of Rayleigh numbers.  On the other hand, this constant increases with the number of vertical modes only, suggesting the vertical resolution plays a decisive role in obtaining better approximations.  
		
		\begin{proof}
			Note that in Fourier space the third expression for the Nusselt number is given by
			\begin{equation} \label{NusseltFourier} \mathsf{Nu}^{\truncInd} = 1  - \frac{\ShaThree }{\sqrt{2\pi^{5}}} \sum_{\textbf{n} \in \mathscr{N}^{\truncInd}_{\theta} \cap \mathscr{N}^{*}_{\theta}} m_3 \overline{\theta^{\truncInd,\textbf{n}}}. \end{equation}
			All solutions eventually enter the attracting ellipsoidal ball given in \eqref{ExplicitAttractingBall}, and remain there ever after, hence the infinite time averaged Nusselt number depends only on $\theta^{\truncInd}$ values in this range.  Here one has
			\[ \sum_{\textbf{n}\in \mathscr{N}_{\theta}^{\truncInd} \cap \mathscr{N}_{\theta}^*} \big ( \theta^{\truncInd,\textbf{n}} + \frac{2(2\pi)^{3/2}}{\ShaThree m_3} \big )^2  \leq \frac{16\pi^3 }{\min_{i\leq 3}(\mathsf{k}_i^2)} \Big ( \frac{|\mathscr{N}_{\theta}^{\truncInd} \cap \mathscr{N}^{*}_{\theta}|}{\min(\Pra,1)} + \frac{\pi^3}{6} \Big ) \]
			Using $\tilde{\theta}^{\truncInd,\textbf{n}} = \theta^{\truncInd,\textbf{n}} + \frac{2(2\pi)^{3/2}}{\ShaThree m_3} $ one finds
			\begin{align}
				\mathsf{Nu}^{\truncInd} & = 1 + \frac{4}{\pi} |\mathscr{N}_{\theta}^{\truncInd} \cap \mathscr{N}_{\theta}^{*}| - \frac{\ShaThree }{\sqrt{2\pi^{5}}} \sum_{\textbf{n} \in \mathscr{N}^{\truncInd}_{\theta} \cap \mathscr{N}^{*}_{\theta}} m_3 \overline{\tilde{\theta}^{\truncInd,\textbf{n}}} \notag \\ & \leq 1 + \frac{4}{\pi} |\mathscr{N}_{\theta}^{\truncInd} \cap \mathscr{N}_{\theta}^{*}| + \frac{\ShaThree }{\sqrt{2\pi^{5}}} \Big ( \sum_{\textbf{n} \in \mathscr{N}^{\truncInd}_{\theta} \cap \mathscr{N}^{*}_{\theta}} m_3^2 \Big )^{1/2} \Big ( \sum_{\textbf{n} \in \mathscr{N}^{\truncInd}_{\theta} \cap \mathscr{N}^{*}_{\theta}} \overline{(\tilde{\theta}^{\truncInd,\textbf{n}})^2} \Big )^{1/2} \notag \\ & \leq 1 + \frac{4}{\pi} |\mathscr{N}_{\theta}^{\truncInd} \cap \mathscr{N}_{\theta}^{*}| + \frac{2\sqrt{2} \ShaThree }{\pi \min_{i\leq 3}(\mathsf{k}_i)} \Big ( \sum_{\textbf{n} \in \mathscr{N}^{\truncInd}_{\theta} \cap \mathscr{N}^{*}_{\theta}} m_3^2 \Big )^{1/2} \Big ( \frac{|\mathscr{N}_{\theta}^{\truncInd} \cap \mathscr{N}^{*}_{\theta}|}{\min(\Pra,1)} + \frac{\pi^3}{6} \Big )^{1/2} \notag \\ & \leq 1 + \frac{4}{\pi} |\mathscr{N}_{\theta}^{\truncInd} \cap \mathscr{N}_{\theta}^{*}| + \frac{2\sqrt{2} \ShaThree }{\pi \min_{i\leq 3}(\mathsf{k}_i)} \max_{\textbf{n}\in\mathscr{N}_{\theta}^{\truncInd}\cap \mathscr{N}_{\theta}^*}(m_3) |\mathscr{N}_{\theta}^{\truncInd} \cap \mathscr{N}^{*}_{\theta}|^{1/2} \Big ( \frac{|\mathscr{N}_{\theta}^{\truncInd} \cap \mathscr{N}^{*}_{\theta}|}{\min(\Pra,1)} + \frac{\pi^3}{6} \Big )^{1/2}  \text{ , } \notag 
			\end{align}
			and by applying Young's inequality one finds that the claimed bound holds with
			\[  C = \max \Big ( \frac{\pi^2\sqrt{2}\ShaThree}{6 \min_{i\leq 3}(\mathsf{k}_i)} + 1 , \frac{\sqrt{2}\ShaThree}{\pi \min_{i\leq 3}(\mathsf{k}_i)} \big ( \frac{2}{\min(\sqrt{\Pra},1)} + 1 \big ) + \frac{4}{\pi} \Big ) \text{ . } \]
		\end{proof}
		
		\begin{Remark}
			While the above bound \eqref{TruncatedNusseltBound} increases with $\max_{\textbf{n}\in\mathscr{N}_{\theta}^{\truncInd}\cap \mathscr{N}_{\theta}^*}(m_3)$, this is likely an artifact of the proof and the upper bound probably depends only on $|\mathscr{N}_{\theta}^{\truncInd} \cap \mathscr{N}^{*}_{\theta}|$.  Indeed, rather than using the bound \eqref{ExplicitAttractingBall} one might pursue an $H^1$ bound, and \eqref{NusseltFourier} would then imply a bound independent of $\max_{\textbf{n}\in\mathscr{N}_{\theta}^{\truncInd}\cap \mathscr{N}_{\theta}^*}(m_3)$.  We did not pursue this here, although we did find that by scaling the Lorenz system to general $m_3$, rather than $m_3 =1$, one actually sees a decreasing Nusselt number bound as $m_3$ increases. 
		\end{Remark}
		
		\subsection{The $\ell^{\infty}$, $\ell^{1}$ and HKC hierarchies of truncated models}
		
		\label{subsec:ConcreteHierarchies}
		
		We consider now concrete choices of $\mathscr{N}_{\textbf{u}}^{\truncInd}$, $\mathscr{N}_{\theta}^{\truncInd}$.  Until now the truncation index $\truncInd$ has been used only to indicate that some kind of truncation has been done, but we define this index here via $\truncInd = (\tau_1,\tau_2)$, where $\tau_1$ specifies which hierarchy is considered and $\tau_2$ specifies a model within a given hierarchy.  In general to approximate a solution of the PDE \eqref{EvolutionPDE}, one must consider a sequence of solutions generated from a hierarchy of index sets $\mathscr{N}_{\textbf{u}}^{\truncInd}$, $\mathscr{N}_{\theta}^{\truncInd}$ for $\tau_2 = 1,2,...$ such that $\mathscr{N}_{\textbf{u}}^{\truncInd}$, $\mathscr{N}_{\theta}^{\truncInd}$ increase to eventually include all Fourier modes, i.e. one should have $\mathscr{N}_{\textbf{u}}^{(\tau_1,\tau_2)} \subseteq \mathscr{N}_{\textbf{u}}^{(\tau_1,\tilde{\tau}_2)}$ for $\tau_2 < \tilde{\tau}_2$ and $\cup_{\tau_2} \mathscr{N}_{\textbf{u}}^{(\tau_1,\tau_2)} = \mathscr{N}_{\textbf{u}}$.  Clearly the mode selection Criteria \ref{Crit:EnergyCrit}, \ref{Crit:VortCrit} are desirable when choosing these sets, since they not only ensure the physically realistic balances \eqref{Balance_PotEnergy}, \eqref{Balance_Vort} but also imply the existence of a compact attractor.  Even under the constraints of Criteria \ref{Crit:EnergyCrit}, \ref{Crit:VortCrit} there remains a good deal of flexibility for how these hierarchies could be chosen.
		
		Perhaps the most obvious choice of hierarchy is found by simply choosing the $\tau_2^{th}$ model to include all variables with wave numbers less than or equal to $\tau_2$.  We refer to this as the "Fourier-box" hierarchy, hence set $\tau_1 = $ `box'.  This hierarchy is defined explicitly as follows in terms of the $\ell^{\infty}$ norm:
		\begin{equation} \label{HierarchyDef_FourierBox} \mathscr{N}_{\textbf{u}}^{(box,\tau_2)} = \{ \textbf{n}\in \mathscr{N}_{\textbf{u}} : |\textbf{m}|_{\infty} \leq \tau_2 \} \hspace{.5 cm} \text{ , } \hspace{.5 cm} \mathscr{N}_{\theta}^{(box,\tau_2)} = \{ \textbf{n}\in \mathscr{N}_{\theta} : |\textbf{m}|_{\infty} \leq \tau_2 \} \text{ . }  \end{equation} 
		However, this naive choice fails to satisfy the Criteria \ref{Crit:EnergyCrit}, \ref{Crit:VortCrit} since for example the modes $(0,0,m_3)$, $m_3\in \{ \tau_2 +1,\ldots,2\tau_2 \}$ are missing from $\mathscr{N}_{\theta}^{\truncInd}$ (For some consequences, see \S\ref{sec:EnergyInconsis}).  One can easily find the minimally extended index sets such that Criteria \ref{Crit:EnergyCrit}, \ref{Crit:VortCrit} are satisfied:
		\begin{align} \label{HierarchyDef_Linf_Full} \mathscr{N}_{\textbf{u}}^{(full,\tau_2)} & := \{ \textbf{n} \in \mathscr{N}_{\textbf{u}} : |\textbf{m}|_{\infty} \leq \tau_2 \} \cup \{ \textbf{n} \in \mathscr{N}_{\textbf{u}}^* : \tau_2 < m_3 \leq 2\tau_2 -1 \} \text{ , } \\ \mathscr{N}_{\theta}^{(full,\tau_2)} & := \{ \textbf{n} \in \mathscr{N}_{\theta} : |\textbf{m}|_{\infty} \leq \tau_2 \} \cup \{ \textbf{n} \in \mathscr{N}_{\theta}^* : \tau_2 < m_3 \leq 2\tau_2 \} \text{ . } \notag \end{align}
		We refer to this hierarchy as the "full" $\ell^{\infty}$ hierarchy, due to its definition in terms of the $\ell^{\infty}$ norm and the fact that all variables are included.  This hierarchy is very general since a very broad class of solutions of \eqref{EvolutionPDE} could be approximated in this way.  However, note that the $\tau_2^{th}$ model chosen in this way satisfies
		\[ |\mathscr{N}_{\textbf{u}}^{(full,\tau_2)}| = 8\tau_2^3 + 12 \tau_2^2 + 8 \tau_2 \hspace{.5 cm} \text{ , } \hspace{.5 cm}  |\mathscr{N}_{\theta}^{(full,\tau_2)}| = 4\tau_2^3 + 4 \tau_2^2 + 4 \tau_2 \text{ , }  \]
		hence the $\tau_2^{th}$ model in this hierarchy is an ODE of dimension $12\tau_2^3 + 16 \tau_2^2 + 12\tau_2$.  The computational cost of such models therefore grows very rapidly, and quickly becomes intractable.  
		
		Rather than consider fully general solutions, one can consider special classes of solutions which have symmetries, namely that they lie in an invariant subspace of the full evolution.  One can deduce the existence of many invariant subspaces of the Boussinesq-Coriolis system from the compatibility conditions \eqref{WaveCompatibilityCond}, \eqref{PhaseCompatibilityCond}.  One easy example is the fully 2d planar convection system, which can be seen to be invariant when the rotation is set to zero by checking the different nonlinear couplings.  Similarly any sublattice of $\mathbb{L} \subset \mathbb{Z}^{3}_{\geq 0}$ corresponds to an invariant subspaces in which only Fourier variables with wave numbers $\textbf{m} \in \mathbb{L}$ are non-zero.  In this paper we focus on invariant subspaces where the phases $\textbf{p}$ lie in some proper subset of $\mathscr{P}_{\textbf{u}}^{\textbf{m}}$,$\mathscr{P}_{\theta}^{\textbf{m}}$, a condition we refer to as phase locking. 
		
		We can therefore define the three main energetically consistent hierarchies used herein.  The first is a straightforward phase locked reduction of the above $\ell^{\infty}$ hierarchy.  The index sets for this hierarchy are given explicitly as follows:
		\begin{align} \label{HierarchyDef_Linf_Lock} \mathscr{N}_{\textbf{u}}^{(1,\tau_2)} & := \{ \textbf{n} \in \mathscr{N}_{\textbf{u}} : |\textbf{m}|_{\infty} \leq \tau_2 \text{ , } p_1 = p_2 \}  \text{ , } \\ \mathscr{N}_{\theta}^{(1,\tau_2)} & := \{ \textbf{n} \in \mathscr{N}_{\theta} : |\textbf{m}|_{\infty} \leq \tau_2  \text{ , } p_1 = p_2 \} \cup \{ \textbf{n} \in \mathscr{N}_{\theta}^* : \tau_2 < m_3 \leq 2\tau_2 \text{ , } \textbf{p} = \textbf{0} \} \text{ . } \notag \end{align}
		Since we only consider phase locked hierarchies in the following, we simply refer to this as the $\ell^{\infty}$ hierarchy.  Due to the phase lock condition the vorticity balance is trivially satisfied, and one has 
		\[ |\mathscr{N}_{\textbf{u}}^{(1,\tau_2)}| = 4\tau_2^3 + 6 \tau_2^2 + 2 \tau_2 \hspace{.5 cm} \text{ , } \hspace{.5 cm}  |\mathscr{N}_{\theta}^{(1,\tau_2)}| = 2\tau_2^3 + 2 \tau_2^2 + 2 \tau_2 \text{ , }  \]
		hence in this case the $\tau_2^{th}$ model in the hierarchy is an ODE of dimension $6\tau_2^3 + 8 \tau_2^2 + 4\tau_2$.  The phase locked $\ell^{\infty}$ hierarchy was chosen because it is a simple choice that perhaps many researchers would consider first.  However, in the numerical work below we found that another hierarchy is also interesting to consider.   Rather than the $\ell^{\infty }$ norm, the $\ell^{1}$ norm is used, and the phase locking  condition $|\textbf{p}|_1 = |\textbf{m}|_1$ (as always with phase indices holding mod 2) is chosen to give the best analogy with the HKC hierarchy from \cite{Welter2025Rotating}.  The index sets for this hierarchy are given explicitly as follows: 
		\begin{align} \label{HierarchyDef_L1_Lock} \mathscr{N}_{\textbf{u}}^{(2,\tau_2)} & := \{ \textbf{n} \in \mathscr{N}_{\textbf{u}} : |\textbf{m}|_{1} \leq \tau_2 \text{ , } |\textbf{p}|_1 = |\textbf{m}|_1 \} \cup \{ \textbf{n} \in \mathscr{N}_{\textbf{u}}^* :  m_3 \leq 2\tau_2-1 \text{ , } |\textbf{p}|_1 = |\textbf{m}|_1 \} \text{ , } \\ \mathscr{N}_{\theta}^{(2,\tau_2)} & := \{ \textbf{n} \in \mathscr{N}_{\theta} : |\textbf{m}|_{1} \leq \tau_2  \text{ , } |\textbf{p}|_1 = |\textbf{m}|_1 \} \cup \{ \textbf{n} \in \mathscr{N}_{\theta}^* : m_3 \leq 2(\tau_2-1) \text{ , } |\textbf{p}|_1 = |\textbf{m}|_1 \} \text{ . } \notag \end{align}
		We refer to this as the $\ell^{1}$ hierarchy.  Note that one has 
		\[ |\mathscr{N}_{\textbf{u}}^{(2,\tau_2)}| = \frac{2}{3} \tau_2^3 + \tau_2^2 + \frac{1}{3} \tau_2 + 2 \max (1, \tau_2-1) \hspace{.5 cm} \text{ , } \hspace{.5 cm}  |\mathscr{N}_{\theta}^{(2,\tau_2)}| = \frac{1}{3} \tau_2^3 + \frac{2}{3}\tau_2 -1 \text{ , }  \]
		hence in this case the $\tau_2^{th}$ model in the hierarchy is an ODE of dimension $\tau_2^3+\tau_2^2+\tau_2+2\max(1,\tau_2-1) -1$.  Note that this hierarchy increases in dimension more slowly, enabling one to include higher vertically stratified modes, which is important for the Nusselt number as shown in \S\ref{s:Nusselt}.  Finally, for comparison with the 2d case we also consider models from the HKC hierarchy studied in \cite{Welter2025Rotating}, which are defined explicitly via
		\begin{align} \label{HierarchyDef_HKC} \mathscr{N}_{\textbf{u}}^{(3,\tau_2)} & := \cup_{j \leq 3} \mathscr{N}_{\textbf{u}}^{(3,\tau_2),j} \hspace{.5 cm} \text{ where } \hspace{.5 cm} \mathscr{N}_{\textbf{u}}^{(3,\tau_2),1} := \{ \textbf{n} \in \mathscr{N}_{\textbf{u}} : m_2 = 0 \text{ , }  m_3 > 0 \text{ , } |\textbf{m}|_{1} \leq \tau_2 \text{ , } |\textbf{p}|_1 = |\textbf{m}|_1 \} \text{ , } \\  \mathscr{N}_{\textbf{u}}^{(3,\tau_2),2} & := \{ (m_1,0,0,m_1+1,1,1) \in \mathscr{N}_{\textbf{u}} : m_{1} \leq \tau_2 -1  \} \hspace{.25 cm} \text{ , } \hspace{.25 cm} \mathscr{N}_{\textbf{u}}^{(3,\tau_2),3} := \{ \textbf{n} \in \mathscr{N}_{\textbf{u}}^* :  m_3 \leq 2\tau_2-1 \text{ , } |\textbf{p}|_1 = |\textbf{m}|_1 \} \text{ , } \notag \\ \mathscr{N}_{\theta}^{(3,\tau_2)} & := \{ \textbf{n} \in \mathscr{N}_{\theta} : m_2 = 0 \text{ , } |\textbf{m}|_{1} \leq \tau_2  \text{ , } |\textbf{p}|_1 = |\textbf{m}|_1 \} \cup \{ \textbf{n} \in \mathscr{N}_{\theta}^* : m_3 \leq 2\tau_2 \text{ , } |\textbf{p}|_1 = |\textbf{m}|_1 \} \text{ . } \notag \end{align}
		One sees that the definition is the restriction of the $\ell^1$ model to the the invariant subspace where $m_2 =0$, except \eqref{HierarchyDef_HKC} includes one fewer mode with $m_3 = 0$ than \eqref{HierarchyDef_L1_Lock}.  This choice was made so the lowest model in this hierarchy is the well-studied Lorenz-Stenflo model.  Note that the definition given here differs slightly from that in \cite{Welter2025Rotating}: since the variables are rescaled, one must take $\ShaTwo = \frac{1}{\ShaOne}$ and the models increase by adding all modes with $|\textbf{m}|_1 \leq \tau_2$, rather than adjoining wave vectors one at a time.

		\section{Effects of energetic inconsistency}
		
		\label{sec:EnergyInconsis}
		
		Criteria \ref{Crit:EnergyCrit}, \ref{Crit:VortCrit} guarantee that the potential energy and vorticity balances \eqref{Balance_PotEnergy}, \eqref{Balance_Vort} hold, and in turn \eqref{Balance_PotEnergy} guarantees the existence of a compact global attractor.  However, one can very easily violate these Criteria when choosing a truncated model, for example with the naive Fourier box model in \eqref{HierarchyDef_FourierBox}.  In this section we consider the dynamics of energetically inconsistent models to underscore the significance of the special class of energetically consistent ones.  We distinguish two types of the latter: For the first type, we are able to prove that these models suffer from `runaway modes' for sufficiently large Rayleigh number, i.e. solutions which grow unboundedly at an exponential rate. We also give bounds on the Rayleigh numbers for which such an inconsistent model still possesses a global attractor and is in this sense usable.  Nonlinear effects play a more prominant role for the second type of inconsistency and it seems difficult to make general statements.  For this we discuss a sample case which does not suffer from runaway, and possesses a global attractor for any Rayleigh number. In order to isolate the effects of energetic inconsistency, in this section we always consider models which fulfill the vorticity balance Criteria \ref{Crit:VortCrit}.
		
		\subsection{Analysis of type one inconsistent models}
		
		The first type of energetic inconsistency is defined as follows:
		
		\begin{Definition}
			\label{Def:EnergyInconsis_Type1}
			A truncated model $\mathscr{N}_{\textbf{u}}^{\truncInd}, \mathscr{N}_{\theta}^{\truncInd}$ has a type 1 inconsistency iff there exists an index $\textbf{n} \in \mathscr{N}_{\textbf{u}}^{\truncInd}$ such that:
			\begin{enumerate}[label=(\roman*)]
				\item $(\textbf{m},\textbf{p},1) \in\mathscr{N}_{\theta}^{\truncInd}$ and either $\ShaOne m_1 \neq \ShaTwo m_2 $ or c = 2.  
				\item For any $(\tilde{\textbf{m}},\textbf{0},\tilde{c}) \in \mathscr{N}_{\textbf{u}}^{\truncInd} \cup \mathscr{N}_{\theta}^{\truncInd}$ there exists at least one wave number $\tilde{m}_k > 0$ such that $\tilde{m}_k \neq 2m_k$.
			\end{enumerate}
		\end{Definition}
		Assumption (i) guarantees the existence of a pair $u^{\textbf{n}}, \theta^{(\textbf{m},\textbf{p},1)}$ which interact nontrivially via the buoyancy force.  As we shall see this produces a linear instability when the Rayleigh number is sufficiently large.  Assumption (ii) implies that the linear subspace on which only variables with wave vector $\textbf{m}$ are nonzero is invariant under the truncated evolution.  For example, in the context of a Fourier box model $\mathscr{N}_{\textbf{u}}^{(box,\tau_2)}, \mathscr{N}_{\theta}^{(box,\tau_2)}$ in \eqref{HierarchyDef_FourierBox}, due to the convolution condition \eqref{WaveCompatibilityCond} all modes $\textbf{n}$ for which $m_3 > \lfloor \frac{\tau_2}{2} \rfloor$ are only coupled to modes $\textbf{n}',\textbf{n}''$ with $m_3', m_3'' < \lfloor \frac{\tau_2}{2} \rfloor$, hence if these modes are initially zero they remain zero for all time, and one can show that the linear dynamics produce an instability.  However, we emphasize that Definition \ref{Def:EnergyInconsis_Type1} is more general and goes far beyond the Fourier box models.  
		
		\subsubsection{Runaway modes for type 1 inconsistent models}\label{s:runaway}
		
		We begin by proving that all type 1 inconsistent models admit runaway modes at sufficiently large Rayleigh numbers:
		
		\begin{Theorem}\label{thm:runaway}
			Suppose a truncated model $\mathscr{N}_{\textbf{u}}^{\truncInd}, \mathscr{N}_{\theta}^{\truncInd}$ has a type 1 inconsistency via an index $\textbf{n} \in \mathscr{N}_{\textbf{u}}^{\truncInd}$.  Runway modes occur, for any fixed value of $\Rot$, for all $\Ray$ sufficiently large of order $|\mathcal{K}\textbf{m}|^4$.
		\end{Theorem}
		
		\begin{proof}  We first note that there are subspaces $\mathscr{U}^*= \mathscr{U}^*_{\textbf{u}} \times \mathscr{U}^*_\theta$ 
			of $\textbf{L}^{2,\truncInd}_{\sigma} \times L^{2,\truncInd}_0$ such that any $(\textbf{v},\theta) \in \mathscr{U}^*$ satisfies 
			\begin{equation}\label{e:lindyn}
				\mathcal{P}_{\textbf{u}}^{\truncInd} \big [ (\textbf{v}\cdot\nabla)\textbf{v} \big ] =0 \hspace{.5 cm} \text{ , } \hspace{.5 cm} \mathcal{P}_{\theta}^{\truncInd}\big [ (\textbf{v}\cdot\nabla)\theta \big ] =0.
			\end{equation}
			Let $\mathscr{M}_\varphi^{\truncInd} \subset \mathbb{Z}^3$, $\varphi\in\{\textbf{u}, \theta\}$ denote the wave vector selections in 
			$\mathscr{N}_{\theta}^{\truncInd},\mathscr{N}_{\textbf{u}}^{\truncInd}$, and $\mathscr{M}^{*}_\varphi \subset \mathscr{M}_\varphi$ those in $\mathscr{U}^*_\varphi$. 
			To enforce \eqref{e:lindyn} we require that for both $\varphi\in\{\textbf{u}, \theta\}$ and any $\textbf{m},\textbf{m}' \in \mathscr{M}^*_\varphi$, we have $\mathrm{diag}(\sigma_1,\sigma_2,\sigma_3) \textbf{m}+ \mathrm{diag}(\sigma_1',\sigma_2',\sigma_3') \textbf{m}'\notin \mathscr{M}_\varphi$ in the component-wise sense of \eqref{WaveCompatibilityCond} with possibly different choice of signs $\sigma_k, \sigma_k'\in\{1,-1\}$  for different components $k=1,2,3$. This ensures that nonlinear combinations from $\mathscr{U}^*$ do not lie in the truncated model, i.e.\ \eqref{e:lindyn} holds. Hence, if $\mathscr{U}^*$ is invariant under the linear dynamics, initial data that is nonzero only on $\mathscr{U}_*$ yields a linear invariant subspace of the nonlinear system \eqref{TruncatedEvolutionPDE} with linear dynamics. 
			
			The simplest example is $\mathscr{M}^*_\varphi = \{\textbf{m}\}$  for some index $\textbf{n}$, where $2m_{k}$ does not occur for any $k$-th component in $\mathscr{M}_\varphi^{\truncInd}$ for $k=1,2,3$ and both $\varphi\in\{\textbf{u}, \theta\}$. In that case at most six components are linearly coupled, cf.\ Figure~\ref{fig:linear_coupling}.  We first note that in case of a truncation to a box of Fourier modes, any $\textbf{m}$ can be chosen for which all components larger than or equal $\lfloor \frac{\tau_2}{2}\rfloor+1$. For the case of a box these are actually the only options to guarantee \eqref{e:lindyn} a priori since two different wave vectors from the box always form a triad within the box: the absolute value of the difference of at least one component lies in $(0,\tau_2)$. 
			
			We next discuss all possible cases that are relevant for Theorem~\ref{thm:runaway}, starting with the simplest $\Rot=0$ and $m_{1}=0$ or $m_{2}=0$ so that $\mathscr{C}^\textbf{m}_\textbf{u}=\{1\}$, cf.\ \eqref{PhaseIndexSets}. With $\textbf{n}=(\textbf{m},\textbf{p},1)$ the invariant subsystem reads 
			\begin{equation}\label{e:ODE-2D}
				\frac{d}{dt} 
				\begin{pmatrix}
					u^{\truncInd,\textbf{n}}\\
					\theta^{\truncInd,\textbf{n}}
				\end{pmatrix}=
				\begin{pmatrix}
					-\Pra|\mathcal{K}\textbf{m}|^2 & \Pra \Ray g_1 
					\\  g_1 & -|\mathcal{K}\textbf{m}|^2 
				\end{pmatrix}
				\begin{pmatrix}
					u^{\truncInd,\textbf{n}}\\
					\theta^{\truncInd,\textbf{n}}
				\end{pmatrix},
			\end{equation}
			where $g_1= \frac{\mathcal{G}_{3,3}^{\textbf{n}}\nu_3^{\textbf{n}}}{|\mathcal{G}^{\textbf{n}} \boldsymbol{\nu}^{\textbf{n}}|}$. 
			Concerning stability of this linear system, the trace and determinant of the system matrix are $t=-(\Pra+1)|\mathcal{K}\textbf{m}|^2<0$ and 
			$d= \Pra\left( |\mathcal{K}\textbf{m}|^4 - \Ray  g_1^2\right)$, respectively. 
			Hence, the stability properties depend in particular on $\Ray$ and on whether $\nu_3^{\textbf{n}}, \mathcal{G}_{3,3}^{\textbf{n}}$ is zero or not. 
			Here $\mathcal{G}_{3,3}^{\textbf{n}}= k_2 m_{2}-k_1 m_{1}$ and $\nu_3^{\textbf{n}}=\mathrm{sgn}(m_{1}^{p_1} m_{2}^{p_2} m_{3})\geq 0$, with the convention $0^0=1$. Hence, $\nu_3^{\textbf{n}}=0$ if $m_{3}=0$ or $p_j=1$ and $m_{j}=0$ for $j=1$ or $j=2$, and in all these cases we have exponential stability of the origin in the linear system. Otherwise  $g_1\neq0$ for $m_{2}+m_{1}>0$ and there is a unique critical value $\Ray_\mathrm{crit,1}=|\mathcal{K}\textbf{m}|^4/g_1^2$ of $\Ray>0$ such that for $\Ray=\Ray_\mathrm{crit,1}$ we have $d=0$ and thus a kernel of the matrix. This forms an unbounded line of equilibria, thus precluding the possibility of a compact attractor, and any larger value of $\Ray$ yields unbounded exponential growth, i.e., runaway modes. A concrete example is $m_{2}=p_1=p_2=0$, $m_{1}=m_{3}=\tau_2$. For $\tau_2=1$ this results from the classical Lorenz '63 case \eqref{e:Lorenz63} discussed above, when removing the wave vector $(0,0,2)$ from $\theta$. 
			
			However, $\Ray_\mathrm{crit,1}$ grows as $|\mathcal{K}\textbf{m}|^4$ since $|g_1|\leq 1$. In fact, for any linear subdynamics, independent of the number of modes involved, runway modes require $\Ray$ at least of order  $|\mathcal{K}\textbf{m}|^2$ (for fixed $\Pra$): Indeed, the diagonal entries in \eqref{GeneralBoussinesqODE_Vel}, \eqref{GeneralBoussinesqODE_Temp} are negative and scale as $|\mathcal{K}\textbf{m}|^2$, while all off-diagonal entries are $O(1)$ with respect to $\mathcal{K}$ and $\mathrm{m}$. Hence, the claim follows from Gershgorin's circle theorem, which states that all eigenvalues must lie in the union of circles around the diagonal entries of radius the sum of absolute values of the other entries in the corresponding row of column. 
			
			For $\Rot\neq 0$, according to \eqref{GeneralBoussinesqODE_Vel}, \eqref{GeneralBoussinesqODE_Temp} the linear coupling, cf.\ Figure~\ref{fig:linear_coupling}, introduces $\tilde{\textbf{n}}=(\textbf{m},\textbf{p}+(1,1),1)$ for $\theta$ and $u$. But  $\nu_3^{\tilde{\textbf{n}}}=0$ so that in the reduced system, $\theta^{\truncInd,\tilde{\textbf{n}}}$ is decoupled from the rest, thus we can reduce to the three-dimensional system \begin{equation}\label{e:ODE-3DS}
				\frac{d}{dt} 
				\begin{pmatrix}    
					u^{\truncInd,\textbf{n}}\\
					u^{\truncInd,\tilde{\textbf{n}}}\\
					\theta^{\truncInd,\textbf{n}}
				\end{pmatrix}=
				\begin{pmatrix}
					-\Pra|\mathcal{K}\textbf{m}|^2 & \Pra\Rot g_2 & \Pra \Ray g_1 
					\\ 
					\Pra\Rot g_3
					& -\Pra|\mathcal{K}\textbf{m}|^2 & 0 
					\\
					g_1 & 0 &
					-|\mathcal{K}\textbf{m}|^2  
				\end{pmatrix}
				\begin{pmatrix}    
					u^{\truncInd,\textbf{n}}\\
					u^{\truncInd,\tilde{\textbf{n}}}\\
					\theta^{\truncInd,\textbf{n}}
				\end{pmatrix}
			\end{equation}
			with $g_2= (-1)^{|\textbf{p}|_1} \frac{\mathcal{G}_{1,1}^{\tilde{\textbf{n}}} \mathcal{G}_{2,2}^{\textbf{n}} \nu_2^{\textbf{n}} - \mathcal{G}_{2,2}^{\tilde{\textbf{n}}} \mathcal{G}_{1,1}^{\textbf{n}} \nu_1^{\textbf{n}} }{|\mathcal{G}^{\tilde{\textbf{n}}} \boldsymbol{\nu}^{\tilde{\textbf{n}}}||\mathcal{G}^{\textbf{n}} \boldsymbol{\nu}^{\textbf{n}}| }$, $g_3= (-1)^{|\textbf{p}|_1} \frac{\mathcal{G}_{1,1}^{\textbf{n}} \mathcal{G}_{2,2}^{\tilde{\textbf{n}}} \nu_2^{\tilde{\textbf{n}}} - \mathcal{G}_{2,2}^{\textbf{n}} \mathcal{G}_{1,1}^{\tilde{\textbf{n}}} \nu_1^{\tilde{\textbf{n}}} }{|\mathcal{G}^{\tilde{\textbf{n}}} \boldsymbol{\nu}^{\tilde{\textbf{n}}}||\mathcal{G}^{\textbf{n}} \boldsymbol{\nu}^{\textbf{n}}| }$ ; the stability of this was also discussed in \cite{Welter2025Rotating}. Since we assume that exactly one of $m_{1},m_{2}$ is zero, inspecting $\nu_j^{\tilde{\textbf{n}}}$, $\nu_j^{\textbf{n}}$,  we find $g_2=-g_3\neq 0$ so that $g_2g_3<0$. 
			By the Routh-Hurwitz criteria, an unstable eigenvalue requires $\Pra(|\mathcal{K}\textbf{m}|^4-g_1^2 |\mathcal{K}\textbf{m}|^2 \Ray + g_2^2 |\mathcal{K}\textbf{m}|^2 \Rot^2) <0$ or $2 (1 + \Pra)^2 |\mathcal{K}\textbf{m}|^4 - g_1^2 (1 + \Pra) \Ray + 2 g_2^2 \Pra^2 \Rot^2<0$. For fixed $\Rot$, the above analysis shows under the given conditions that $g_1\neq 0$. Hence, one or both of these conditions are satisfied for sufficiently large $\Ray$ of order $|\mathcal{K}\textbf{m}|^4$. We also see that increasing $\Rot$ has a stabilizing effect. See also \cite{Welter2025Rotating}.
			
			\bigskip
			Next we consider $\textbf{m}>0$. For $\Rot=0$ the linear coupling, cf.\ Figure~\ref{fig:linear_coupling}, augments \eqref{e:ODE-2D} by $\textbf{n}' = (\textbf{m},\textbf{p},2)$, i.e. component index $2$. Hence, we obtain the three-dimensional system
			\[
			\frac{d}{dt} 
			\begin{pmatrix}    
				u^{\truncInd,\textbf{n}}\\
				u^{\truncInd,\textbf{n}'} \\
				\theta^{\truncInd,\textbf{n}}
			\end{pmatrix}=
			\begin{pmatrix}
				-\Pra|\mathcal{K}\textbf{m}|^2 & 0 & \Pra \Ray g_1 \\
				0 & -\Pra|\mathcal{K}\textbf{m}|^2 & \Pra\Ray g_1' \\
				g_1 & g_1'  & -|\mathcal{K}\textbf{m}|^2 
			\end{pmatrix}
			\begin{pmatrix}    
				u^{\truncInd,\textbf{n}}\\
				u^{\truncInd,\textbf{n}'} \\
				\theta^{\truncInd,\textbf{n}}
			\end{pmatrix}
			\]
			with $g_1'=\frac{\mathcal{G}_{3,3}^{\textbf{n}'}}{|\mathcal{G}^{\textbf{n}'} \boldsymbol{\nu}^{\textbf{n}'}|}$, since $\boldsymbol{\nu}^{\textbf{n}'} = (1,1,1)$. We have $\mathcal{G}_{3,3}^{\textbf{n}'} = -(k_1^3 k_2 m_{1}^2 + k_1 k_2^3 m_{2}^2 + k_1 m_{1} m_{3} + k_2 m_{2} m_{3})/(k_1 k_2))<0$, so that $g_1'<0$. The Routh-Hurwitz criteria in this case imply instability precisely in case $|\mathcal{K}\textbf{m}|^4 - (g_1^2 + (g_1')^2) \Ray)<0$, i.e., for $\Ray$ of order $|\mathcal{K}\textbf{m}|^4$ as above, but at a smaller threshold since $(g_1')^2>0$. 
			
			Finally, in case $\textbf{m}>0$, $\Rot\neq 0$ we need to additionally augment $\tilde{\textbf{n}}' = (\textbf{m},\textbf{p}+(1,1),2)$ and $\theta^{\truncInd,\tilde{\textbf{n}}}$ is no longer decoupled so that we obtain the six-dimensional system
			\begin{equation} \label{LinMat_Full3d}
				\frac{d}{dt} 
				\begin{pmatrix}    
					u^{\truncInd,\textbf{n}}\\
					u^{\truncInd,\tilde{\textbf{n}}'} \\
					u^{\truncInd,\tilde{\textbf{n}}}\\
					u^{\truncInd,\textbf{n}'}\\
					\theta^{\truncInd,\textbf{n}}\\
					\theta^{\truncInd,\tilde{\textbf{n}}}
				\end{pmatrix}=
				\begin{pmatrix}
					-\Pra|\mathcal{K}\textbf{m}|^2 & \Pra\Rot g_2' & 0 & 0 & \Pra \Ray g_1 & 0  \\
					-\Pra\Rot g_2' & -\Pra|\mathcal{K}\textbf{m}|^2 & 0 & 0 & 0 & \Pra \Ray g_1'  \\
					0 & 0 &-\Pra|\mathcal{K}\textbf{m}|^2 & \Pra\Rot g_2' &  0 & \Pra \Ray g_1'  \\
					0 & 0 & -\Pra\Rot g_2' & -\Pra|\mathcal{K}\textbf{m}|^2 & \Pra \Ray g_1 & 0 \\
					g_1 & 0 & 0 & g_1 & -|\mathcal{K}\textbf{m}|^2 & 0 \\
					0 & g_1' & g_1' & 0 & 0 & -|\mathcal{K}\textbf{m}|^2 
				\end{pmatrix}
				\begin{pmatrix}    
					u^{\truncInd,\textbf{n}}\\
					u^{\truncInd,\tilde{\textbf{n}}'} \\
					u^{\truncInd,\tilde{\textbf{n}}}\\
					u^{\truncInd,\textbf{n}'}\\
					\theta^{\truncInd,\textbf{n}}\\
					\theta^{\truncInd,\tilde{\textbf{n}}}
				\end{pmatrix}.
			\end{equation}
			with $g_2'= (-1)^{|\textbf{p}|_1} \frac{\mathcal{G}_{1,1}^{\tilde{\textbf{n}}'} \mathcal{G}_{2,2}^{\textbf{n}} - \mathcal{G}_{2,2}^{\tilde{\textbf{n}}'} \mathcal{G}_{1,1}^{\textbf{n}} }{|\mathcal{G}^{\tilde{\textbf{n}}'} \boldsymbol{\nu}^{\tilde{\textbf{n}}'}||\mathcal{G}^{\textbf{n}} \boldsymbol{\nu}^{\textbf{n}}| }$.  Here we used that various coefficients in the entries vanish for $\textbf{m}>0$, are equal or negatives of each other.  The determinant of the system matrix reads 
			\[ |\mathcal{K}\textbf{m}|^4 \Pra^4 (|\mathcal{K}\textbf{m}|^4 - g_1^2 \Ray + (g_2')^2 \Rot^2)(|\mathcal{K}\textbf{m}|^4 - (g_1')^2 \Ray + (g_2')^2 \Rot^2) \] 
			giving a kernel at the critical Rayleigh number $\Ray_\mathrm{crit}:= \min(\frac{|\mathcal{K}\textbf{m}|^4 + (g_2')^2 \Rot^2}{g_1^2},\frac{|\mathcal{K}\textbf{m}|^4 + (g_2')^2 \Rot^2}{(g_1')^2})$. This threshold illustrates the again the stabilizing effect of rotation, although it might not be the smallest in the complicated Routh-Hurwitz criteria.
			
		\end{proof}

		\subsubsection{Bounds for partially consistent type 1 models}
		
		
		While a type 1 inconsistency will inevitably lead to runaway modes for $\Ray$ sufficiently large, partial energetic consistency can extend the range of Rayleigh numbers for which the model still admits bounded dynamics.  In general, without assuming potential energy balance, following the steps in the proof of Proposition~\ref{prop:BalancePreservation} yields the following analogue of \eqref{ODE_AttractingBall}:
		\begin{equation}
			\frac{1}{2} \frac{d}{dt} \Big \langle \frac{|\textbf{u}^{\truncInd}|^2}{\Pra \Ray}  + \big ( \theta^{\truncInd} + \frac{2\pi}{\ShaThree} \ell^{\truncInd} \big )^2 \Big \rangle = 
			- \Big \langle  \frac{|\nabla \textbf{u}^{\truncInd}|^2}{\Ray} +  | \nabla \theta^{\truncInd} |^2\Big \rangle 
			+ \frac{2\pi}{\ShaThree} \langle \ell^{\truncInd} \partial_{x_3}^2 \theta^{\truncInd} \rangle + 2\imOp(u_3^{\truncInd} \theta^{\truncInd}), \label{e:enbal1}
		\end{equation}
		where the imbalance operator $\imOp$ is defined 
		\begin{equation}
			\imOp g:= \big \langle \left(1  + \frac{\pi}{\ShaThree}  \partial_{x_3} \ell^{\truncInd}\right) g  \big \rangle. 
		\end{equation}
		In case of potential energy balance, $g=u_3^{\truncInd} \theta^{\truncInd}$ lies in the kernel $\mathsf{ker} (\imOp)$ for any $\textbf{u}^{\truncInd} \in \textbf{L}^{2,\truncInd}_{\sigma}$, $\theta^{\truncInd}\in L^{2,\truncInd}_{0}$, cf.\ \eqref{TruncatedODESpaces}. In particular, $u_3^{\truncInd} \theta^{\truncInd}\in \mathsf{ker} (\imOp)$ is equivalent to \eqref{EnerConsis_Condition}.  More generally, $u_3^{\truncInd} \theta^{\truncInd} \in \mathsf{ker} (\imOp)$ holds only on a union of subspaces of $\textbf{L}^{\truncInd}:=\textbf{L}^{2,\truncInd}_{\sigma} \times L^{2,\truncInd}_{0}$ that depend on the truncation.  Following the proof of Proposition~\ref{prop:BalancePreservation}, a violation of Criterion~\ref{Crit:EnergyCrit} for a pair $(\textbf{n}', \textbf{n}'')$ means that only either the left hand side or the right hand side of \eqref{EnerConsis_Cancel} is present in the truncation so that \eqref{EnerConsis_Condition} fails for $(\textbf{n}', \textbf{n}'')$, which means $\imOp g\neq 0$ with $g=\textbf{v}^{\textbf{n}'}f^{\textbf{n}''}$. Hence, the energy equation is insufficient to provide a bound on the subspace of $\textbf{L}^{\truncInd}$ spanned by $(\textbf{v}^{\textbf{n}'},0), (0,f^{\textbf{n}''})$. From Lemma~\ref{lem:ODE_Existence} we know that $\Ray< (\lambda_1^{\truncInd})^4 $ ensures boundedness, but then the long-term dynamics is trivial, and this bound does not increase when adding high wave number modes. The latter should provide an improvement of the model accuracy, in particular for the full Fourier box truncation. 
		
		As a proof of principle, the following Lemma applies to the full Fourier box truncation and provides a rough threshold for $\Ray$ that increases with the number of high wave number modes.  For the formulation we define the following `imbalance' sets  $\mathscr{N}_{\rm imb ,1}^{\truncInd}, \mathscr{N}_{\rm imb ,2}^{\truncInd} \subset\mathscr{N}_{\textbf{u}}^{\truncInd} \times \mathscr{N}_{\theta}^{\truncInd}$ of mode pairs $(\textbf{n}',\textbf{n}'')$ satisfying $m'_1=m''_1, m'_2=m''_2, \textbf{p}'=\textbf{p}''$ and 
		\begin{itemize}
			\setlength{\itemsep}{0mm}
			\item[(i)] $(\textbf{n}',\textbf{n}'') \in \mathscr{N}_{\rm imb ,1}^{\truncInd}$ if $m_3'=m''_3$ and $(0,0,2m_3')$ is not a wave vector in $\mathscr{N}_{\theta}^{\truncInd}$,
			\item[(ii)] $(\textbf{n}',\textbf{n}'') \in \mathscr{N}_{\rm imb ,2}^{\truncInd}$ if $m_3'\neq m''_3$ and either $(0,0,|m_3'-m_3''|)$ is a wave vector in $\mathscr{N}_{\theta}^{\truncInd}$ but $(0,0,m'_3+m''_3)$ is not, or $(0,0,m_3'+m_3'')$ is a wave vector in $\mathscr{N}_{\theta}^{\truncInd}$ but $(0,0,|m'_3-m''_3|)$ is not.
		\end{itemize}
		Furthermore, let $\mathscr{N}_{\rm imb }^{\truncInd} := \mathscr{N}_{\rm imb ,1}^{\truncInd} \cup \mathscr{N}_{\rm imb ,2}^{\truncInd}$.  The Lemma is then as follows:
		\begin{Lemma}\label{lem:inconsRay}
			Consider any truncation (denoted with $\tilde \truncInd$) for which $\mathscr{N}_{\rm imb}^{\tilde \truncInd}$ is a subset of $\mathscr{N}_{\rm imb}^{(box,\tau_2)}$ for the full Fourier box truncation $\mathscr{N}_{\textbf{u}}^{(box,\tau_2)}$, $\mathscr{N}_{\theta}^{(box,\tau_2)}$, i.e. the truncation has the same or fewer imbalances. If $\Ray< \Ray^{\truncInd}:= \ShaThree^2 \tau_2$, then  the semi-group $\mathcal{S}^{\tilde \truncInd}(t)$ admits a compact global attractor.
		\end{Lemma}
		
		In particular, for the Fourier box truncation, the range of Rayleigh numbers for which the the semi-group $\mathcal{S}^{\truncInd}(t)$ admits a global attractor can be extended by increasing $\tau_2$. The bound $\Ray^{\truncInd}$ is not sharp (see the discussion at the end of this section) and it scales linearly in $\tau_2$, which is very far from the threshold of order $\tau_2^4$ of Theorem~\ref{thm:runaway}, beyond which runaway modes occur. We note that in \cite{HowardKrishnamurti_1986}, for the six component model studied there, it was shown that these thresholds coincide.
		
		\begin{proof}
			We show that under the given conditions, the right hand side of \eqref{e:enbal1} is negative for all large $|\textbf{u}^{\truncInd}|^2+|\theta^{\truncInd}|^2$, which is sufficient for the claim, analogous to the proof of, e.g.\ Proposition \ref{prop:BalancePreservation}.  From \eqref{EnerConsis_LHS} the cancellations \eqref{EnerConsis_Cancel} for modes that are in potential energy balance gives the following expression for the imbalance operator in Fourier space:
			\begin{align}
				\imOp (u_3^{\truncInd} \theta^{\truncInd}) = 2 \sum_{(\textbf{n}', \textbf{n}'') \in \mathscr{N}_{\rm imb }^{\truncInd}  } u^{\truncInd,\textbf{n}'} \theta^{\truncInd,\textbf{n}''} \langle  \cos(\ShaThree m_3 x_3) v^{\textbf{n}'}_3 f^{\textbf{n}''} \rangle\label{e:imb2}
			\end{align}
			where one notes that $m_3 = m_3(\textbf{n}', \textbf{n}'')$ in \eqref{e:imb2} is uniquely defined due to the definition of $\mathscr{N}_{\rm imb}^{\truncInd}$.  By the Peter-Paul case of Young's inequality we have $2 u^{\truncInd,\textbf{n}'} \theta^{\truncInd,\textbf{n}''} \leq \varepsilon_{\textbf{n}',\textbf{n}''}^{-1} (u^{\truncInd,\textbf{n}'})^2 + \varepsilon_{\textbf{n}',\textbf{n}''} (\theta^{\truncInd,\textbf{n}''})^2$ for any choice of weights $\varepsilon_{\textbf{n}',\textbf{n}''}>0$, and using $ \langle  \cos(\ShaThree m_3 x_3) v^{\textbf{n}'}_3 f^{\textbf{n}''} \rangle \leq 1$, we can estimate from \eqref{e:imb2} that 
			\begin{align*}
				\imOp (u_3^{\truncInd} \theta^{\truncInd}) \leq  \sum_{(\textbf{n}', \textbf{n}'') \in \mathscr{N}_{\rm imb}^{\truncInd}} \varepsilon_{\textbf{n}',\textbf{n}''}^{-1} (u^{\truncInd,\textbf{n}'})^2 + \varepsilon_{\textbf{n}',\textbf{n}''}(\theta^{\truncInd,\textbf{n}''})^2.
			\end{align*}
			Next, using \eqref{ConductStateFourier} one has 
			\[ \frac{2\pi}{\ShaThree} \langle \ell^{\truncInd} \partial_{x_3}^2 \theta^{\truncInd} \rangle = \frac{2\pi}{\ShaThree} \sum_{\textbf{n} \in \mathscr{N}_{\theta}^{\truncInd} \cap \mathscr{N}_{\theta}^{*}} - \ShaThree^2 m_3^2 \frac{2 \sqrt{2\pi}}{m_3} \theta^{\truncInd,\textbf{n}} \leq 16 \pi^3 |\mathscr{N}_{\theta}^{\truncInd} \cap \mathscr{N}_{\theta}^{*}| + \frac{1}{2} \sum_{\textbf{n} \in \mathscr{N}_{\theta}^{\truncInd} \cap \mathscr{N}_{\theta}^{*}} (\ShaThree m_3 \theta^{\truncInd,\textbf{n}} )^2 \text{ , } \]
			where we used the identity $\frac{\pi^2}{\ShaThree^2} \langle |\nabla \ell^{\truncInd}|^2 \rangle = 8\pi^3 |\mathscr{N}_{\theta}^{\truncInd} \cap \mathscr{N}^{*}_{\theta}|$ noted before \eqref{ODE_AttractingBall2}.  Applying these bounds to \eqref{e:enbal1} one obtains
			\[ \frac{1}{2} \frac{d}{dt} \Big \langle \frac{|\textbf{u}^{\truncInd}|^2}{\Pra \Ray}  + \big ( \theta^{\truncInd} + \frac{2\pi}{\ShaThree} \ell^{\truncInd} \big )^2 \Big \rangle \leq \mathcal{Q}^{\truncInd} (\textbf{u}^{\truncInd},\theta^{\truncInd}) + 16 \pi^3 |\mathscr{N}_{\theta}^{\truncInd} \cap \mathscr{N}^{*}_{\theta}|  ,  \] 
			in which the quadratic form is defined by
			\begin{align*}
				\mathcal{Q}^{\truncInd}(\textbf{u}^{\truncInd},\theta^{\truncInd}) := \mathcal{Q}^{\truncInd}_{\rm bal}(\textbf{u}^{\truncInd},\theta^{\truncInd}) + \mathcal{Q}^{\truncInd}_{\rm imb}(\textbf{u}^{\truncInd},\theta^{\truncInd}) , 
			\end{align*}
			in which 
			\begin{align*}
				\mathcal{Q}_{\rm bal}^{\truncInd} (\textbf{u}^{\truncInd}, \theta^{\truncInd}) & := -\sum_{\textbf{n}\in \mathscr{N}_{\textbf{u}}^{\truncInd}\setminus \mathscr{N}_{\rm imb, \textbf{u}}^{\truncInd}} \frac{|\mathcal{K}\textbf{m}|^2}{\Ray} (u^{\truncInd,\textbf{n}})^2  
				-\sum_{\textbf{n}\in \mathscr{N}_\theta^{\truncInd} \setminus ( \mathscr{N}_{\theta}^{*} \cup \mathscr{N}_{\rm imb, \theta}^{\truncInd} ) } |\mathcal{K}\textbf{m}|^2 (\theta^{\truncInd,\textbf{n}})^2 
				-\frac{1}{2} \sum_{\textbf{n}\in \mathscr{N}_\theta^{\truncInd} \cap \mathscr{N}_{\theta}^{*} } \ShaThree^2 m_3^2 (\theta^{\truncInd,\textbf{n}})^2  \text{ , } \\ \mathcal{Q}^{\truncInd}_{\rm imb}(\textbf{u}^{\truncInd},\theta^{\truncInd}) & := \sum_{\textbf{n}\in \mathscr{N}_{\rm imb,\textbf{u}}^{\truncInd}} \mathcal{Q}^{\truncInd, \textbf{n}}_{\textbf{u}}(\textbf{u}^{\truncInd,\textbf{n}})^2 +  \sum_{\textbf{n}\in \mathscr{N}_{\rm imb,\theta}^{\truncInd}} \mathcal{Q}^{\truncInd,\textbf{n}}_{\theta}(\theta^{\truncInd,\textbf{n}})^2 , \qquad \mathcal{Q}^{\truncInd,\textbf{n}}_{\textbf{u}} := - \frac{|\mathcal{K}\textbf{m}|^2}{\Ray} + 
				\sum_{\textbf{n}'' \in \mathscr{N}_{\theta}^{\truncInd}} 
				\chi_{ \mathscr{N}_{\rm imb}^{\truncInd}} (\textbf{n},\textbf{n}'')
				\varepsilon_{\textbf{n},\textbf{n}''}^{-1}, \\ \mathcal{Q}^{\truncInd,\textbf{n}}_{\theta} & :=
				- |\mathcal{K}\textbf{m}|^2 + 
				\sum_{\textbf{n}' \in \mathscr{N}_{\textbf{u}}^{\truncInd}} 
				\chi_{ \mathscr{N}_{\rm imb}^{\truncInd}} (\textbf{n}',\textbf{n})
				\varepsilon_{\textbf{n}',\textbf{n}}, 
			\end{align*}
			where the marginal sets $\mathscr{N}_{\rm imb, \textbf{u}}^{\truncInd},\mathscr{N}_{\rm imb, \theta}^{\truncInd}$ are given by 
			\[ \mathscr{N}_{\rm imb, \textbf{u}}^{\truncInd} = \{ \textbf{n} \in \mathscr{N}_{\textbf{u}}^{\truncInd} : \exists \text{ } \textbf{n}'' \in \mathscr{N}_{\theta}^{\truncInd} \text{ with } (\textbf{n},\textbf{n}'') \in \mathscr{N}_{\rm imb}^{\truncInd}  \} \hspace{.25 cm} \text{ , } \hspace{.25 cm} \mathscr{N}_{\rm imb, \theta}^{\truncInd} = \{ \textbf{n} \in \mathscr{N}_{\theta}^{\truncInd} : \exists \text{ } \textbf{n}' \in \mathscr{N}_{\textbf{u}}^{\truncInd} \text{ with } (\textbf{n}',\textbf{n}) \in \mathscr{N}_{\rm imb}^{\truncInd}  \} \text{ , }  \]
			and where $\chi_A$ is the characteristic function for a set $A$.
			
			In order to finish the proof, it is sufficient to find a critical Rayleigh number $\Ray^{\truncInd}$ and weights $\varepsilon_{\textbf{n}',\textbf{n}}$ such that the quadratic form $\mathcal{Q}^{\truncInd}$ is negative definite for $\Ray\leq \Ray^{\truncInd}$.  Note that $\mathcal{Q}^{\truncInd}$ is given by the direct sum $\mathcal{Q}_{\rm bal}^{\truncInd} \oplus \mathcal{Q}_{\rm imb}^{\truncInd}$ and $\mathcal{Q}_{\rm bal}^{\truncInd}$ is clearly negative definite on the kernel of $\mathcal{Q}_{\rm imb}^{\truncInd}$.  For each $\textbf{n} \in \mathscr{N}_{\rm imb, \textbf{u}}^{\truncInd}$ one has a threshold 
			\[ \Ray^{\truncInd,\textbf{n}} := |\mathcal{K}\textbf{m}|^2 \left(\sum_{\textbf{n}'' \in \mathscr{N}_{\theta}^{\truncInd}} \chi_{ \mathscr{N}_{\rm imb}^{\truncInd}} (\textbf{n},\textbf{n}'') \varepsilon_{\textbf{n},\textbf{n}''}^{-1}\right)^{-1}. \]
			such that $\mathcal{Q}_{\textbf{u}}^{\truncInd,\textbf{n}} < 0$ for $\Ray < \Ray^{\truncInd,\textbf{n}}$, so the goal is to choose the weights $\varepsilon_{\textbf{n}',\textbf{n}}$ to maximize 
			\[ \Ray^{\truncInd} := \min_{\textbf{n} \in \mathscr{N}_{\rm imb, \textbf{u}}^{\truncInd}} \Ray^{\truncInd,\textbf{n}} \text{ , } \]
			subject to the constraint $\mathcal{Q}^{\truncInd,\textbf{n}}_{\theta}\leq 0$ for all indices.  Note that only indices $\textbf{n}',\textbf{n}''$ with $\textbf{m}_h' =( m_1',m_2') = (m_1'',m_2'') = \textbf{m}_h'',\textbf{p}'=\textbf{p}''$ are coupled, so one can find the critical Rayleigh number $\Ray^{\truncInd,\textbf{m}_h,\textbf{p}}$ for each subsystem and take the minimum.  Thus we henceforth drop $\textbf{m}_h$ and $\textbf{p}$ and use only $m_3$ as an index, since we consider only the subsystems.  However, in general there are no further simplifications that can be made and there are many cases to consider.  On the other hand, one can make a comparison principle.  Consider any other truncation, denoted with $\tilde \truncInd$, with the property $\mathscr{N}_{\rm imb}^{\tilde \truncInd}\subset \mathscr{N}_{\rm imb}^{\truncInd}$. Then one has the monotonicity property that $\mathcal{Q}^{\tilde{\truncInd},\textbf{n}}_{\textbf{u}} \leq \mathcal{Q}^{\truncInd,\textbf{n}}_{\textbf{u}}$ for each choice of weights, and the monotonicity of $\Ray^{\truncInd,\textbf{m}_h,\textbf{p}} \leq \Ray^{\tilde{\truncInd},\textbf{m}_h,\textbf{p}}$ follows. Hence, the bound on $\Ray^{\truncInd}$ obtained from the truncation $\truncInd$ applies to the truncation $\tilde \truncInd$. To conclude the proof it thus remains to consider the Fourier box truncation. 
			
			Rather than pursue optimal bounds, the main goal of this theorem is to obtain qualitative bounds which grow unboundedly with $\tau_2$.  To this end, we implement the likely suboptimal choice $\varepsilon_{\textbf{n},\textbf{n}''} = \varepsilon_{\textbf{n}}$ so that 
			\[ \Ray^{\truncInd,\textbf{n}} = \frac{|\mathcal{K}\textbf{m}|^2\varepsilon_{\textbf{n}}}{|\mathscr{N}_{\rm imb}^{\truncInd}(\textbf{n},\cdot)|}\;,\qquad
			\mathcal{Q}_{\theta}^{\truncInd,\textbf{n}} := - |\mathcal{K}\textbf{m}|^2 + 
			\sum_{\textbf{n}' \in \mathscr{N}_{\textbf{u}}^{\truncInd}} 
			\chi_{ \mathscr{N}_{\rm imb}^{\truncInd}} (\textbf{n}',\textbf{n})
			\varepsilon_{\textbf{n}'} ,
			\]
			and specialize to the full Fourier box truncation.  As above $\mathscr{N}_{\rm imb}^{\truncInd}$ contains the pairs of indices whose wave vectors $\textbf{m}, \textbf{m}'$ which satisfy $\textbf{m}_h=\textbf{m}_h', \textbf{p}=\textbf{p}'$, hence we consider a generic subsystem, define $\kappa_{m_3} = \ShaOne^2 m_1^2 + \ShaTwo^2 m_2^2 + \ShaThree^2 m_3^2$ and replace the index $\textbf{n}$ by $m_3$.  For the Fourier box truncation the wave vectors must satisfy $m_3+m'_3>\tau_2$.  Hence, $|\mathscr{N}_{\rm imb}^{\truncInd}(\textbf{n},\cdot)| = m_3$, the $\mathcal{Q}_{\theta}^{\truncInd,\textbf{n}}$ are given by
			\[ \mathcal{Q}^{\truncInd,m_3}_{\theta}= - \kappa_{m_3}  + 
			\sum_{j=1}^{m_3} \varepsilon_{\tau_2-m_3+j},
			\]
			and straightforward computation shows that the linear system of equations $\mathcal{Q}^{\truncInd,m_3}_{\theta}=0$, $m_3=1,\ldots, \tau_2$, is solved by 
			$\varepsilon_{\tau_2} = \kappa_{1}$ and, for $j=1,\ldots,\tau_2-1$, $\varepsilon_{\tau_2-m_3+j} = \kappa_{m_3-j+1} - \kappa_{m_3-j}$, or equivalently 
			\[ \varepsilon_{\tau_2} = \kappa_{1}, \quad \varepsilon_{m_3} = \kappa_{\tau_2-m_3+1} - \kappa_{\tau_2-m_3} , \; m_3=1,\ldots,\tau_2-1. \]
			Substitution into $\Ray^{\truncInd,\textbf{n}}=\Ray^{\truncInd,m_3}$,  $m_3=1,\ldots, \tau_2$,  gives the conditions 
			\[
			\Ray< \Ray^{\truncInd,m_3} = \frac{\kappa_{m_3}\varepsilon_{m_3}}{m_3} = 
			\begin{cases}
				m_3=\tau_2 : & \frac{\kappa_{\tau_2}}{\tau_2}\kappa_1\\
				m_3=1,\ldots, \tau_2-1: & \frac{\kappa_{m_3}}{m_3}(\kappa_{\tau_2-m_3+1}-\kappa_{\tau_2-m_3}),
			\end{cases}
			\]
			From these explicit formulas one obtains the bounds $\Ray^{(box,\tau_2),m_3} \geq C_{m_3} \tau_2 \ShaThree^2$ for constants $C_{m_3}$, since for $m_3 = \tau_2$ it is obvious and 
			for $m_3=1,\ldots,\tau_2-1$, one has 
			\[
			\Ray^{(box,\tau_2),m_3} \geq \; \ShaThree^2 m_3 \left((\tau_2-m_3+1)^2- (\tau_2-m_3)^2 \right)  
			=  \ShaThree^2 m_3\left( 2(\tau_2-m_3)+1 \right)  > 2\ShaThree^2(\tau_2-1)  .
			\]
			Finally, the above is sufficient for the claim that $\Ray< \ShaThree^2 \tau_2$ ensures a negative definite quadratic form, since for $\Ray < \Ray^{\truncInd}$ one can choose the a perturbation of the above weights such that $\mathcal{Q}_{\theta}^{\truncInd,\textbf{n}} < 0$. 
		\end{proof}
		
		We finish this section by illustrating that $\Ray^{\truncInd}$ is generally far from optimal, in particular for `small' deviations from an energetically consistent truncation. Specifically, we consider the $\ell^{\infty}$ hierarchy, whose index sets are given explicitly in \eqref{HierarchyDef_Linf_Lock}. A minimal deviation from potential energy balance is to remove the wave vector $(0,0,2\tau_2)$ from $\mathscr{N}_{\theta}^{(1,\tau_2)}$ so that 
		\[ \mathscr{N}_{\rm imb}^{\truncInd} = \{(\textbf{n}',\textbf{n}'') \in \mathscr{N}_{\textbf{u}}^{(1,\tau_2)} \times \mathscr{N}_{\theta}^{(1,\tau_2)}: \textbf{p}' = \textbf{p}'', \textbf{m}''=(m_1,m_2,\tau_2), \textbf{m}'=(m_1,m_2,\tau_2), m_1,m_2=1,\ldots,\tau_2\}.
		\]
		Hence, for each $m_1,m_2$, as in the proof, we can use $m_3$ as an index and only need to consider $m_3=\tau_2$ in $\mathcal{Q}^{\truncInd,m_3}_{\theta}$, $\Ray^{\truncInd,m_3}$.  We then have $\mathcal{Q}^{\truncInd,\tau_2}_{\theta} = -\kappa_{\tau_2} + \varepsilon_{\tau_2}$ so that we choose $\varepsilon_{\tau_2}=\kappa_{\tau_2}$, which gives $\Ray^{\truncInd,\tau_2}=\kappa_{\tau_2}\varepsilon_{\tau_2} = \kappa_{\tau_2}^2 \geq \ShaThree^4 \tau_2^4 $. Notably, this scales as the runaway threshold from Theorem~\ref{thm:runaway} and the optimal bounds might coincide in this case. Similar computations yield an order $\tau_2^4$ scaling of the threshold when removing modes $(0,0,2\tau_2-j)$ for $j$ independent of $\tau_2$. However, it seems this fails when $j$ is order $\tau_2$, although it appears that an order $\tau_2^2$ scaling could be retained when optimizing the choice of weights. 
		
		Lastly, we again point out that the approach via the energy as a quadratic form is convenient since the nonlinear terms do not explicitly contribute. However, due to the non-normality of the linear part this approach cannot be expected to yield optimal thresholds. 
		
		\subsection{Analysis of a simple type two inconsistent model}
		
		We will say that a model $\mathscr{N}_{\textbf{u}}^{\truncInd}, \mathscr{N}_{\textbf{u}}^{\truncInd}$ has an energetic inconsistency of type 2 when it fails to fulfill Criterion \ref{Crit:EnergyCrit}, but does not have any inconsistency of type 1.  While type 1 energetic inconsistencies have a linear structure which one can easily exploit to prove the existence of unbounded dynamics, type 2 models have a nonlinear structure and it is not as clear what happens.  We will let $\tau_1 = 4$ denote a hierarchy of type 2 inconsistent models, although we are not really interested in studying the behavior of a full hierarchy of such models, but rather just a few illustrative examples.   
		
		The first type 2 inconsistent model we consider is chosen to be as simple as possible, so we fix all of the phases $\textbf{p} = \textbf{0}$.  We choose the model by selecting a mode $\textbf{n}^1 = (1,1,1,\textbf{p},1)$ and we deliberately break the consistency Criterion \ref{Crit:EnergyCrit} by not including the mode $(0,0,2,\textbf{p},1)$ in $\mathscr{N}_{\theta}^{\truncInd}$.  We want to include some nonlinear behavior, hence we include the mode $\textbf{n}^2 = (2,0,2,\textbf{p},1)$ which clearly satisfies the compatibility conditions \eqref{WaveCompatibilityCond}, \eqref{PhaseCompatibilityCond}, and one can check that $I_{\textbf{u}}^{(\textbf{n}^1,\textbf{n}^1,\textbf{n}^2)},I_{\theta}^{(\textbf{n}^1,\textbf{n}^1,\textbf{n}^2)} \neq 0$ in \eqref{def:NonlinearCoefs} .  If we include nothing more, then the mode $\textbf{n}^2$ exhibits a type 1 inconsistency, so we already know that it will exhibit runaway modes.  We therefore include the mode $\textbf{n}^3 = (0,0,4,\textbf{p},1)$ in $\mathscr{N}_{\theta}^{\truncInd}$, hence $\textbf{n}^2$ fulfills its consistency Criterion \ref{Crit:EnergyCrit}, and the model is given by
		\[ \mathscr{N}_{\textbf{u}}^{(4,1)} = \{ \textbf{n}^1 , \textbf{n}^2 \} \hspace{.5 cm} \text{ , } \hspace{.5 cm}  \mathscr{N}_{\theta}^{(4,1)} = \{ \textbf{n}^1 , \textbf{n}^2 , \textbf{n}^3 \}.  \]
		We remark here that as we developed this model it became clear that type 1 inconsistencies are abundant and perhaps the more common type of energy inconsistency.  Indeed, the only way to have a model which contains an energetically inconsistent mode $\textbf{n}$ and for which the linear subspace spanned by $\textbf{n}$ is not invariant is to ensure that either $(u^{\textbf{n}})^2$ or $u^{\textbf{n}}\theta^{\textbf{n}}$ appears as a nonlinearity for one of the other variables.  There are a limited number of available choices due to the compatibility conditions \eqref{WaveCompatibilityCond}, \eqref{PhaseCompatibilityCond}, and for example the mode $\textbf{n}^1$ was chosen to have fully 3 dimensional behavior because the above construction process does not work for 2 dimensional flows.  In 2d flows if one wants to include a nonlinear coupling for the mode $\textbf{n} = (m_1,0,m_3,\textbf{p},1)$, then the only options are to include either $\textbf{n}' = (0,0,2m_3,\textbf{p},1)$ or  $\textbf{n}'' = (2m_1,0,2m_3,\textbf{p},1)$.  The mode $\textbf{n}'$ is exactly the mode required for consistency, so if one wants an inconsistency this should be excluded.  However, the mode $\textbf{n}''$ satisfies $I_{\textbf{u}}^{(\textbf{n},\textbf{n},\textbf{n}'')} = I_{\theta}^{(\textbf{n},\textbf{n},\textbf{n}'')} = 0$, so there is no nonlinear coupling and the inconsistency would be of type 1.
		
		It turns out that despite the energetic inconsistency the model admits a compact attractor.  For the remainder of this subsection, the superscript $(4,1)$ will be dropped, and it is understood that $\textbf{X} = (u^{\textbf{n}^1},u^{\textbf{n}^2},\theta^{\textbf{n}^1},\theta^{\textbf{n}^2},\theta^{\textbf{n}^3})$ refers to this model.  We write the ODE somewhat abstractly as 
		\begin{equation}\label{e:type2}
			\frac{d}{dt} \textbf{X} = F(\textbf{X}):=\begin{pmatrix}
				-\Pra \kappa_1 & 0 & \Pra \Ray g_1 & 0 & 0\\ 
				0 & -\Pra \kappa_2 & 0 & \Pra \Ray g_2 & 0\\
				g_1 & 0 & -\kappa_1 & 0 & 0\\
				0 & g_2 & 0 & -\kappa_2 & 0\\
				0 & 0 & 0 & 0 & -\kappa_3
			\end{pmatrix} \textbf{X}
			+
			\begin{pmatrix}
				I_1 X_1 X_2 \\ -I_1 X_1^2\\ I_2 X_1 X_4 \\ -I_2 X_1 X_3 - I_3 X_2 X_5 \\ I_3 X_2 X_4
			\end{pmatrix},
		\end{equation}
		where inspecting the formulas for the specific terms gives in particular $\kappa_1,\kappa_2, \kappa_3, I_1>0$, $g_1,I_3\neq 0$, which is sufficient for the existence of a global attractor as in the lemma below.  To prove this, we will find a Lyapunov function for large $|\textbf{X}|$ of the form
		\[
		V(\textbf{X})= \frac 1 2 \sum_{j=1}^5 a_j (X_j - c_j)^2,
		\]
		where $a_j>0$ (so that $V$ possesses compact sublevel sets) and $\frac{d}{dt}V(\textbf{X}(t)) = \nabla V(\textbf{X}(t))\cdot \dot{\textbf{X}}(t)<0$ for $|\textbf{X}|\gg 1$. 
		
		In the proof we provide some details on how to find values of $a_j,c_j$, $j=1,\ldots,5$ since the approach might  apply more broadly to type 2 models. This in particular leverages the structure of the nonlinearity in its energy conserving nature. As noted before, for any truncation the resulting nonlinearity in the ODE format $\textbf{N}(\textbf{X})$ satisfies $\textbf{N}(\textbf{X})\cdot \textbf{X}=0$. The details of this cancellation depend on the way in which structure of $\textbf{N}$ is coupled to the components, i.e., the triads. For the nonlinear part $\textbf{N}$ of \eqref{e:type2} we have $(N_1, N_2)\cdot (X_1,X_2)=0$ and $(N_3, N_4,N_5)\cdot (X_3,X_4,X_5)=0$, i.e., cancellation among the velocity and temperature components, respectively. Hence, $\textbf{N} \cdot (a X_1, aX_2, bX_3, b X_4, b X_5)=0$  for any $a,b\in\mathbb{R}$, which gives an additional degree of freedom when seeking bounds. Such sub-cancellations occur more generally, but depend strongly on the chosen truncation, and the usefulness hinges on the relation to the $\Ray$-dependency of the linear part. 
		
		\begin{Lemma}\label{lem:type2sample}
			The ODE \eqref{e:type2} possesses a compact global attractor for any $\Ray\in\mathbb{R}$, if $\kappa_1,\kappa_2, \kappa_3, I_1, \Pra>0$, $g_1,I_3\neq 0$.
		\end{Lemma}
		
		\begin{proof}
			We show that there are $c_j$, $a_j>0$, $j=1,\ldots,5$ such that 
			\[ W(\textbf{X}^{(4,1)}):=\nabla V \cdot F(\textbf{X}^{(4,1)})<0, \;\ |\textbf{X}^{(4,1)}|\gg 1. \]
			Direct computations show that the cubic terms in $W$ cancel if and only if $a_1=a_2$, $a_3=a_4=a_5$. In order to cancel quadratic terms, we choose $c_1=c_3=c_4=0$ and $a_5\neq 0$ to be determined lated, and set (since $I_3\neq 0$)
			\[
			c_5 = \frac{a_5 + a_2 \Pra \Ray}{a_5 I_3}g_2. 
			\]
			Thus,  
			\[
			W(X) = a_2 (c_2 I_1 -  \kappa_1 \Pra ) X_1^2 + (a_5 g_1 +  a_2 g_1 \Pra \Ray) X_1 X_3 - a_5 \kappa_1 X_3^2 - a_2 \kappa_2 \Pra X_2^2 - a_5 \kappa_2 X_4^2 - a_5 \kappa_3 X_5^2,
			\]
			and only the first two terms are critical. Computing the Hessian of $W$ gives the negative eigenvalues $-2 a_2 \kappa_2 \Pra,-2 a_5 \kappa_2, -2 a_5 \kappa_3$, as well as the critical ones 
			\begin{align*}
				e_\pm := - a_5 \kappa_1 - E_1 \pm \sqrt{E_2}\;, \qquad & E_1:=a_2(\kappa_1 \Pra-c_2 I_1),\\
				&E_2:= E_1^2 -2 a_5\kappa_1 E_1 + a_5^2 (g_1^2 + \kappa_1^2) + 2 a_2 a_5  g_1^2 \Pra \Ray  + 
				a_2^2  g_1^2 {\Pra}^2 \Ray^2.
			\end{align*}
			The goal is to find $c_2, a_2, a_5$ such that $e_+<0$, and since $- a_5 \kappa_1<0$ it suffices to solve $E_1= \sqrt{E_2}$. For this it suffices to solve 
			$E_1^2=E_2$ with $E_1>0$, for which it suffices to have $c_2<0$. (Note that $E_2>0$ since eigenvalues of the Hessian are real.) We solve $E_1^2= E_2$ using $c_2$, which gives
			\[
			c_2^\pm := -\frac{1}{5a_2I_1}\left((4 a_5 - 5 a_2\Pra) \kappa_1  \pm \sqrt{E_3}\right)\; , \qquad 
			E_3 = 4 a_5^2 (5 g_1^2 + 9 \kappa_1^2) - 5a_2 g_1^2(8a_5 + a_2  {\Pra} \Ray)\Pra\Ray,
			\]
			and $c_2^\pm$ are real valued if $E_3>0$. This fails for fixed $a_2, a_5$ and sufficiently large $\Ray$, but solving $E_3=0$ in terms of $\Ray$ gives 
			\[
			\Ray_\pm = \frac{2 a_5}{
				a_2 \Pra} \left(-2 \pm \sqrt{5} \sqrt{1 + \frac {9  \kappa_1^2}{25 g_1^2}}\right),
			\]
			and $\Ray_+$ is monotonically increasing in $\kappa_1$, which gives $\Ray_+ > \frac{2a_5}{a_2 \Pra}(\sqrt{5}-2) $. Hence, $\Ray_+>0$ and it is linear in $a_5$. Notably, for given large $\Ray$, realizing $\Ray_+>\Ray$ requires $a_5\neq a_2$. Specifically, for any given $\Ray$, choosing 
			\[
			a_5 > \max\left\{ \frac{a_2\Pra\Ray}{2}  \left(\sqrt{5} \sqrt{1 + \frac {9  \kappa_1^2}{25 g_1^2})}-2\right)^{-1},\; \frac{5}{4} a_2\Pra \right\}
			\]
			gives $E_3=0$ and $c_2^\pm<0$ since $I_1>0$. Hence, with these choices we have $e_+<0$ and thus $\dot V<0$ along solutions. 
		\end{proof}
		
		We suspect that for similar bounds in type-2 inconsistent models, the remaining triads need to suitable relate with a linear dissipative mode. Conversely, we suspect that runaway modes can occur in type-2 models if this is not the case. 
		
		%
		%
		%
		
		\section{Numerical analysis}\label{s:numerics}
		
		This section summarizes some numerical investigations into heat transport in the Boussinesq Coriolis model via the HKC models.  First, we describe our code repository, in particular how the workflow was chosen to facilitate heat transport computations.  The results of several numerical studies regarding heat transport and energetic consistency are then presented.  Note that all of the computations were done using MATLAB on a standard laptop or desktop, and as mentioned above the codes are available on GitHub.
		
		\subsection{Code description}\label{s:code}
		
		We saw in Section \ref{sec:EnergyInconsis} that energetically inconsistent models can exhibit infinite Nusselt numbers, a violation of proven bounds in the 2D case and likely a violation in the 3D case as well.  On the other hand Lemma \ref{lem:TruncatedNusseltBound} shows that the Nusselt number for any fixed energetically consistent spectral truncation is bounded above by a constant independent of $\Ray$.  This too is in disagreement with the expected behavior for the PDE; numerical evidence and scaling laws suggest the Nusselt number increases unboundedly via $\mathsf{Nu} \sim \Ray^{\frac{1}{3}}$ or $\Ray^{\frac{5}{12}}$.  On the other hand, one sees that the bound in Lemma \ref{lem:TruncatedNusseltBound}  increases with the number of vertical modes included in the model, hence as we increase the Rayleigh number we need to generate hierarchies of higher resolution energetically consistent spectral models.  We generally favor the idea that a model with strictly higher resolution will give a more accurate Nusselt number, or more precisely
		\[ \mathscr{N}_{\textbf{u}}^{\tilde{\truncInd}} \subseteq \mathscr{N}_{\textbf{u}}^{\truncInd} \text{ and } \mathscr{N}_{\theta}^{\tilde{\truncInd}} \subseteq \mathscr{N}_{\theta}^{\truncInd} \hspace{.5 cm} \Rightarrow \hspace{.5 cm} |\mathsf{Nu}^{\truncInd} - \mathsf{Nu} | \leq |\mathsf{Nu}^{\tilde{\truncInd}} - \mathsf{Nu} | \text{ , } \]
		although we do not have any such monotonic convergence results analytically.  However, it is unclear which Fourier modes are most important for an accurate heat transport calculation, so for instance a comparison of Nusselt numbers from the $\ell^1$ vs the $\ell^{\infty}$ hierarchies is of interest.  Furthermore it is of interest whether fully three dimensional flows exhibit substantially different Nusselt numbers than two dimensional flows.
		
		With these things in mind, we specifically designed our codes to streamline the construction of arbitrary models.  This was done by separating the Fourier mode selection process from the model construction process.  In Figure \ref{fig:CodeRepo} below, we include a diagram which indicates the intended workflow and dependencies for our code repository.  From the user's perspective, the workflow begins in the folder "ModeSelection" contained in "ModelConstruction".  This folder contains the function $\mathtt{ModelSelector.m}$ which accepts as input a truncation index $\truncInd$ and produces as output the sets $\mathscr{N}_{\textbf{u}}^{\truncInd}, \mathscr{N}_{\theta}^{\truncInd}$.  This function is already built to handle the values $\tau_1 = 1,2,3$ corresponding to the $\ell^{\infty},\ell^{1}$ and HKC hierarchies above, but we have in mind that the user could create their own hierarchies here if desired.  With a specific hierarchy in mind, the user can generate any such model using the script $\mathtt{ModelConstructor.m}$ in the parent folder "ModelConstruction".  When the user provides a truncation number $\truncInd$ and box parameters $\ShaOne,\ShaTwo$, this script writes a file which corresponds to the right hand side of \ref{GeneralBoussinesqODE_Vel}, \ref{GeneralBoussinesqODE_Temp}, and saves this file in the folder "TimeSteppers".  Note that by providing the box parameters at this stage, one can evaluate all of the nonlinear coefficients \eqref{def:NonlinearCoefs} in advance of computing solutions.  While the drawback is that a model file must be generated for each box parameter of interest, one sees a dramatic increase in computational efficiency, and indeed it is often not of interest to adjust the box parameters dynamically over the course of a solution.
		
		\begin{figure}[H]
			\begin{centering}
				\begin{tikzpicture}[
					sqnode/.style={rectangle, draw=yellow!60, fill=yellow!5, very thick, minimum size=5mm},filenode/.style={rectangle, draw=blue!60, fill=blue!2, very thick, minimum size=5mm}
					,tips=proper]
					\node(CodeRepo)[sqnode] at (0,4)  {\includegraphics[width=14pt]{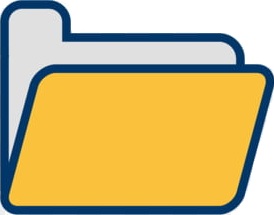} SpectralCodeRepo};
					
					\node(Col1NW)  at (-9,2.5) {};
					\node(Col1NE)  at (-5,2.5) {};
					\node(Col1SW)  at (-9,-0.5) {};
					\node(Col1SE)  at (-5,-0.5) {};
					\node(Col1ref)  at (-7,2.5) {};
					\node(ModCon)[sqnode]   at (-7,3)  {\includegraphics[width=14pt]{folderIcon.jpg} ModelConstruction};
					\node(AuxFunc)[sqnode]  at (-7,2)  {\includegraphics[width=14pt]{folderIcon.jpg} AuxiliaryFunctions};
					\node(ModSel)[sqnode]   at (-7,1)  {\includegraphics[width=14pt]{folderIcon.jpg} ModeSelection};
					\node(AuxFuncInv)       at (-7,1.5)  {};
					\node(AuxFuncInv2)      at (-8.75,1.5)  {};
					\node(AuxFuncInv3)      at (-8.75,.575)  {};
					\node(AuxFuncInv4)      at (-7.5,.575)  {};
					\node(AuxFuncInv5)      at (-7.5,0.25)  {};
					\node(Func_ModCon)[filenode]      at (-7,0)  {\includegraphics[width=10pt]{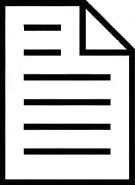} ModelConstructor.m};
					\node(Func_ModConInv)  at (-9.25,0)  {};
					
					\node(DiffEq)         at (-4.35,-.75)  {\makecell[c]{\includegraphics[width=12pt]{fileIcon.jpg} \\ Model.m}};
					\node(DiffEqInv)      at (-7,-.75)  { };
					\node(DiffEqInv2)      at (-4.35,2)  { };
					
					
					\node(Col2NW)  at (-3.65,2.5) {};
					\node(Col2NE)  at (-.35,2.5) {};
					\node(Col2SW)  at (-3.65,-0.5) {};
					\node(Col2SE)  at (-.35,-0.5) {};
					\node(Col2ref)  at (-2,2.5) {};
					\node(FluSol)[sqnode]   at (-2,3)  {\includegraphics[width=14pt]{folderIcon.jpg} FluidSolver};
					\node(TimSte)[sqnode]   at (-2,2)  {\includegraphics[width=14pt]{folderIcon.jpg} TimeSteppers};
					\node(TimSteInv)       at (-2,1.5)  {};
					\node(TimSteInv2)      at (-3.4,1.5)  {};
					\node(TimSteInv3)      at (-3.4,.575)  {};
					\node(TimSteInv4)      at (-2.5,.575)  {};
					\node(TimSteInv5)      at (-2.5,0.25)  {};
					\node(Init)[filenode]  at (-2,1)  {\includegraphics[width=14pt]{folderIcon.jpg} InitCond.m};
					\node(Func_FluSol)[filenode]      at (-2,0)  {\includegraphics[width=10pt]{fileIcon.jpg} FluidSolver.m};
					\node(Func_FluSolInv)   at (-2,-.75)  {};
					\node(Func_FluSolInv2)   at (4.5,-.75)  {};
					\node(Func_FluSolInv3)   at (4.5,-.25)  {};
					\node(Func_FluSolInv4)   at (5.65,-.25)  {};
					
					
					\node(Col3NW)  at (.5,1.5) {};
					\node(Col3NE)  at (3.5,1.5) {};
					\node(Col3SW)  at (.5,-0.5) {};
					\node(Col3SE)  at (3.5,-0.5) {};
					\node(Col3ref)  at (2,1.5) {};
					\node(HeaTra)[sqnode]   at (2,3)  {\includegraphics[width=14pt]{folderIcon.jpg} HeatTransport};
					\node(Func_HeaIte)[filenode]      at (2,0)  {\includegraphics[width=10pt]{fileIcon.jpg} Iterator.m};
					\node(Func_HeaAss)[filenode]      at (2,1)  {\includegraphics[width=10pt]{fileIcon.jpg} Assembler.m};
					\node(Func_HeaAssInv)         at (3.35,.8)      {};
					
					
					\node(Col4NW)  at (5.25,1.5) {};
					\node(Col4NE)  at (8.75,1.5) {};
					\node(Col4SW)  at (5.25,-0.5) {};
					\node(Col4SE)  at (8.75,-0.5) {};
					\node(Col4ref)  at (7,1.5) {};
					\node(SavDat)[sqnode]   at (7,3)  {\includegraphics[width=14pt]{folderIcon.jpg} Saved Data};
					\node(HieNam)[sqnode]  at (7,2)  {\includegraphics[width=14pt]{folderIcon.jpg} HierarchyName};
					\node(HeaTraSav)[sqnode]  at (7,1)  {\includegraphics[width=14pt]{folderIcon.jpg} HeatTransport};
					\node(TrajSav)[sqnode]  at (7,0)  {\includegraphics[width=14pt]{folderIcon.jpg} Trajectories};
					
					\draw[-] (CodeRepo.south) edge (ModCon.north);
					\draw[-] (CodeRepo.south) edge (FluSol.north);
					\draw[-] (CodeRepo.south) edge (HeaTra.north);
					\draw[-] (CodeRepo.south) edge (SavDat.north);
					
					\draw[-] (ModCon.south) edge (Col1ref.center);
					\draw[-] (FluSol.south) edge (Col2ref.center);
					\draw[-] (HeaTra.south) edge (Col3ref.center);
					\draw[-] (SavDat.south) edge (HieNam.north);
					\draw[-] (HieNam.south) edge (Col4ref.center);
					
					\draw[-] (Col1NW.center) edge (Col1NE.center) (Col1NE.center) edge (Col1SE.center) (Col1SE.center) edge (Col1SW.center) (Col1SW.center) edge (Col1NW.center);
					\draw[-] (Col2NW.center) edge (Col2NE.center) (Col2NE.center) edge (Col2SE.center) (Col2SE.center) edge (Col2SW.center) (Col2SW.center) edge (Col2NW.center);
					\draw[-] (Col3NW.center) edge (Col3NE.center) (Col3NE.center) edge (Col3SE.center) (Col3SE.center) edge (Col3SW.center) (Col3SW.center) edge (Col3NW.center);
					\draw[-] (Col4NW.center) edge (Col4NE.center) (Col4NE.center) edge (Col4SE.center) (Col4SE.center) edge (Col4SW.center) (Col4SW.center) edge (Col4NW.center);

					
					\draw[-, red, very thick] (AuxFunc.south) edge (AuxFuncInv.center) (AuxFuncInv.center) edge (AuxFuncInv2.center) (AuxFuncInv2.center) edge (AuxFuncInv3.center) (AuxFuncInv3.center) edge (AuxFuncInv4.center) ;
					\draw[->, red, very thick] (AuxFuncInv4.center) edge (AuxFuncInv5.north) ;
					\draw[->, red, very thick] (ModSel.south) edge (Func_ModCon.north) ;
					\draw[-, red, very thick] (TimSte.south) edge (TimSteInv.center) (TimSteInv.center) edge (TimSteInv2.center) (TimSteInv2.center) edge (TimSteInv3.center) (TimSteInv3.center) edge (TimSteInv4.center) ;
					\draw[->, red, very thick] (TimSteInv4.center) edge (TimSteInv5.north) ;
					\draw[->, red, very thick] (Init.south) edge (Func_FluSol.north) ;
					\draw[->, red, very thick] (Func_FluSol.east) edge (Func_HeaIte.west) ;
					\draw[->, red, very thick] (TrajSav.west) edge (Func_HeaAssInv.center) ;
					
					
					\draw[-, to path={-| (\tikztotarget)}, blue, very thick] (Func_ModCon.south) edge (DiffEqInv.center) ;
					\draw[->, blue, very thick] (DiffEqInv.center) edge (DiffEq.west) ;
					\draw[-, to path={-| (\tikztotarget)}, blue, very thick] (DiffEq.north) edge (DiffEqInv2.center) (DiffEqInv2.center) edge (TimSte.west) ;
					\draw[->, blue, very thick] (DiffEqInv2.center) edge (TimSte.west) ;
					\draw[-, blue, very thick] (Func_FluSol.south) edge (Func_FluSolInv.center) (Func_FluSolInv.center) edge (Func_FluSolInv2.center) (Func_FluSolInv2.center) edge (Func_FluSolInv3.center);
					\draw[->, blue, very thick] (Func_FluSolInv3.center) edge (Func_FluSolInv4.center) ;
					\draw[->, blue, very thick] (Func_HeaAss.east) edge (HeaTraSav.west) ;
					\draw[->, blue, very thick] (Func_HeaIte.east) edge (TrajSav.west) ;
					
				\end{tikzpicture}
			\end{centering}
			\caption{A diagram depicting the dependencies and work flow structure of the code repository.  Yellow boxes indicate folders and blue boxes indicate scripts or functions.  Black lines indicate folder locations, red lines indicate code dependencies and blue lines indicate output locations.}
			\label{fig:CodeRepo}
		\end{figure}
		
		When the user has the time stepper for an ODE model of interest, they can move to the folder "FluidSolver".  The function $\mathtt{FluidSolver.m}$ takes a vector of physical parameters $\parameters$, a truncation index $\truncInd$ and a vector of computational parameters $\compParam$ then uses the standard MATLAB routine $\mathtt{ode45}$ together with the time stepper for the model $\truncInd$ to generate a numerical solution $\textbf{X}^{\truncInd}$ of that ODE, which is then saved to the appropriate "Trajectories" folder.  The computational parameters which must be provided are as follows:
		\begin{align} \compParam := ( \mathtt{tInc} , \mathtt{Tf}, \mathtt{errTol}, \mathtt{runTime} ) \hspace{.25 cm } \text{ , } \hspace{.25 cm } \mathtt{tInc} & = \text{time Increment } \hspace{.25 cm } \text{ , } \hspace{.25 cm } \mathtt{Tf} = \text{final time , } \notag \\  \mathtt{errTol} = \text{error tolerance } \hspace{.25 cm } & \text{ , } \hspace{.25 cm } \mathtt{runTime} = \text{cumulative run time } \text{ . }  \notag \end{align}
		The function $\mathtt{FluidSolver.m}$ thus computes an approximation for the solution $\textbf{X}^{\truncInd}$ at times $t_j = j\cdot \mathtt{tInc}$ for $j = 0,...,\frac{\mathtt{Tf}}{\mathtt{tInc}}$.  $\mathtt{FluidSolver.m}$ calls the function $\mathtt{InitCond.m}$ to provide the initial conditions $\textbf{u}_0^{\truncInd},\theta_0^{\truncInd}$.  The initial conditions which are of primary interest for this paper are random perturbations of the uniform initial state $(\textbf{u}_0(\textbf{x}),T_0(\textbf{x})) = (0,1/2)$, although by modifying $\mathtt{InitCond.m}$ the user could study solutions beginning from the initial condition of their choice.  On the other hand, $\mathtt{FluidSolver.m}$ can be called with additional arguments.  If one calls $\mathtt{FluidSolver.m}$ with a previously computed trajectory $\mathtt{Xprev}$ and a scalar $\mathtt{addTime}$, then $\mathtt{FluidSolver.m}$ extends the solution from the time interval $[0,\mathtt{Tf} ]$ to the interval $[0,\mathtt{Tf} + \mathtt{addTime}]$.   
		
		For heat transport computations one is interested in solutions for many different parameter values $\parameters$, so it does not make sense to generate trajectories one at a time.  Moving to the folder "HeatTransport", the user can run the script $\mathtt{Iterator.m}$ to compute Nusselt number approximations for a whole range of parameter values.  For each parameter value in the desired range, $\mathtt{Iterator.m}$ calls $\mathtt{FluidSolver.m}$ to generate a trajectory on a user specified fixed time interval $[0,\mathtt{initTime}]$.  $\mathtt{Iterator.m}$ then computes a finite time Nusselt number approximation via the Fourier analogue of the third expression in \eqref{NusseltNumExpressions}, also used in  \eqref{NusseltFourier}, as follows:
		\[ \widetilde{\mathsf{Nu}}^{\truncInd}(\mathtt{Tf}) = 1  - \frac{\ShaThree }{\sqrt{2\pi^{5}}} \frac{\mathtt{tInc}}{\mathtt{Tf}} \sum_{j \leq \frac{\mathtt{Tf}}{\mathtt{tInc}}} \sum_{\textbf{n} \in \mathscr{N}^{\truncInd}_{\theta} \cap \mathscr{N}^{*}_{\theta}} m_3 \theta^{\truncInd,\textbf{n}}(  t_j ) \text{ , }  \] 
		with $t_j = j \cdot \mathtt{tInc}$.  In order to check that the finite integration time is sufficiently long, $\mathtt{Iterator.m}$ then computes the standard deviation of $\mathsf{Nu}^{\truncInd}(t)$ over the second half of the time interval, ie $[\mathtt{initTime}/2, \mathtt{initTime}]$.  If this is below a user specified threshold ($2\%$ for the results herein), then it saves the trajectory to the appropriate "Trajectories" folder.  If the standard deviation is not below this threshold, then $\mathtt{Iterator.m}$ repeatedly calls $\mathtt{FluidSolver.m}$ to extend the trajectory until either the threshold is met or a maximum number of extensions is reached.  After this loop is completed, we compute another finite time approximation for the Nusselt number via the Fourier space representation of the first expression in \eqref{NusseltNumExpressions} as follows:
		\[ \mathsf{Nu}^{\truncInd}(\mathtt{Tf}) := 1 + \frac{\mathtt{tInc}}{4\pi^4 \mathtt{Tf}} \sum_{j \leq \frac{\mathtt{Tf}}{\mathtt{tInc}}} \sum_{\textbf{n} \in \mathscr{N}^{\truncInd}_{\theta}} \sum_{\substack{\textbf{n}' \in \mathscr{N}^{\truncInd}_{\textbf{u}} \\ \textbf{m}' = \textbf{m} \\ \textbf{p}' = \textbf{p} } } \frac{\mathcal{G}_{3,3}^{\textbf{n}}}{|\mathcal{G}^{\textbf{n}}\boldsymbol{\nu}^{\textbf{n}}|}  u^{\truncInd,\textbf{n}'} ( t_j ) \theta^{\truncInd,\textbf{n}} (  t_j ) \text{ . }\]
		Indeed, while one expects that the infinite time averages in \eqref{NusseltNumExpressions} coincide for energetically consistent models, comparing the finite time averages $\mathsf{Nu}^{\truncInd}(\mathtt{Tf}),\widetilde{\mathsf{Nu}}^{\truncInd}(\mathtt{Tf})$ can serve as another check for sufficient integration time.  Finally, when trajectories have been computed for a whole range of parameter values, the user can run the script $\mathtt{Assembler.m}$, which collects the Nusselt number approximations for the range of parameter values and compiles them into a single variable, which is then saved in the appropriate "HeatTransport" folder.
		
		\subsection{Heat transport in energetically consistent models}
		
\begin{figure}[H]
			\begin{center}
				\begin{tabular}{cc}
					\includegraphics[height=60mm]{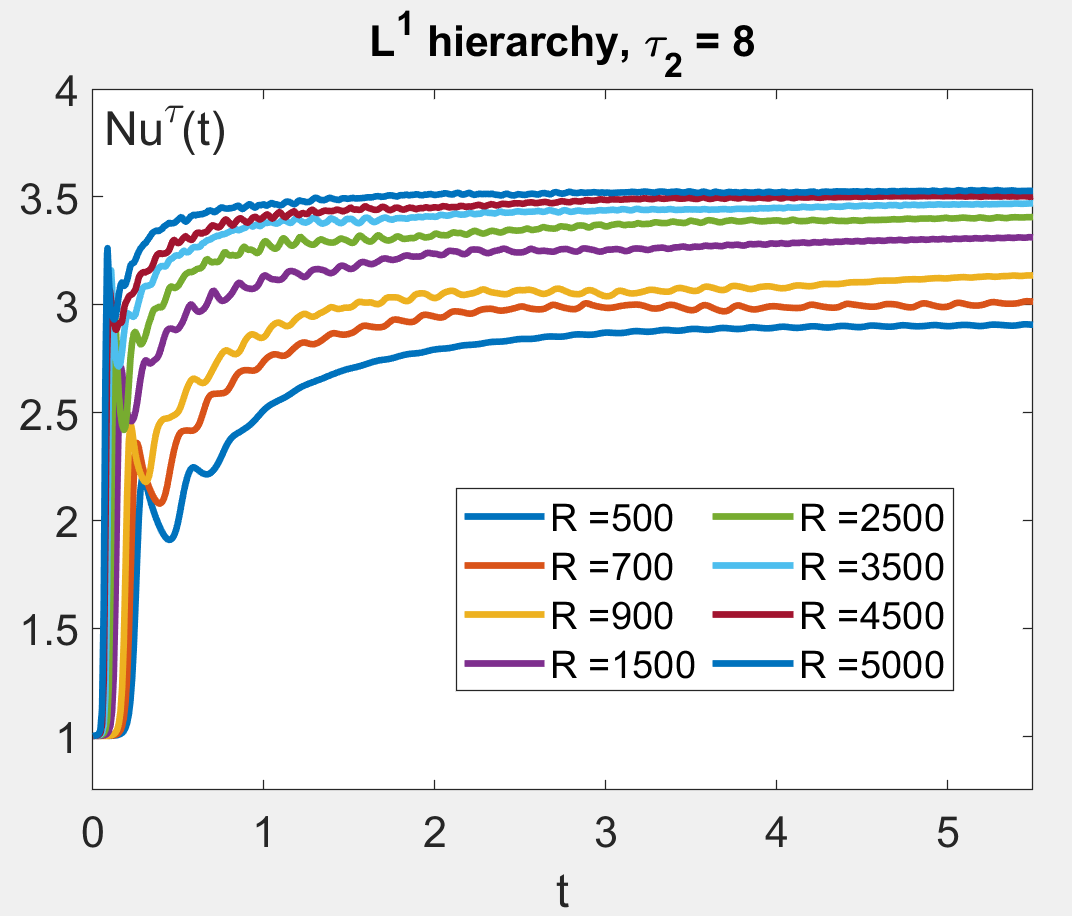}
					&
					\includegraphics[height=60mm]{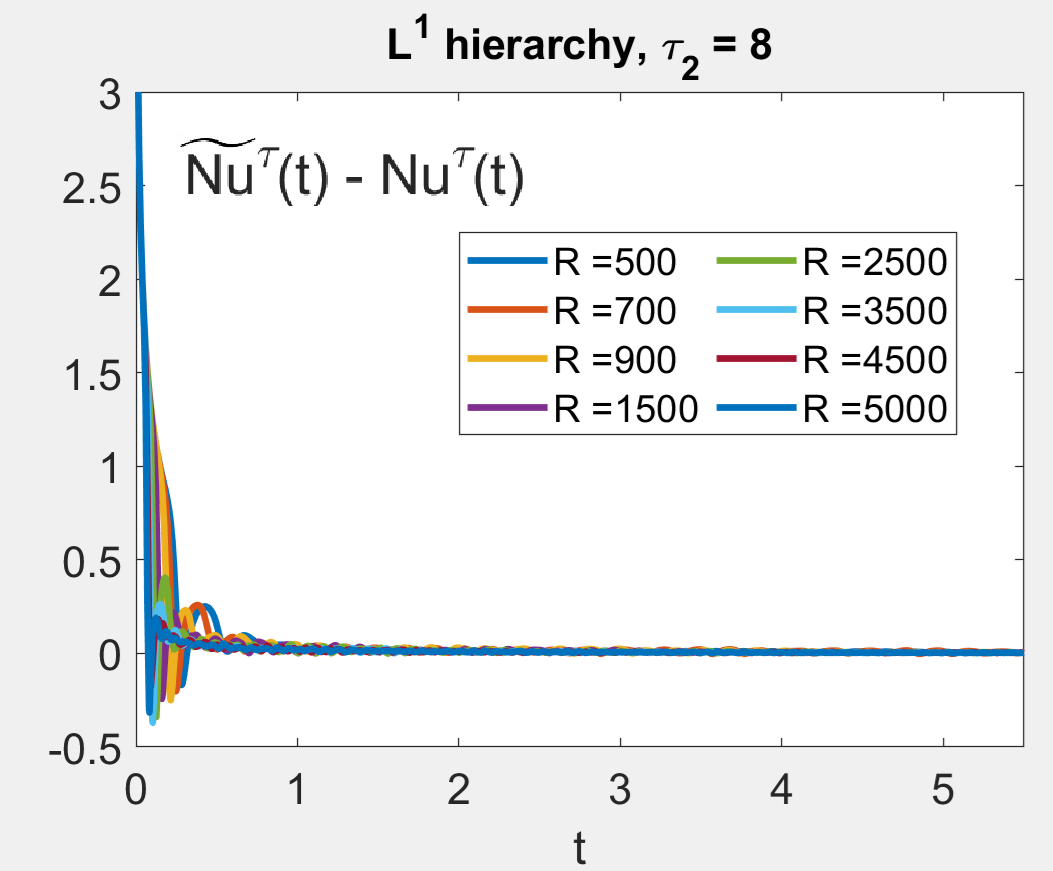}  \\
					(a) & (b)
				\end{tabular}
				\caption{Convergence of the finite time Nusselt numbers as a function of time for the $8^{th}$ model in the $\ell^1$ hierarchy at several Rayleigh numbers. }
				\label{fig:HeatTransportConvergence}
			\end{center}
		\end{figure}

		In order to study the convergence of the Nusselt numbers from the ODE models toward a candidate Nusselt number for the PDE, trajectories were generated using $\mathtt{Iterator.m}$ for each the following models in the $\ell^{\infty}$ and $\ell^1$ hierarchies defined in \eqref{HierarchyDef_Linf_Lock}, \eqref{HierarchyDef_L1_Lock}, beginning from a random perturbation of the uniform initial state $(\textbf{u}_0(\textbf{x}),T_0(\textbf{x})) = (0,1/2)$:
		\begin{equation} \label{NumericalStudy_ModelNumbers} \ell^{\infty} :  \tau_2 = 1 \text{ , } 2 \text{ , } 3 \text{ , } 4 \hspace{.75 cm} \text{ , } \hspace{.75 cm} \ell^1 :  \tau_2 = 3 \text{ , } 4 \text{ , } 5 \text{ , } 6 \text{ , } 7 \text{ , } 8 \text{ . } \end{equation}
		For each model, a trajectory was generated  with $\Pra = 10,$ $\ShaOne = \ShaTwo = 1$, for each pair of $\Ray,\Rot$ in the following range:
		\begin{equation} \label{HKC_ModelComparison_ParameterRange}  \Ray = [ 1 , 50:50:500 , 600:100:1000 , 1500:1000:5000  ]  \hspace{.5 cm} \text{ , } \hspace{.5 cm}  \Rot =  0:50:400  . \end{equation}
		where $x_0 : \Delta x: x_f$ is the MATLAB notation denoting the set of all numbers beginning from $x_0$, incrementing by $\Delta x$ and ending with $x_f$.  For these trajectories, time increment and error tolerance were set $10^{-4}$ and $10^{-3}$ respectively, and the trajectories were initially integrated for $5\cdot 10^{4}$ time increments, then $10^{3}$ more time increments until the fluctuations in heat transport met the 2\% error threshold.  Furthermore, the expressions $\mathsf{Nu}^{\truncInd}(\mathtt{Tf}), \widetilde{\mathsf{Nu}}^{\truncInd}(\mathtt{Tf})$ were checked against each other, and it was always found that the difference was less than $1\%$ of the value of $\mathsf{Nu}^{\truncInd}(\mathtt{Tf})$.  Thus this choice of finite integration time seemed to give good convergence results, for instance Figure \ref{fig:HeatTransportConvergence} displays the convergence of the finite time Nusselt number for the $8^{th}$ model in the $\ell^1$ hierarchy for each of the different Rayleigh numbers in \eqref{NumericalStudy_ModelNumbers}, as well as the asymptotic agreement between $\mathsf{Nu}^{\truncInd}(\mathtt{Tf}), \widetilde{\mathsf{Nu}}^{\truncInd}(\mathtt{Tf})$.  Similar temporal convergence results held for all models above.
		
\begin{figure}[H]
			\begin{center}
				\begin{tabular}{cc}
					\includegraphics[height=68mm]{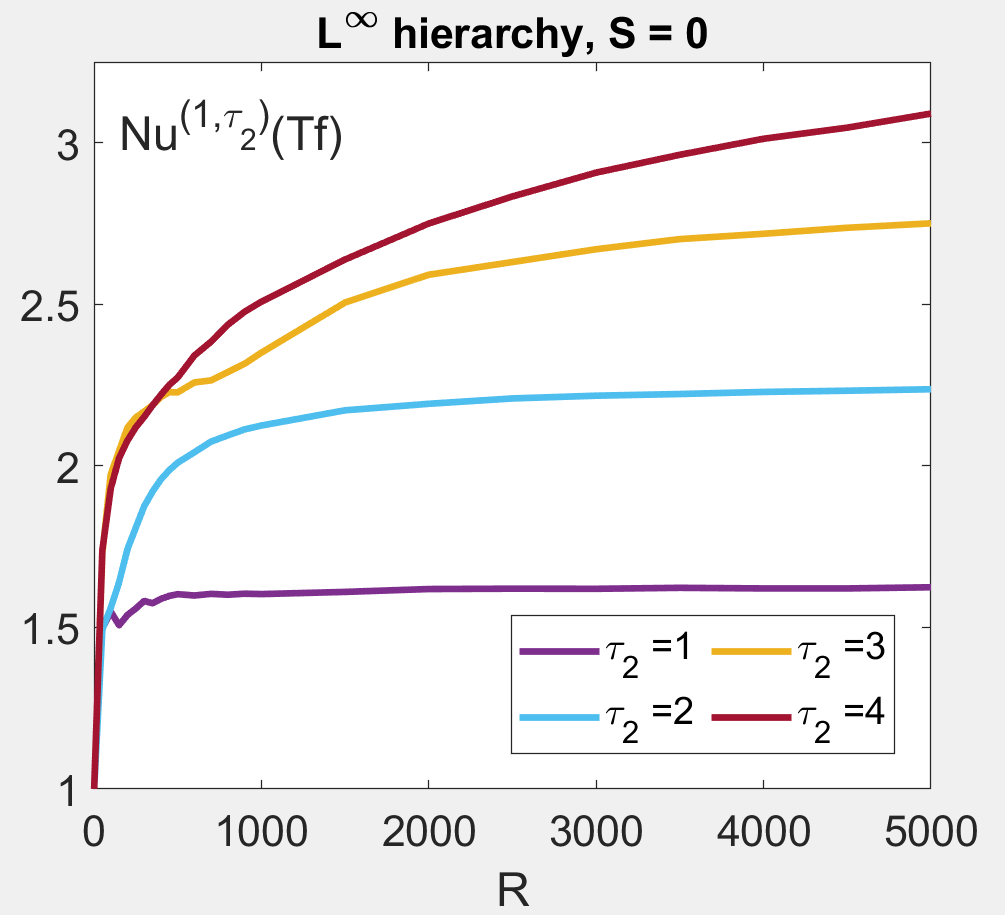}
					&\includegraphics[height=68mm]{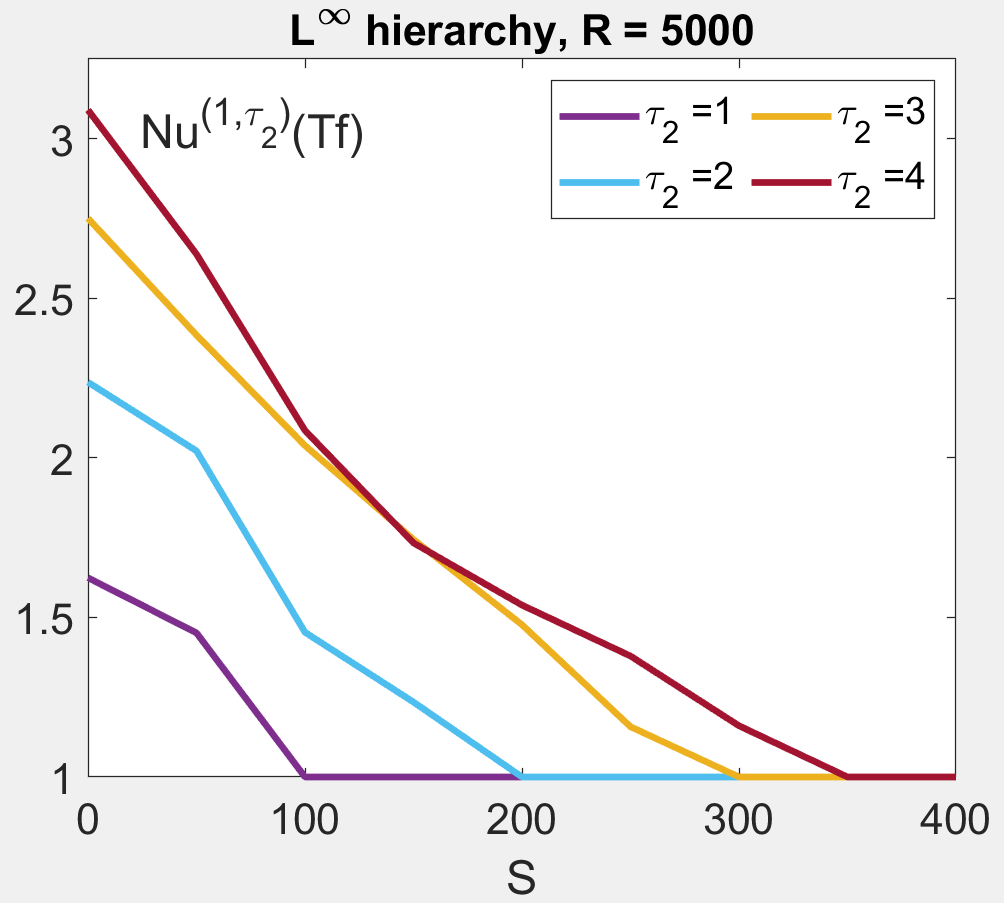}  \\
					(a) & (b)
				\end{tabular}
				\caption{Nusselt comparison for the $\ell^{\infty}$ hierarchy (a) for $0 \leq \Ray \leq 5000, \Rot = 0$ and (b) for $\Ray = 5000$, $0 \leq \Rot \leq 400$.}
				\label{fig:HeatTransport_Linf}
			\end{center}
		\end{figure}

\begin{figure}[H]
			\begin{center}
				\begin{tabular}{cc}
					\includegraphics[height=68mm]{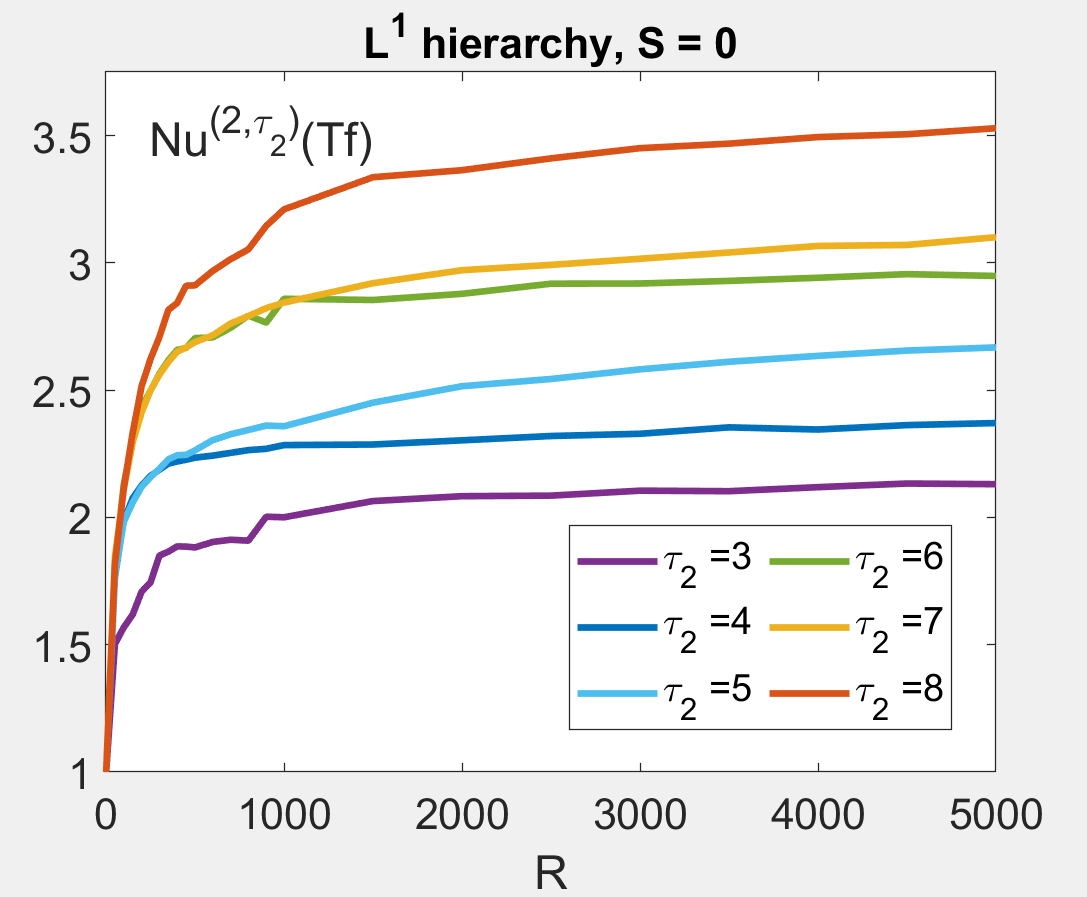}
					&\includegraphics[height=68mm]{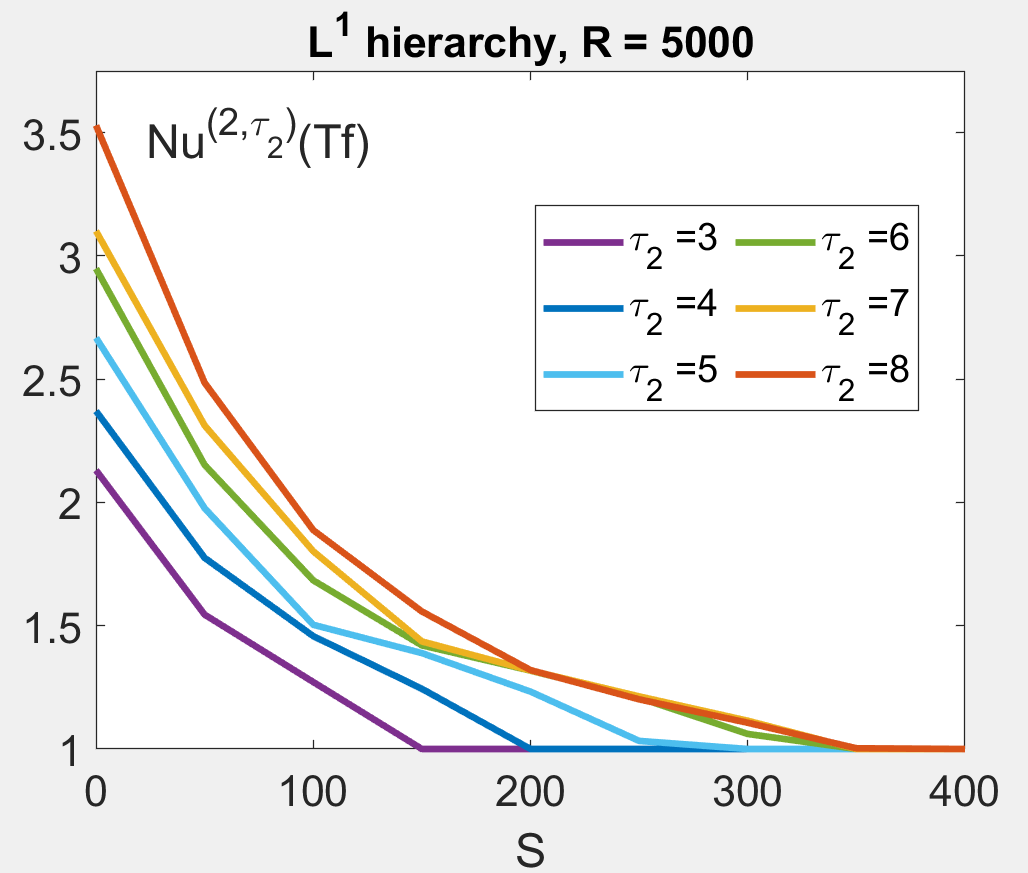}  \\
					(a) & (b)
				\end{tabular}
				\caption{Nusselt comparison for the $\ell^{1}$ hierarchy, (a) for $0 \leq \Ray \leq 5000, \Rot = 0$ and (b) for $\Ray = 5000$, $0 \leq \Rot \leq 400$.}
				\label{fig:HeatTransport_L1}
			\end{center}
		\end{figure}

		The results of the heat transport computations for all models \eqref{NumericalStudy_ModelNumbers} in the $\ell^{\infty}$ and $\ell^1$ hierarchies are shown in Figures \ref{fig:HeatTransport_Linf} and \ref{fig:HeatTransport_L1}, respectively.  Panel (a) in each Figure displays the dependence of the heat transport on $\Ray$ for $\Rot = 0$, whereas panel (b) in each Figure shows the dependence of the heat transport on the Coriolis number for $\Ray = 5000$.  
		
		One sees in Figures \ref{fig:HeatTransport_Linf} and \ref{fig:HeatTransport_L1} that in each model the heat transport increases rather rapidly for low Rayleigh numbers, then the rate of increase levels off for higher Rayleigh numbers.  This increase is fairly monotonic, although with some fluctuations.  In agreement with Lemma \ref{lem:TruncatedNusseltBound}, the higher dimensional models within a hierarchy almost always exhibit more heat transport than the lower dimensional models.  The dependence of the approximate Nusselt number on the model resolution is already very clear at Rayleigh numbers as low as $300$ or so, and can produce significant differences, where for instance the heat transport exhibited by the $\truncInd = (2,3)$ model is less than half of that of of the $\truncInd = (2,8)$ model.  On the other hand, one sees that the heat transport decreases with increasing rotation for all models, although more slowly for the higher dimensional models.  This decrease appears to be monotonic, although one sees that the slopes may not be strictly increasing.
		
\begin{figure}[H]
			\begin{center}
				\begin{tabular}{cc}
					\includegraphics[height=68mm]{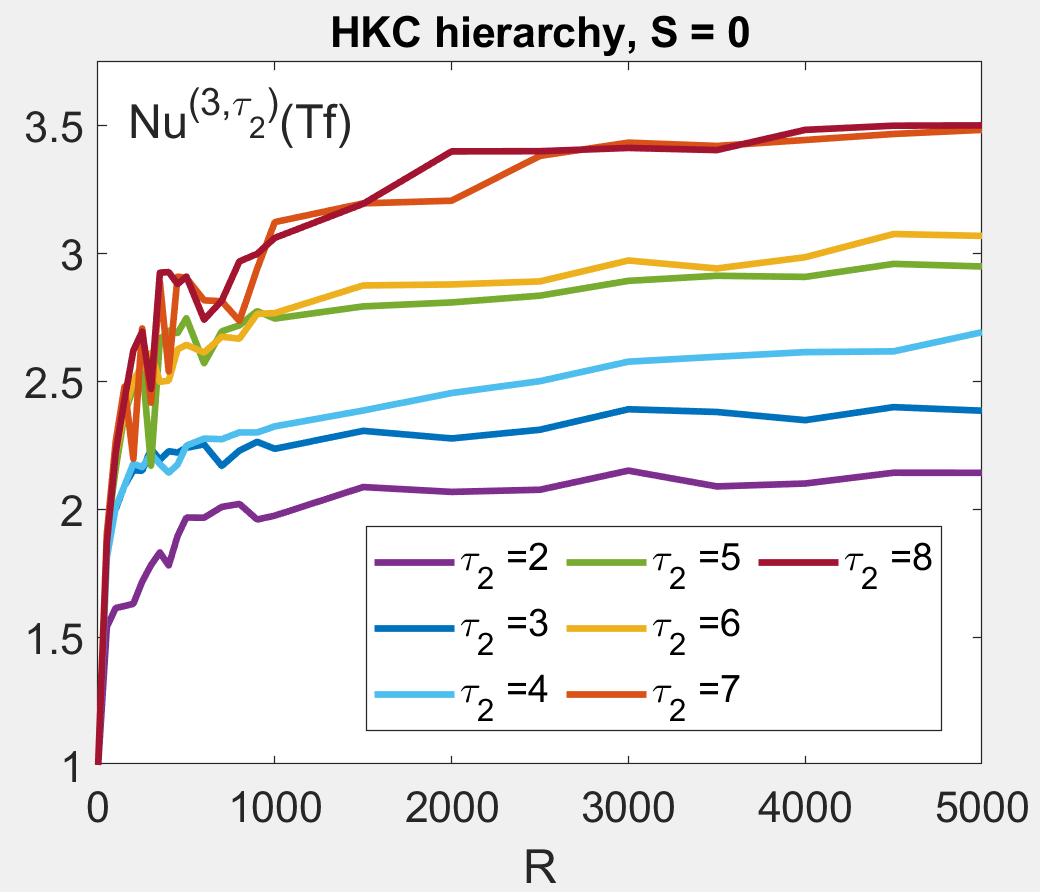}
					&\includegraphics[height=68mm]{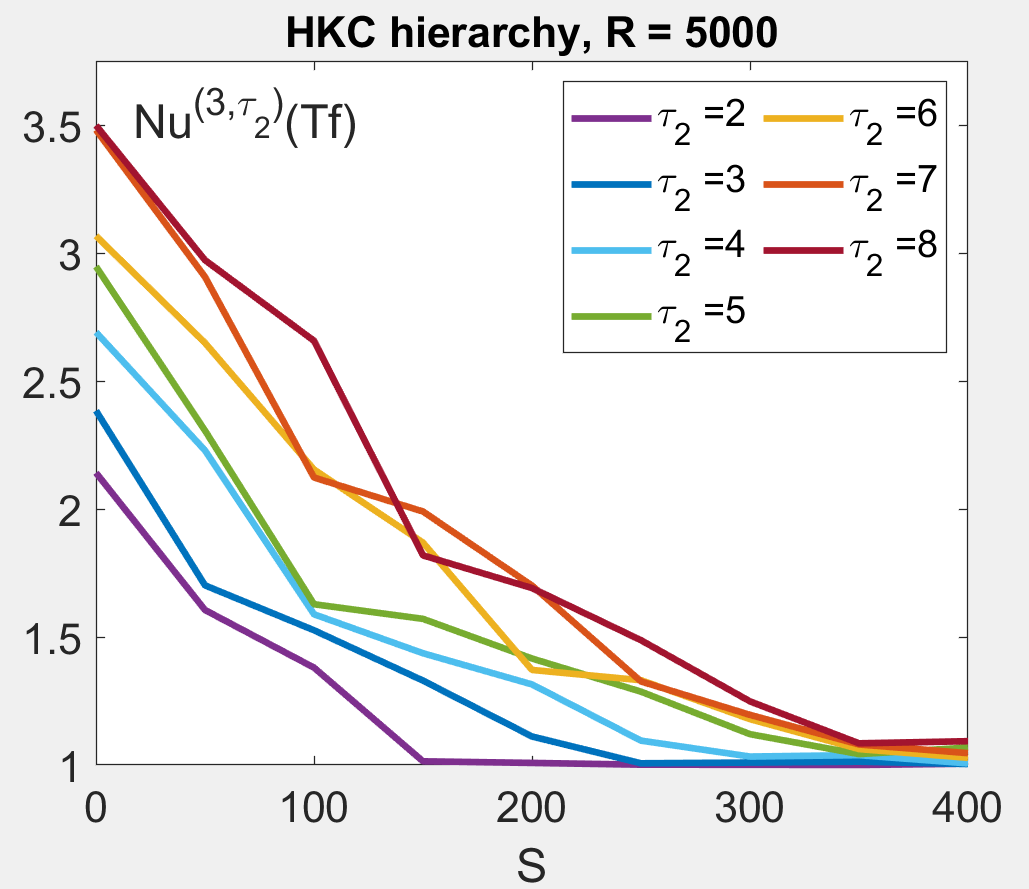}  \\
					(a) & (b)
				\end{tabular}
				\caption{Nusselt comparison for the HKC hierarchy, (a) for $0 \leq \Ray \leq 5000, \Rot = 0$ and (b) for $\Ray = 5000$, $0 \leq \Rot \leq 400$.}
				\label{fig:HeatTransport_HKC}
			\end{center}
		\end{figure}

		In order to compare these results to the 2d case, we include in Figure \ref{fig:HeatTransport_HKC} the results of the heat transport computations for several models in the HKC hierarchy defined in \eqref{HierarchyDef_HKC}.  In particular, HKC models were chosen with the same number of vertically stratified temperature modes as for the models in the $\ell^{\infty}$ and $\ell^1$ hierarchies, which are indicated by the colors in Figures \ref{fig:HeatTransport_Linf}, \ref{fig:HeatTransport_L1} and \ref{fig:HeatTransport_HKC}. 
		
		One sees that the results for models with the same number of vertically stratified modes largely agree across the $\ell^1$ and HKC hierarchies, whereas the models from the $\ell^{\infty}$ hierarchy exhibit somewhat less heat transport.  Notably the $\ell^1$ and HKC hierarchies contain the exact same vertically stratified Fourier modes, namely the modes $(0,0,2m_3,0,0,1)$ for $0 \leq m_3 \leq \tau_2-1$, whereas the $\ell^{\infty}$ hierarchy contains the modes $(0,0,m_3,0,0,1)$ for $0 \leq m_3 \leq 2\tau_2$.  Hence the models from $\ell^1$ and HKC hierarchies reach smaller scales than those from the $\ell^{\infty}$ hierarchy.  In Figure \ref{fig:HeatTransport_HKC} one sees a more jagged dependence of the Nusselt number on the Rayleigh number, but indeed this is likely due to the presence of multi-stability discussed in \cite{Welter2025Rotating}, whereby random initial conditions can lead to different Nusselt numbers at the same parameter value.  
		
\begin{figure}[H]
			\begin{center}
				\begin{tabular}{cc}
					\includegraphics[height=65mm]{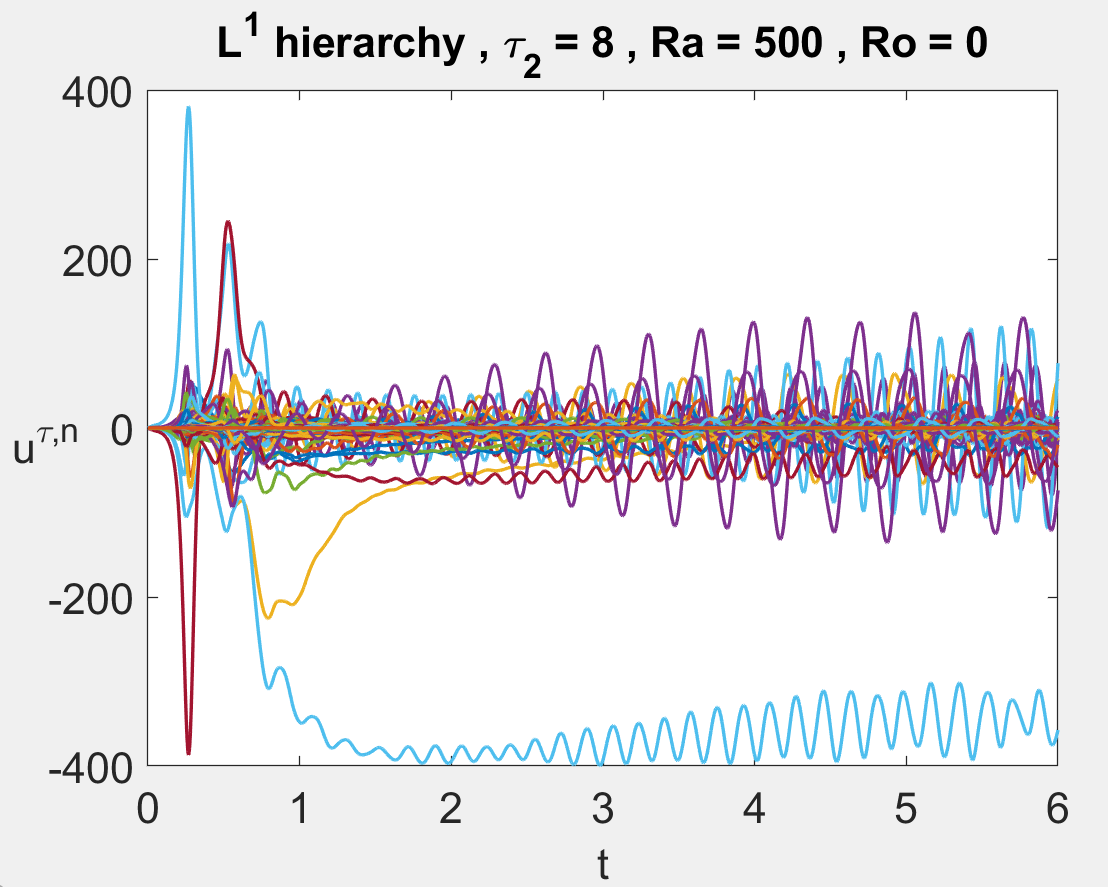}
					&\includegraphics[height=65mm]{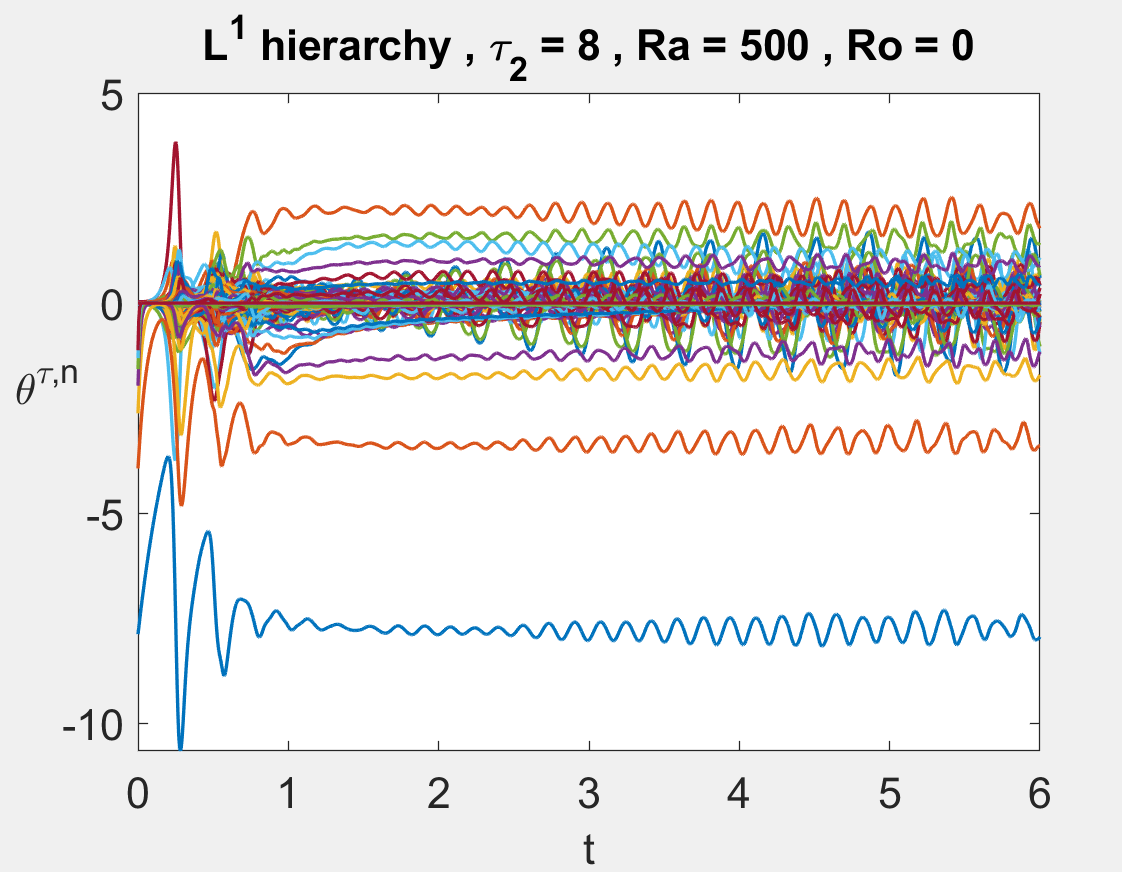}  \\
					(a) & (b)
				\end{tabular}
				\caption{Time evolution of the Fourier modes for a solution of the $\truncInd = (2,8)$ model, (a) the modes $u^{\truncInd,\textbf{n}}(t)$ for all $\textbf{n} \in \mathscr{N}_{\textbf{u}}^{\truncInd}$ (here light blue corresponds to mode $\textbf{n} = (1,0,1,0,0,1)$) and (b) the modes $\theta^{\truncInd,\textbf{n}}(t)$ for all $\textbf{n} \in \mathscr{N}_{\theta}^{\truncInd}$. }
				\label{fig:L1hier_Vars}
			\end{center}
		\end{figure}

		While higher dimensional models were also considered, the computational cost begins to increase rapidly, so while we could reasonably generate solutions at a fixed parameter value as in Figure \ref{fig:FluidVisualization}, computing long time averages at many parameter values can take an unreasonable amount of time on a laptop.  For example the models $\truncInd = (1,5)$ and $\truncInd = (2,9)$ have dimension $970$ and $834$, respectively, and we were able to compute $3000$ time steps in roughly 20 hours.  By contrast the HKC 24 model (i.e. \eqref{HierarchyDef_HKC} with $\tau_2 = 24$) has dimension $995$ and we were able to compute $10^5$ time steps in 20 minutes.  The fully 3d models from the $\ell^{\infty}$ and $\ell^1$ hierarchies have many more nonlinear couplings than the 2d models from the HKC hierarchy, leading to larger files for the differential equations.  For fixed dimension it might be possible that one could retain accuracy by trading horizontal resolution for more vertical resolution.  For example, Figure \ref{fig:L1hier_Vars} depicts the time evolution of the Fourier modes from a solution of the $\truncInd = (2,8)$ model, and one notices that a small subset of the modes account for much of the $L^2$ norm.  Panel (a) depicts the evolution of the 422 velocity modes, and one sees the mode $u^{\truncInd,\textbf{n}}(t)$ for $\textbf{n} = (1,0,1,0,0,1)$ accounts for $86\%$ of the squared $L^2$ norm $\|\textbf{u}^{\truncInd}(t)\|_{\textbf{L}^2}^2$ for $t$ sufficiently large.  Panel (b) depicts the 175 temperature modes, and one sees that the following five temperature modes $\theta^{\textbf{n}}(t)$ account for $89\%$ of the squared $L^2$ norm $\|\theta^{\truncInd}(t)\|_{L^2}^2$ for $t$ sufficiently large:
		\[ \textbf{n} = (0,0,2,0,0,1) \text{ , } (0,0,4,0,0,1)  \text{ , } (0,0,6,0,0,1) \text{ , } (1,0,1,0,0,1) \text{ , } (2,0,1,1,0,1)  \text{ . }  \]
		
		\subsection{Numerical analysis of energetically inconsistent models}
		
		While in Section \ref{sec:EnergyInconsis} it was proven that type 1 models exhibit unbounded, exponentially increasing trajectories, it is unclear whether said trajectories occur only on the proper subspaces identified there, or whether this model failure occurs for more general initial conditions.  To this end we studied the first four models from the Fourier box hierarchy defined in \eqref{HierarchyDef_FourierBox} by computing ten different trajectories for a range of Rayleigh numbers.  In order to explore a wider range of the phase space every variable $u^{\truncInd,\textbf{n}},\theta^{\truncInd,\textbf{n}}$ was initialized via a normal random variable with standard deviation 10.  A time increment $\mathtt{tInc} = 10^{-4}$ was chosen and solutions were initially integrated for $5 \cdot 10^{4}$ time steps.  For each Fourier box $\tau_2 = 1,2,3,4$ one can find the critical Rayleigh numbers at which the eigenvalues of the matrices in \eqref{e:ODE-3DS}, \eqref{LinMat_Full3d} cross the imaginary axis.  Theorem \ref{thm:runaway} guarantees the existence of runaway modes for all Rayleigh numbers above the smallest such critical number corresponding to the linear invariant subspaces, hence we examined each model starting from Rayleigh numbers slightly below said smallest critical Rayleigh number.  
		
		\begin{figure}[H]
			\begin{center}
				\begin{tabular}{ccc}
					\includegraphics[height=45mm]{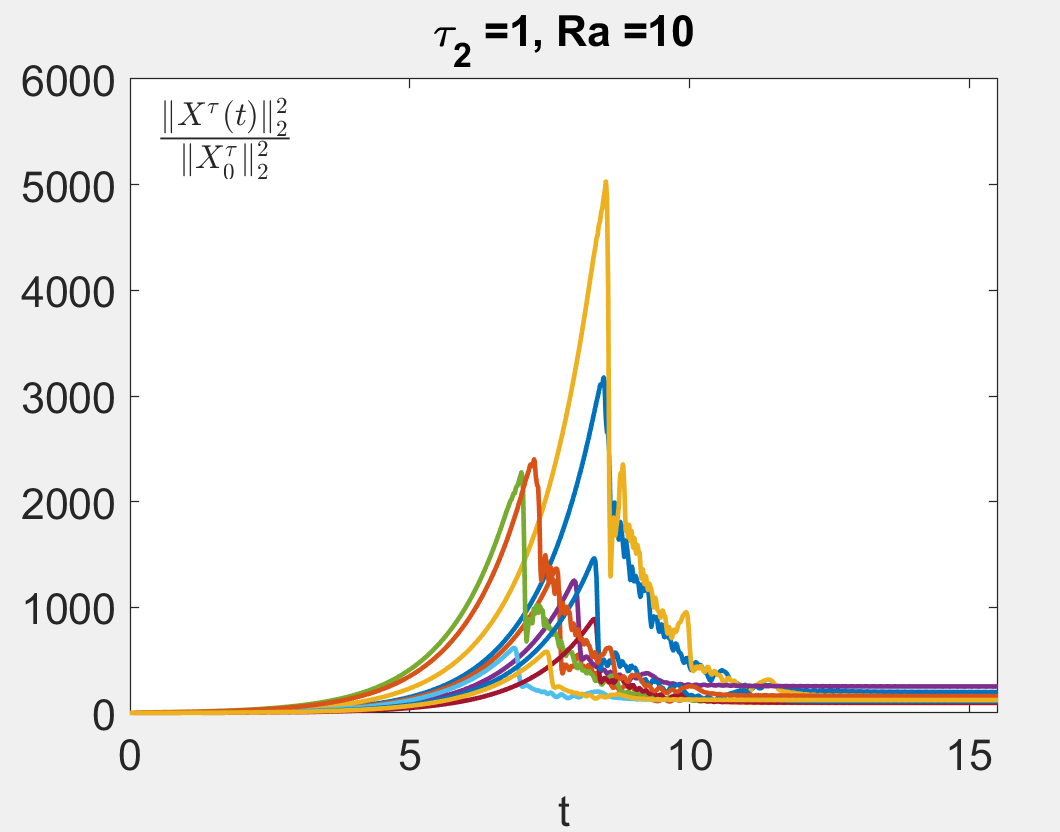}
					& \includegraphics[height=45mm]{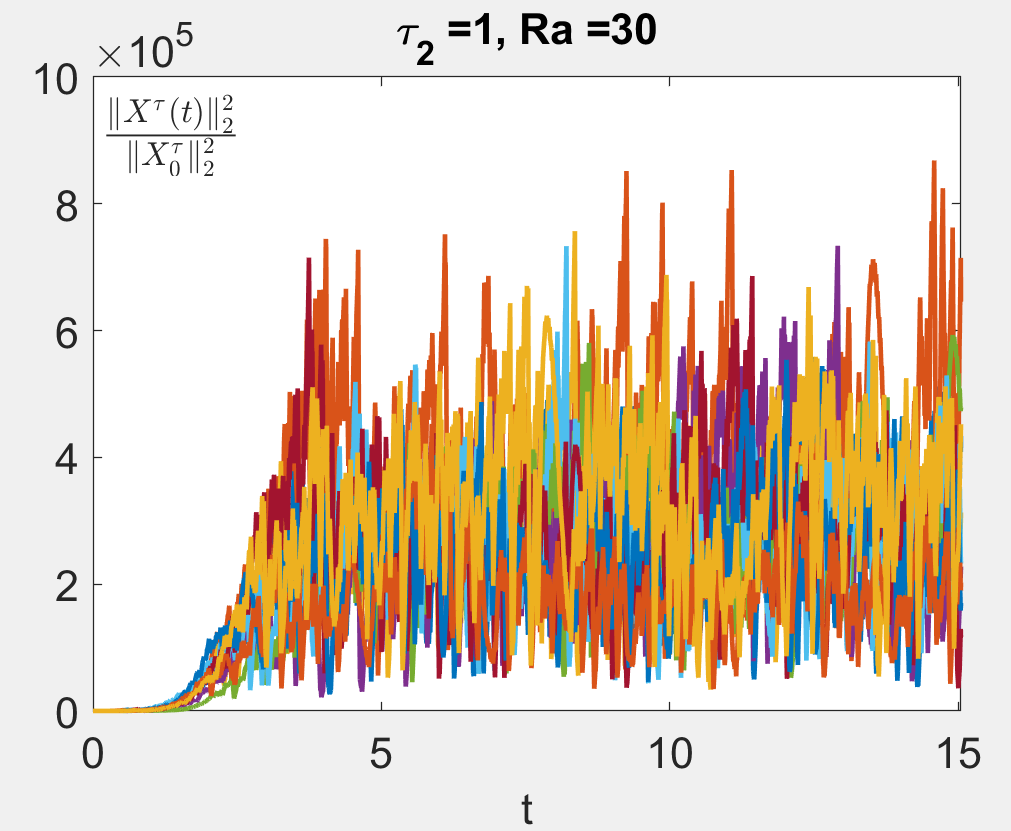} & \includegraphics[height=45mm]{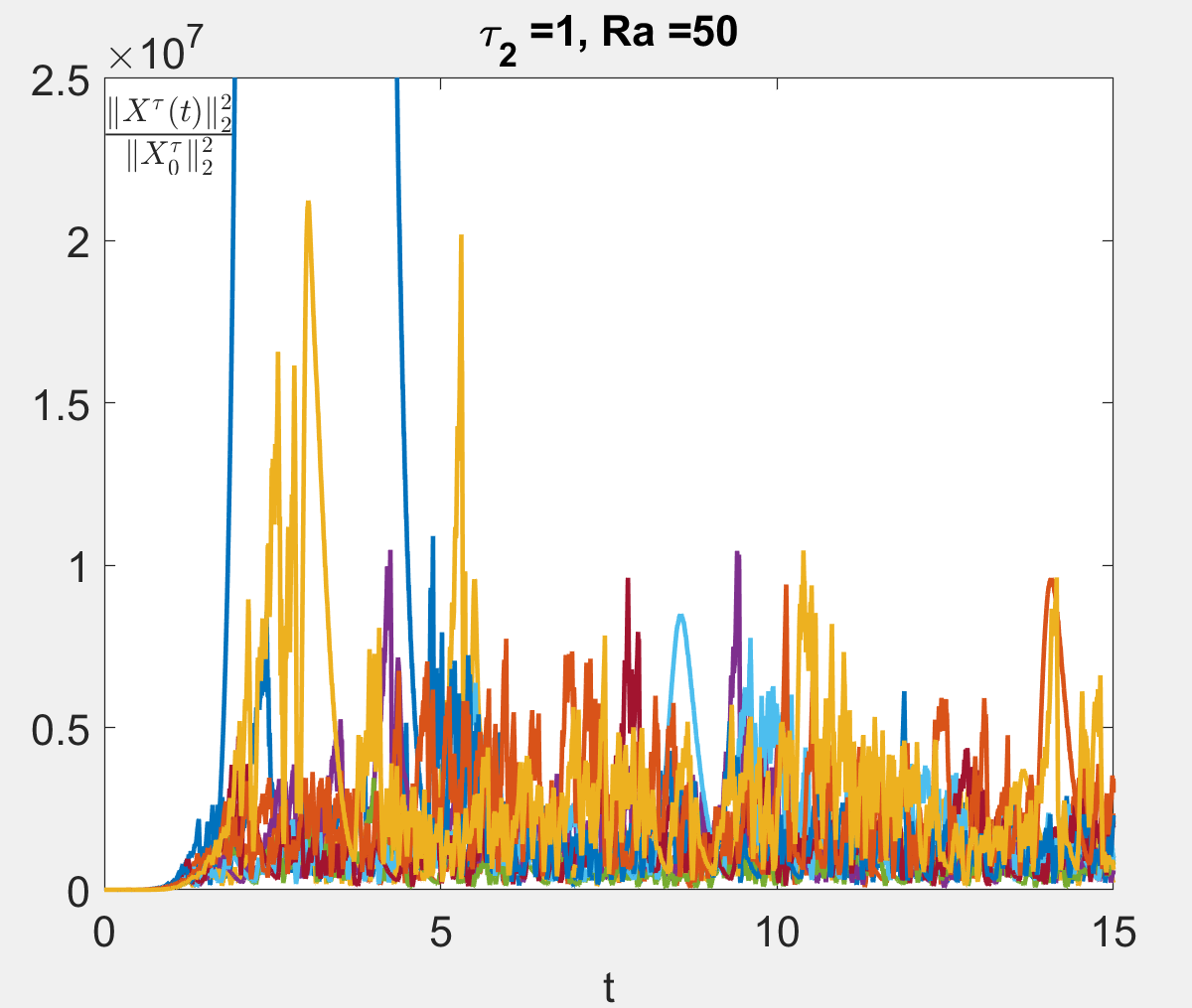}  \\
					(a) & (b) & (c)
				\end{tabular}
				\caption{Plots of $\frac{\|\textbf{X}^{\truncInd}(t)\|_2^2}{\|\textbf{X}^{\truncInd}_0\|_2^2}$ vs time for 10 solutions of the 1st Fourier box model, with randomly selected initial conditions and with parameters $\ShaOne = \ShaTwo = 1$, $\Rot = 0$, and (a) $\Ray = 10$, (b) $\Ray = 30$ and (c) $\Ray = 50$ .}
				\label{fig:FourierBox_L2Norms}
			\end{center}
		\end{figure}
		
		Figure \ref{fig:FourierBox_L2Norms} depicts the $L^2$ norms of such trajectories, normalized by the $L^2$ norm of their initial condition, for the first Fourier box model $\tau_2 = 1$ for several Rayleigh numbers beginning from slightly above the critical Rayleigh number of $9/4$.  Slightly above the critical Rayleigh number all such trajectories appear to converge to the origin, but by $\Ray = 10$ exponential growth is clearly observed for all ten trajectories by time $t = 5$.  To determine whether this growth was merely transient, these trajectories were then integrated for a much longer time out to $\mathtt{Tf} = 20$ (though only displayed for $t < 15$).  One sees in panel (a) that the exponential growth indeed seems to break after sufficiently long time, and thereafter solutions tended to some steady states.  Increasing the Rayleigh number further seemed to increase the exponential growth rate and the amplitudes which are achieved, although significant fluctuations start to be observed.  At higher Rayleigh numbers, for example at $\Ray = 30$ in panel (b), the exponential growth seems to no longer exhibit the "breaking" behavior, but instead solutions eventually seem to fluctuate wildly about some very large asymptotic mean value.  On the other hand, increasing the Rayleigh number further to $\Ray = 50$, we see in panel (c) another solution which exhibited exponential growth up to a maximum of around $2.5 \cdot 10^9$ before breaking, and all other solutions tended to wildly fluctuate about mean values on the order of $10^{6}$.  
		
		\begin{figure}[H]
			\begin{center}
				\begin{tabular}{ccc}
					\includegraphics[height=48mm]{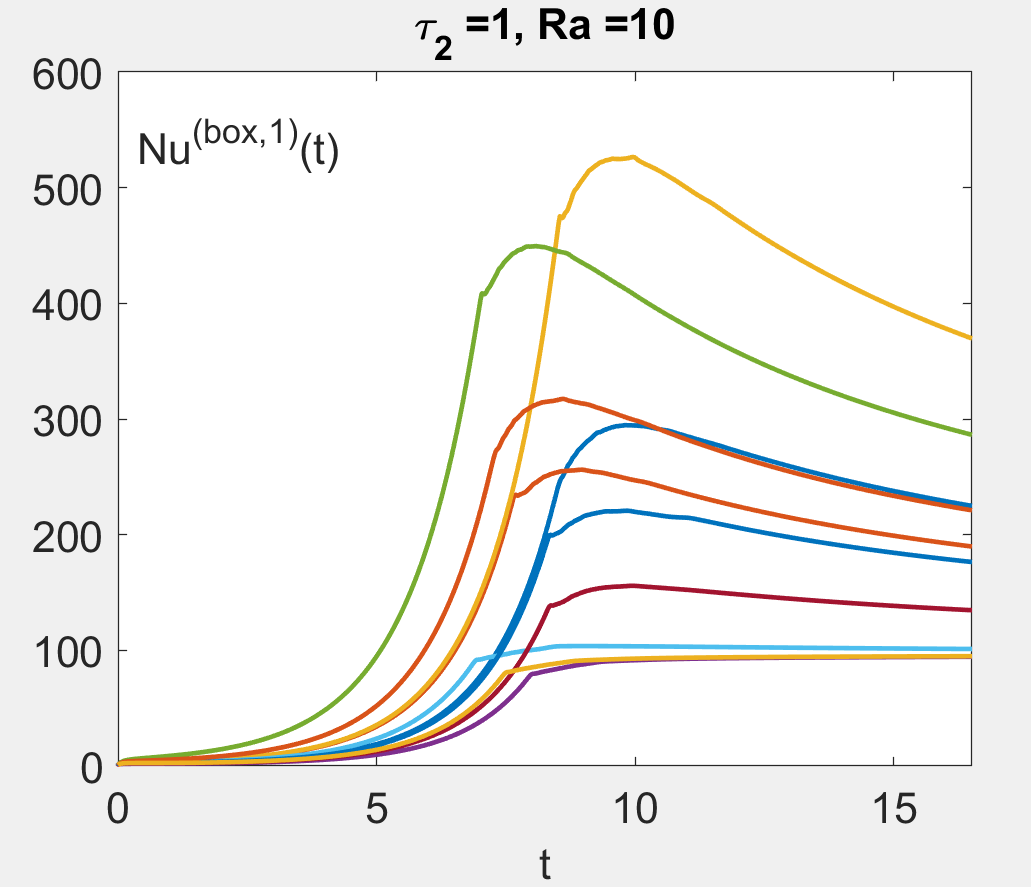} & \includegraphics[height=48mm]{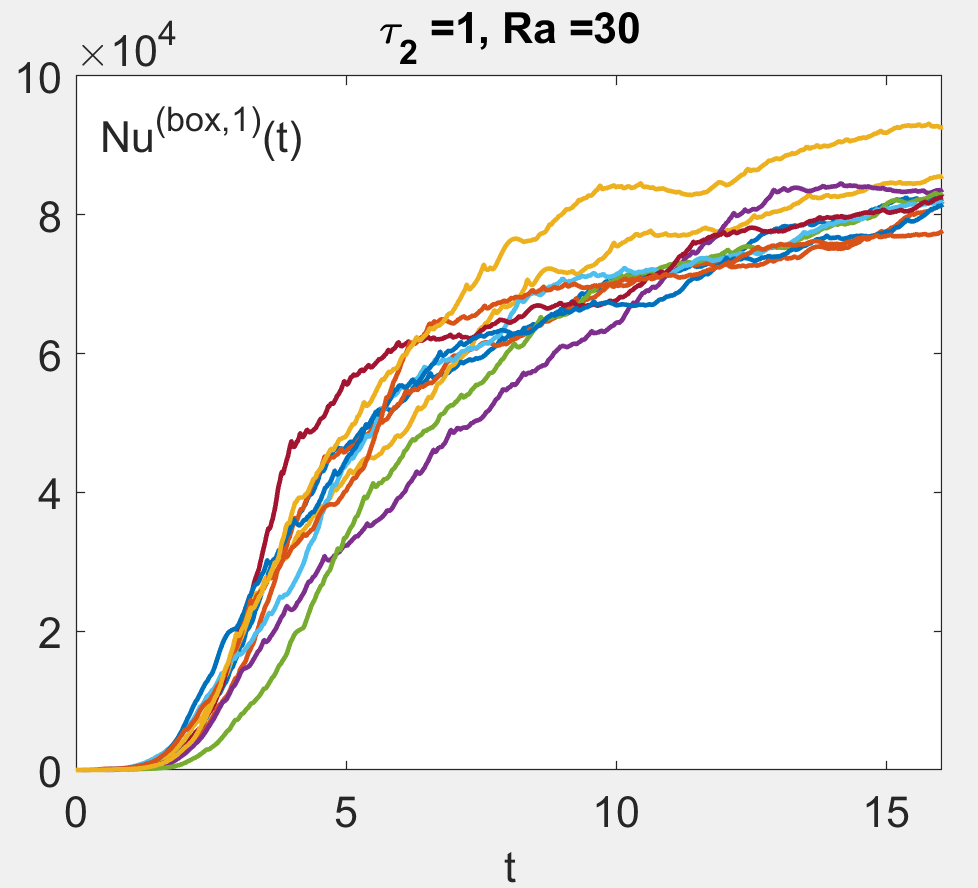} & \includegraphics[height=48mm]{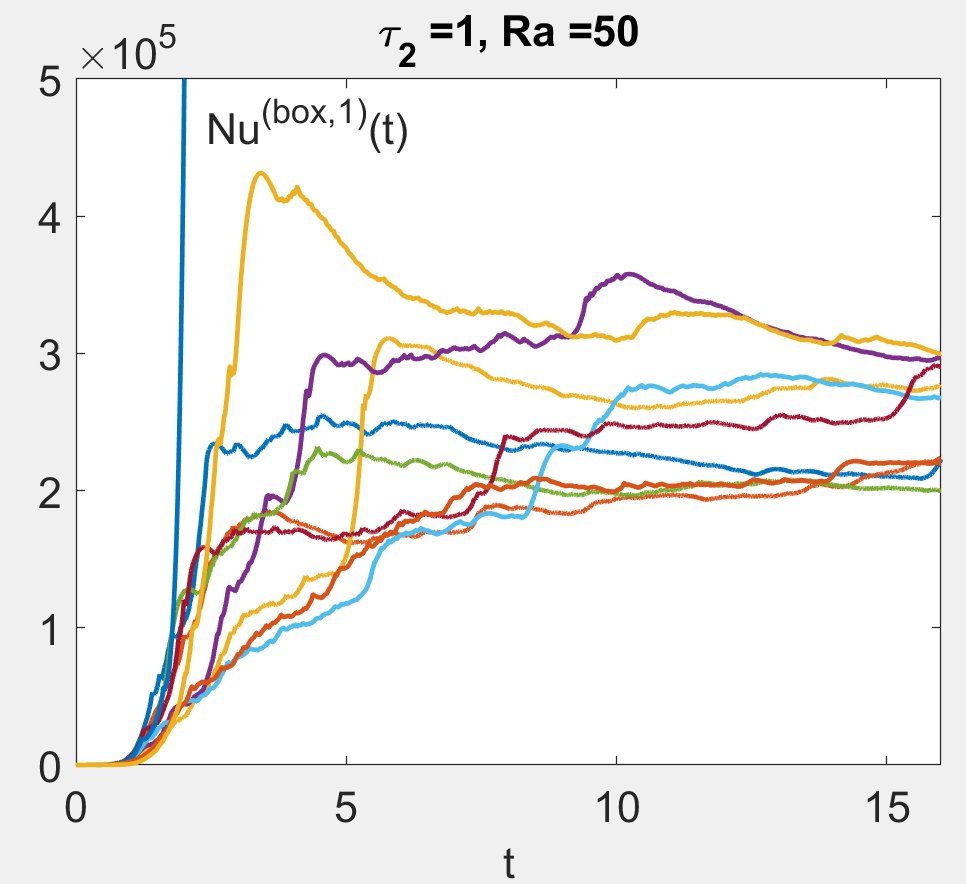}  \\
					(a) & (b) & (c)
				\end{tabular}
				\caption{Plots of $\mathsf{Nu}^{\truncInd}(t)$ vs time for 10 solutions of the 1st Fourier box model, with randomly selected initial conditions and with parameters $\ShaOne = \ShaTwo = 1$, $\Rot = 0$, and (a) $\Ray = 10$, (b) $\Ray = 30$ and (c) $\Ray = 50$ .}
				\label{fig:FourierBox_HeatTransport}
			\end{center}
		\end{figure}
		
		The corresponding finite time Nusselt number approximations for the above trajectories are depicted in Figure \ref{fig:FourierBox_HeatTransport}.  One sees the kind of the temporal convergence of the finite time Nusselt numbers that might be expected given the above description of the behavior of the solutions.  For $\Ray = 10$, the Nusselt numbers exhibit exponential growth for a long time, but after the exponential growth breaks the Nusselt numbers exhibit a slow (algebraic) relaxation to some apparent long time average.  For $\Ray = 30$ the Nusselt numbers exhibit exponential growth with a large number of fluctuations, but eventually seem to level off.  For $\Ray = 50$ one sees a similar story; the trajectory mentioned above exhibiting large exponential growth does indeed reach a maximum Nusselt number of around $10^8$, but the relaxation is too slow to be seen again in panel (c).  Note that these Nusselt number values differ dramatically from those obtained via energy consistent hierarchies.  In particular, the models from $\ell^{\infty}$ hierarchy are comparable to those from the Fourier box since the model of matching $\tau_2$ contains all of the same wave vectors, and in Figure \ref{fig:HeatTransport_Linf} one sees Nusselt numbers less than about $3$ on the whole range $\Ray \leq 5000$, as opposed to $\mathsf{Nu} \approx 10^{5}$ by $\Ray = 50$.  It is also notable how quickly the Nusselt numbers increase with respect to the Rayleigh number, increasing over five orders of magnitude from $\Ray \approx 3$ to $\Ray = 50$, which seems starkly at odds with known scaling laws for the PDE such as $\mathsf{Nu} \lesssim 1 + \Ray^{\frac{5}{12}}$.  We also report that the two expressions $\mathsf{Nu}^{\truncInd}(\mathtt{Tf}),\widetilde{\mathsf{Nu}}^{\truncInd}(\mathtt{Tf})$ defined above differed significantly for these models, where the difference between them was often on the order of $60\%$ of the value of $\mathsf{Nu}^{\truncInd}(\mathtt{Tf})$, and sometimes on the order of more than $100\%$. Recall that the equality of these two different expressions for the Nusselt number were derived in \eqref{NusseltNumExpressions} by explicitly making use of the potential energy balance, so it is sensible that we see such large differences in the inconsistent Fourier box model. 
		
		The exponential growth and abnormally fast growth to large Nusselt numbers above also appear for the Fourier boxes $\tau_2 = 2, 3$ and $4$.  While these effects appear very early for $\tau_2 = 1$,  they seem to be shifted to higher Rayleigh numbers for higher dimensional models.  Whereas exponential growth of the solutions with $\tau_2 = 1$ was observed at around $\Ray = 10$, exponential growth for $\tau_2 = 2,3$ is not observed until around $\Ray = 160$.  For $\tau_2 = 4$, the exponential growth does not seem to appear until Rayleigh numbers around $\Ray = 1000$, and it is more subtle than for $\tau_2 < 4$ because the amplitudes are already relatively large by this point.  Additionally, while the finite time Nusselt numbers for $\tau_2 = 1$ increased to the order of $10^5$ by $\Ray = 50$, the finite time Nusselt numbers for $\tau_2 = 2$ increased to $10^5$ by $\Ray = 1.2\cdot 10^5$.  Interestingly, the finite time Nusselt numbers for $\tau_2 = 3$ increased to $10^5$ by $\Ray = 5\cdot 10^4$, and hence the shift of the effects of inconsistency toward higher Rayleigh numbers is apparently non-monotonic.  We did not pursue the Rayleigh number threshold at which the Nusselt numbers reached $10^5$ for the $\tau_2 = 4$ case.  The observed monotonicity seems to make sense because as one moves from the first Fourier box to the second, all of the lowest modes (i.e. with maximal wavenumber equal to one) become stabilized by the newly adjoined $\theta^{(0,0,2,0,0,1)}$ mode.  There are more Fourier modes which fail their consistency criterion in the second Fourier box, and all of these modes remain inconsistent in the third Fourier box model.  There are even more Fourier modes which fail their consistency criterion in the third Fourier box than the second, hence one sees larger Nusselt numbers.  However, when moving to the fourth Fourier box all of the modes in the second Fourier box become stabilized by the addition of the $\theta^{(0,0,4,0,0,1)}$ mode, and so on.
		
		Finally, Figure \ref{fig:EnergyInconsistency} below displays a more direct comparison of the Nusselt numbers computed from the Fourier box models to those computed from models from the $\ell^{\infty}$ hierarchy.  In particular, since the finite time Nusselt numbers for the Fourier box models did not appear to converge to a single value, and possibly do not converge as an infinite time average (see Figure \ref{fig:FourierBox_HeatTransport}), we computed an average finite time Nusselt number value over 5 randomly selected initial conditions, where $\mathsf{Tf} = 10$ was chosen as being sufficiently large:
		\[ \mu^{(box,\tau_2)} = \frac{1}{5} \sum_{i\leq 5} \mathsf{Nu}^{(1,\tau_2)}(\textbf{u}_{0,i}^{\truncInd},\theta_{0,i}^{\truncInd}) \text{ . } \]
		The goal in computing this quantity was to gain a sense about the order of magnitude of such finite time Nusselt numbers.  The ratio of the quantity $\mu^{(box,\tau_2)}$ to $\mathsf{Nu}^{(1,\tau_2)}$ grows very fast, hence Figure \ref{fig:EnergyInconsistency} displays the logarithm.  One sees that this ratio is not so large for small Rayleigh numbers, but tends to increase for larger Rayleigh numbers, where the lower dimensional models tend to exhibit much larger ratios, presumably representing larger error.  The increase with Rayleigh number is essentially monotonic; the few apparent exceptions seem to arise from rare trajectories which exhibit extremely large growth as in panel (c) of Figures \ref{fig:FourierBox_L2Norms}, \ref{fig:FourierBox_HeatTransport}.  Interestingly Figure \ref{fig:EnergyInconsistency} seems to show the ratios increasing very dramatically at first, and then leveling off.  While we are not sure about the mechanism underlying this effect, it does not seem that the errors level off to a constant ratio, but rather they just seem to increase more slowly.  
		
		\begin{figure}[H]
			\begin{center}
				
				\includegraphics[height=55mm]{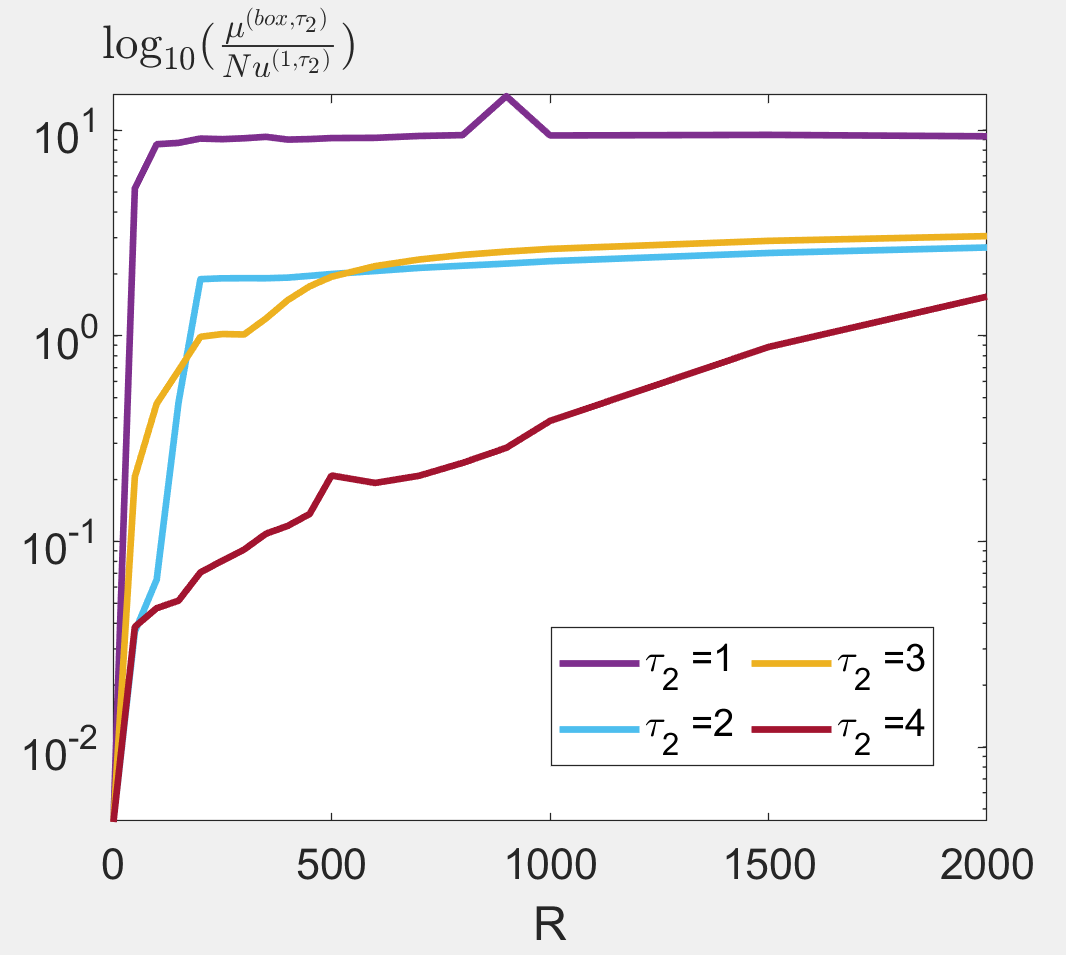} 
				\caption{A plot of $\log_{10} ( \frac{\mu^{(box,\tau_2)}}{\mathsf{Nu}^{(1,\tau_2)}})$ vs Rayleigh number as a comparison of Nusselt numbers from the Fourier box models to those from the $\ell^{\infty}$ hierarchy.  Here $\Pra = 10$, $\ShaOne = \ShaTwo = 1$, $\Rot = 0$.}
				\label{fig:EnergyInconsistency}
			\end{center}
		\end{figure}
		
		\section{Conclusion}\label{s:conclusion}
		
		In this paper we derived and analysed Fourier-Galerkin truncations of the 3D rotating Rayleigh-B\'enard convection problem with free-slip boundary conditions with focus on the preservation or violation of potential energy balance. The setup admitted a rather explicit analytical treatment, allowing precise statements about the structural requirements for energetic consistency and an implementation in MATLAB.  While we do not propose this implementation for high resolution DNS of Rayleigh-B\'enard convection, we were able to explore convergence issues and the effects of energetic consistency out to moderate resolution and parameter values.  Furthermore our analysis revealed two types of energetically inconsistent truncations for arbitrarily high dimension: The first type features one-dimensional invariant subspaces with linear dynamics that generate runaway modes for large enough Rayleigh number, an unphysical phenomenon that highlights deviation from the limiting PDE dynamics. In particular the standard Fourier-box truncations are of this type.  For the second type, which does not possess such invariant subspaces, the consequences of inconsistency are less clear, and these can even possess global attractors for any Rayleigh number.  Besides the infinite Nusselt numbers seen on proper subspaces, a clear violation of known scaling laws, our numerical results indicate energetic inconsistency as a potential source of large errors.  While in low dimensional truncations one sees extremely rapid growth of Nusselt numbers for generic conditions, and resulting values which are orders of magnitude away from those obtained in energetically consistent models, for higher dimensional models these effects appear to be mitigated.  On the other hand, it is perhaps more dangerous if such errors were of the same order of magnitude as those obtained from consistent models, since they might then go unnoticed.
		
		As type-one illustrates, truncations can generate unnatural invariant subspaces and another observation in this direction is the following in the $\ell^\infty$-hierarchy with index $\tau_2\in \mathbb{N}$. Here, a triad-interaction resulting from or among any two wave vectors $\textbf{m}_1:=(\tilde m,0,m), \textbf{m}_2:=(0,0,2m)$ with $\tilde m, m> \frac {\tau_2}{2}$ is either one of $\textbf{m}_1, \textbf{m}_2$, or lies outside the range of wave vectors since these are either the zero vector or a vector $(m_1,m_2,m_3)$ with $m_1>\tau_2$, $m_2>\tau_2$, or $m_3>2\tau_2$. Hence, there is no nonlinear coupling to other components. Since both $\textbf{m}_1, \textbf{m}_2$ have a zero entry, there is only one component $c=1$ involved, so that linear coupling to other phases is possible for $S\neq 0$ only and then is at most a phase shift by $(1,1)$, cf.\ Figure~\ref{fig:linear_coupling}. Analogous to \S\ref{app:ConsProced}, but here for any such $\textbf{m}_1, \textbf{m}_2$, we thus obtain a Lorenz-Stenflo or, for $\Rot=0$, a Lorenz '63 equation as an invariant subsystem. One relevant question would be whether these low-dimensional subspaces are unstable in the truncated model. 
		Notably, the chaotic Lorenz-attractors can occur simultaneously in these invariant subsystems, and also for arbitrary large Rayleigh number, when choosing $m, \tilde m$ sufficiently large. Indeed, it follows as in \S\ref{s:runaway}, that the critical Rayleigh number is of order $(|\mathcal{K}|\tau_2)^4$. Moreover, since $\tilde m, m$ need to be relatively large, this  complex dynamics pertains ``small scales'', in particular purely vertical modes that are needed for potential energy consistency. It would be interesting to understand better what the impact of this is on the accuracy of the truncation.

		Although we focused heavily on potential energy consistency, we also characterized truncations that satisfy the vorticity balance.  Whereas energy consistency led to a compact attractor, it would be interesting to understand the consequences of violating the vorticity balances in a truncation, beyond the mere fact that angular momentum is not properly represented.  One possible direction would be to assess bounds analogous to those for the PDE derived from the background flow method, which makes use of the vorticity balance. 
		
		For other more physically realistic boundary conditions a typically vertical discretization uses Chebychev polynomials \cite{Miquel2021Coral}, which does not admit as much explicit analysis, but the results for free slip provide a direction of analysis that would be interesting to pursue in this case.  More generally, our results indicate the relevance of energetic consistency in truncations and discretizations for the overall dynamics and also for computing long time averages such as the Nusselt number.  While more difficult to assess analytically, it is desirable to study the dynamical effects of energetic consistency using other discretizations, since these more readily adapt to other geometries and parallelize more easily as one moves to higher resolution.  One step in this direction would be to consider discrete Fourier transform rather than the explicit spectral projection we used. However, this entails an aliasing error and it is not clear how to obtain exact balance preservation.  We also note that structure preserving discretisations have a long history, in particular for conserved quantities, e.g., \cite{WIMMER2020109016}. For energy balances we mention the GENERIC framework \cite{Grmela97}, and in the context of Rayleigh-Benard convection the very recent \cite{SANDERSE2025106473}. Also the time-stepping needs to be accounted for, which we neglected in the present work and is considererd in the mentioned references; we also refer to recent developments regarding property-preserving methods in port-Hamiltonian systems \cite{Geisselmann_2024_Energy}, and even for the Navier-Stokes equations \cite{Andrews_2024_Enforcing}.
		
		\appendix{}
		
		\section{Further details on model construction}
		
		\subsection{The general Fourier expansions}
		
		\subsubsection{Explicit formulas for the velocity vector fields}
		
		\label{app:VectorFieldDef_Explicit}
		
		For $\textbf{m} \in \mathbb{Z}^3_{>0}$, the vector fields in \eqref{VectorFieldDef_FreeSlip} for the case $c = 1$ are given more explicitly by
		
		\vspace{-.5 cm}
		\setlength\extrarowheight{2.5pt}
		
		\begin{equation} \label{VectorFieldDef_Explicit} \begin{split}  & \textbf{v}^{(\textbf{m},0,0,1)} := \frac{\eta^{\textbf{m}}}{ |\mathcal{G}^{(\textbf{m},0,0,1)}\boldsymbol{\nu}^{\textbf{n}}| V} \begin{pmatrix} ( \ShaThree m_3 + \ShaTwo m_2) \sin( \ShaOne m_1 x_1 ) \cos (\ShaTwo m_2 x_2) \cos( \ShaThree m_3 x_3 ) \\ -( \ShaOne m_1 + \ShaThree m_3 ) \cos( \ShaOne m_1 x_1 ) \sin (\ShaTwo m_2 x_2) \cos( \ShaThree m_3 x_3 ) \\ ( \ShaTwo m_2 - \ShaOne m_1 ) \cos( \ShaOne m_1 x_1 ) \cos (\ShaTwo m_2 x_2)  \sin( \ShaThree m_3 x_3 ) \end{pmatrix} \text{ , } \\
				& \textbf{v}^{(\textbf{m},1,0,1)} := \frac{\eta^{\textbf{m}}}{ |\mathcal{G}^{(\textbf{m},1,0,1)}\boldsymbol{\nu}^{\textbf{n}}| V} \begin{pmatrix} - ( \ShaThree m_3 + \ShaTwo m_2) \cos( \ShaOne m_1 x_1 ) \cos (\ShaTwo m_2 x_2) \cos( \ShaThree m_3 x_3 ) \\ - ( \ShaOne  m_1 + \ShaThree m_3 ) \sin( \ShaOne m_1 x_1 ) \sin (\ShaTwo m_2 x_2) \cos( \ShaThree m_3 x_3 ) \\ (\ShaTwo m_2 - \ShaOne m_1 )  \sin ( \ShaOne m_1 x_1 ) \cos (\ShaTwo m_2 x_2)  \sin( \ShaThree m_3 x_3 ) \end{pmatrix} \text{ , } \\ 
				& \textbf{v}^{(\textbf{m},0,1,1)} := \frac{\eta^{\textbf{m}}}{ |\mathcal{G}^{(\textbf{m},0,1,1)}\boldsymbol{\nu}^{\textbf{n}}| V} \begin{pmatrix} ( \ShaThree m_3 + \ShaTwo m_2) \sin( \ShaOne m_1 x_1 ) \sin (\ShaTwo m_2 x_2) \cos( \ShaThree m_3 x_3 ) \\ ( \ShaOne m_1 + \ShaThree m_3 ) \cos( \ShaOne m_1 x_1 ) \cos (\ShaTwo m_2 x_2) \cos( \ShaThree m_3 x_3 ) \\ ( \ShaTwo m_2 - \ShaOne m_1 )  \cos( \ShaOne m_1 x_1 ) \sin (\ShaTwo m_2 x_2)  \sin( \ShaThree m_3 x_3 ) \end{pmatrix} \text{ , } \\
				& \textbf{v}^{(\textbf{m},1,1,1)} := \frac{\eta^{\textbf{m}}}{ |\mathcal{G}^{(\textbf{m},1,1,1)}\boldsymbol{\nu}^{\textbf{n}}| V} \begin{pmatrix} - ( \ShaThree m_3 + \ShaTwo m_2) \cos( \ShaOne m_1 x_1 ) \sin (\ShaTwo m_2 x_2) \cos( \ShaThree m_3 x_3 ) \\ ( \ShaOne m_1 + \ShaThree m_3 ) \sin( \ShaOne m_1 x_1 ) \cos (\ShaTwo m_2 x_2) \cos( \ShaThree m_3 x_3 ) \\ (\ShaTwo m_2 - \ShaOne m_1 )  \sin( \ShaOne m_1 x_1 ) \sin (\ShaTwo m_2 x_2)  \sin( \ShaThree m_3 x_3 ) \end{pmatrix} \text{ . } \end{split} \end{equation}
		
		\setlength\extrarowheight{0pt}
		\vspace{-.1 cm}
		\noindent Note that the vector fields are the same when one or more of the components of $\textbf{m}$ are zero, although in that case one must apply $\sin (0) = 0$.  The definitions for $c = 2$ are very similar, with merely different constants in front of the sinusoids.
		
		\subsubsection{Proof of Proposition \ref{prop:DivFreeVectorBasis}}
		
		\label{app:FourierExpDeriv}
		
		Note the vertical sinusoid in \eqref{def:ThetaFourierBasis} is $\sin(\ShaThree m_3 x_3)$, hence the boundary conditions \eqref{BC_Temp} are satisfied, and the orthonormality follows from the orthonormality of the sinusoids in each spatial direction.  Indeed, using the substitutions $y_j = \mathsf{k}_j x_j$ together with $\ShaOne \ShaTwo \ShaThree =1$ one obtains
		\[ \int_{\Omega} f^{\textbf{n}} f^{\tilde{\textbf{m}}} d\textbf{x} = \frac{\eta^{\textbf{m}}\eta^{\tilde{\textbf{m}}}}{4\pi^3} \Big ( \prod_{j \leq 2} \int \nolimits_0^{2\pi} s^{m_j,p_j}(y_j ) s^{\tilde{m}_j,\tilde{p}_j}(y_j ) dy_j \Big ) \Big ( \int \nolimits_0^{\pi} s^{m_3,1}(y_3 ) s^{\tilde{m}_3,1}(y_3 ) dy_3 \Big ) = \delta^{\textbf{m},\tilde{\textbf{m}}} \delta^{\textbf{p},\tilde{\textbf{p}}} \text{ . } \]
		
		One sees in \eqref{VectorFieldDef_FreeSlip} or \eqref{VectorFieldDef_Explicit} that the vertical sinusoids are $\cos (\ShaThree x_3)$ for $v_1^{\textbf{n}},v_2^{\textbf{n}}$ and $\sin (\ShaThree x_3)$ for $v_3^{\textbf{n}}$, hence the boundary conditions \eqref{BC_Velocity} are satisfied.  Note that one has the property $\sum_{i\leq 3} \mathcal{G}_{i,i}^{\textbf{n}} \mathsf{k}_i m_i = 0 $ which easily follows from the definitions of $\mathcal{G}_{i,i}^{\textbf{n}}$ in terms of the cross product in \eqref{VectorFieldDef_MatsVecs}.  One can then check the divergence free property using this property together with \eqref{SinusoidDerivative} as follows:
		\[ \nabla \cdot \textbf{v}^{\textbf{n}} = \frac{\eta^{\textbf{m}}}{|\mathcal{G}^{\textbf{n}}\normVec^{\textbf{n}}|V} \sum_{i \leq 3} \mathcal{G}^{\textbf{n}}_{i,i} \varsigma_{i,i}^{\textbf{p}} \partial_{x_i} \Big [ \prod_{j \leq 3}   s^{m_j,p_j + \delta^{i,j}}(\mathsf{k}_{j} x_j) \Big ] = \frac{\eta^{\textbf{m}} \prod_{j \leq 3}   s^{m_j,p_j}(\mathsf{k}_{j} x_j) }{|\mathcal{G}^{\textbf{n}}\normVec^{\textbf{n}}|V} \sum_{i \leq 3} \mathcal{G}^{\textbf{n}}_{i,i} \mathsf{k}_{i} m_i = 0 . \]
		To check the orthonormality property note that using the substitutions $y_j = \mathsf{k}_j x_j$ one has
		\begin{align} \int_{\Omega} v_i^{\textbf{n}} v_i^{\textbf{n}} d\textbf{x} & = \frac{\eta^{\textbf{m}}\mathcal{G}_{i,i}^{\textbf{n}} \varsigma_{i,i}^{\textbf{p}} \eta^{\tilde{\textbf{m}}} \mathcal{G}_{i,i}^{\tilde{\textbf{n}}} \varsigma_{i,i}^{\tilde{\textbf{p}}} }{|\mathcal{G}^{\textbf{n}}\boldsymbol{\nu}^{\textbf{n}}||\mathcal{G}^{\tilde{\textbf{n}}}\boldsymbol{\nu}^{\tilde{\textbf{n}}}|V^2} \Big ( \prod_{j \leq 2} \int \nolimits_0^{2\pi} s^{m_j,p_j+\delta^{i,j}}(y_j ) s^{\tilde{m}_j,\tilde{p}_j+\delta^{i,j}}(y_j ) dy_j \Big ) \Big ( \int \nolimits_0^{\pi} s^{m_3,\delta^{i,3}}(y_3 ) s^{\tilde{m}_3,\delta^{i,3}}(y_3 ) dy_3 \Big ) \notag \\ & = \frac{ \mathcal{G}_{i,i}^{\textbf{n}} \mathcal{G}_{i,i}^{\tilde{\textbf{n}}} \nu_{i}^{\textbf{n}} }{|\mathcal{G}^{\textbf{n}}\boldsymbol{\nu}^{\textbf{n}}||\mathcal{G}^{\textbf{n}}\boldsymbol{\nu}^{\tilde{\textbf{n}}}|} \delta^{\textbf{m},\tilde{\textbf{m}}} \delta^{\textbf{p},\tilde{\textbf{p}}} \text{ . } \notag  \end{align}
		One has the property $\sum_{i \leq 3} \mathcal{G}_{i,i}^{(\textbf{m},\textbf{p},1)} \mathcal{G}_{i,i}^{(\textbf{m},\textbf{p},2)} = 0$ due to the definitions \eqref{VectorFieldDef_MatsVecs} in terms of the cross products, so since $\nu_{i}^{\textbf{n}} = (\nu_{i}^{\textbf{n}} )^2 \in\{0,1\}$
		it follows that 
		\[ \int_{\Omega} \textbf{v}^{\textbf{n}} \cdot \textbf{v}^{\textbf{n}} d\textbf{x} = \delta^{\textbf{m},\tilde{\textbf{m}}} \delta^{\textbf{p},\tilde{\textbf{p}}} \sum_{i\leq 3} \frac{ \mathcal{G}_{i,i}^{\textbf{n}} \mathcal{G}_{i,i}^{\tilde{\textbf{n}}} \nu_{i}^{\textbf{n}} }{|\mathcal{G}^{\textbf{n}}\boldsymbol{\nu}^{\textbf{n}}||\mathcal{G}^{\textbf{n}}\boldsymbol{\nu}^{\tilde{\textbf{n}}}|} = \delta^{\textbf{m},\tilde{\textbf{m}}} \delta^{\textbf{p},\tilde{\textbf{p}}} \delta^{c,\tilde{c}} \text{ . }  \]

		In order to prove completeness, we artificially extend the domain by defining $\textbf{u}$ for $x_3 \in [-\frac{\pi}{\ShaThree},0)$ by an odd extension for $\theta, u_3$ and an even extension for $u_1,u_2$ in order to satisfy the boundary conditions.  The sinusoidal functions are a complete basis for $L^2([0,2\pi])$, so it follows easily that $\{ f^{\textbf{n}} \}_{\textbf{n}\in \mathscr{N}_{\theta}}$ is complete for $L^2_0$, so one need only consider the velocity by expanding each component of the vector field.  As mentioned above one can assume each component has zero mean, and furthermore due to the even/odd properties above $u_1$, $u_2$ must be given only in terms of $\cos(\ShaThree m_3 x_3)$ and $u_3$ must be given only in terms $\sin(\ShaThree m_3 x_3)$.  Therefore the naive expansion of each component is given as follows, where the phase condition is enforced to avoid zero functions 
		\[ \begin{split} \textbf{u}(\textbf{x},t) = & \sum_{\substack{\textbf{m} \in \mathbb{Z}_{\textbf{u}} \\ \textbf{p} \in \mathscr{P}^{m_1} \times \mathscr{P}^{m_2} }} s^{m_1,p_1}(\ShaOne x_1) s^{m_2,p_2}(\ShaTwo x_2) \Big ( \sum_{\ell \leq 2} \hat{u}^{\textbf{m},\textbf{p}}_{\ell}(t) s^{m_3,0}(\ShaThree x_3 ) \uvec{e}^{\scriptscriptstyle \ell}  + \hat{u}^{\textbf{m},\textbf{p}}_{3}(t) s^{m_3,1}(\ShaThree x_3 ) \uVecThree \Big ) \text{ , } \end{split} \]
		in which the Fourier coefficients are given by
		\[ \hat{u}_{\ell}^{\textbf{m},\textbf{p}}(t) = \frac{(\eta^{\textbf{m}})^2}{V^2} \int_{0}^{\frac{2\pi}{\ShaOne}}\int_{0}^{\frac{2\pi}{\ShaTwo}}\int_{-\frac{\pi}{\ShaThree}}^{\frac{\pi}{\ShaThree}} u_\ell(\textbf{x},t) s^{m_1,p_1}(\ShaOne x_1) s^{m_2,p_2}(\ShaTwo x_2) s^{m_3,\delta^{\ell,3}} ( \ShaThree x_3) d\textbf{x} \text{ . } \]
		
		Since $\textbf{u}(\textbf{x},t)$ is incompressible, one can use the definitions of the $\hat{u}_{\ell}^{\textbf{m},\textbf{p}}$, integrate by parts and apply \eqref{SinusoidDerivative} to establish that for each $\textbf{m},\textbf{p}$ one must have the relation
		\begin{equation} \label{FourierBasis_DivFreeCond} (-1)^{p_1} \ShaOne m_1 \hat{u}^{\textbf{m},(p_1+1,p_2)}_{1} + (-1)^{p_2} \ShaTwo m_2  \hat{u}^{\textbf{m},(p_1,p_2+1)}_{2} + \ShaThree m_3  \hat{u}^{\textbf{m},\textbf{p}}_{3} = 0 . \end{equation}
		This relation then implies that the Fourier coefficients $\hat{u}_{\ell}^{\textbf{m},\textbf{p}}$ cannot be chosen independently, but must fulfill the linear constraint \eqref{FourierBasis_DivFreeCond}.  One must therefore determine how many coefficients can be chosen freely.  For example, when $\textbf{m} \in \mathbb{Z}_{> 0}^3$ each component of the vector field ($\ell \leq 3$) has four Fourier coefficients (one for each phase $\textbf{p} \in \{0,1\}^2$) for a total of twelve coefficients.  However, for each $\textbf{p}$ the condition \ref{FourierBasis_DivFreeCond} implies that one of these Fourier coefficients must be determined by the other two, hence one ends up with eight independent Fourier coefficients.  Therefore there is some transformation carrying the eight corresponding vector fields to the eight vector fields in \eqref{VectorFieldDef_FreeSlip}.  Similar arguments hold for $\textbf{m} \in \mathbb{Z}_{\textbf{u}}$ with some wave numbers equal to zero.

					\subsection{Derivation of the explicit form of the nonlinear terms}
					
					\label{app:NonlinearDerivation}
					
					The superscript $\truncInd$ is dropped for this derivation.  Expanding the expressions \eqref{AbstractNonlinear} using the Fourier expansions \eqref{GeneralFourierExpansion}, one obtains the following:\begin{equation} \notag
						\begin{split}  \int_{\Omega} \textbf{v}^{\textbf{n}} \cdot \big [ \big ( \textbf{u} \cdot \nabla \big )  \textbf{u} \big ]  d\textbf{x} & = \int_{\Omega} \textbf{v}^{\textbf{n}} \cdot \big [ \big ( \sum_{\textbf{n}'} u^{\textbf{n}'} \textbf{v}^{\textbf{n}'}  \cdot \nabla \big )  \sum_{\textbf{n}''} u^{\textbf{n}''} \textbf{v}^{\textbf{n}''}  \big ] d\textbf{x}  = \sum_{\textbf{n}'} \sum_{\textbf{n}''}  u^{\textbf{n}'} u^{\textbf{n}''} \int_{\Omega} \textbf{v}^{\textbf{n}} \cdot \big [ (\textbf{v}^{\textbf{n}'} \cdot \nabla ) \textbf{v}^{\textbf{n}''} \big ] d\textbf{x} , \\ \int_{\Omega} f^{\textbf{n}} \cdot \big [ \textbf{u} \cdot \nabla \theta  \big ] d\textbf{x} & = \int_{\Omega} f^{\textbf{n}} \cdot \big [ \sum_{\textbf{n}'} u^{\textbf{n}'} \textbf{v}^{\textbf{n}'} \cdot \nabla \sum_{\textbf{n}''} \theta^{\textbf{n}''} f^{\textbf{n}''} \big ] d\textbf{x} = \sum_{\textbf{n}'} \sum_{\textbf{n}''}  u^{\textbf{n}'} \theta^{\textbf{n}''} \int_{\Omega} f^{\textbf{n}} \cdot \big [ \textbf{v}^{\textbf{n}'} \cdot \nabla f^{\textbf{n}''} \big ] d\textbf{x} . \end{split}
					\end{equation}
					Having extracted the time dependent terms, denote the (time-independent) integrals via
					\[ I_{\textbf{u}}^{\boldsymbol{\alpha}} = \int_{\Omega} \textbf{v}^{\textbf{n}} \cdot \big [ (\textbf{v}^{\textbf{n}'} \cdot \nabla ) \textbf{v}^{\textbf{n}''} \big ] d\textbf{x} \hspace{.5 cm} \text{ , } \hspace{.5 cm} I_{\theta}^{\boldsymbol{\alpha}} = \int_{\Omega} f^{\textbf{n}} \cdot \big [ \textbf{v}^{\textbf{n}'} \cdot \nabla f^{\textbf{n}''} \big ] d\textbf{x} , \]
					where we have used $\boldsymbol{\alpha} := (\textbf{n},\textbf{n}',\textbf{n}'')$.  Next, we expand the dot products in terms of the vector components via
					\[ I_{\textbf{u}}^{\boldsymbol{\alpha}} =  \sum_{i,j \leq 3 } J_{i,j}^{\boldsymbol{\alpha}} \hspace{.5 cm} \text{ , } \hspace{.5 cm} I_{\theta}^{\boldsymbol{\alpha}} =  \sum_{i \leq 3} \tilde{J}_{i}^{\boldsymbol{\alpha}} \text{ , } \]
					in which
					\begin{equation} \label{NonlinDeriv_ExpDotProd} J_{i,j}^{\boldsymbol{\alpha}} = \int_{\Omega} v_{j}^{\textbf{n}}  \cdot \big [ v^{\textbf{n}'}_{i} \partial_{x_{i}} v^{\textbf{n}''}_{j} \big ] d\textbf{x} \hspace{.5 cm} \text{ , } \hspace{.5 cm} \tilde{J}_{i}^{\boldsymbol{\alpha}} = \int_{\Omega} f^{\textbf{n}} \cdot \big [ v^{\textbf{n}'}_{i} \partial_{x_{i}} f^{\textbf{n}''} \big ] d\textbf{x}  \text{ . } \end{equation}
					We can then pull out all parameter dependence and explicit dependence on $\textbf{m},\textbf{m}',\textbf{m}''$ from the integrals by inserting the formulas for $f^{\textbf{n}}$, $\textbf{v}^{\textbf{n}}$ in \eqref{def:ThetaFourierBasis},\eqref{VectorFieldDef_FreeSlip}:
					\begin{equation} \label{NonlinDeriv_WaveDep} \begin{split} J_{i,j}^{\boldsymbol{\alpha}} = \frac{ \eta^{\textbf{m}}\eta^{\textbf{m}'}\eta^{\textbf{m}''} \mathcal{G}_{j,j}^{\textbf{n}} \mathcal{G}_{i,i}^{\textbf{n}'}\mathcal{G}_{j,j}^{\textbf{n}''} \varsigma^{\textbf{p}}_{j,j} \varsigma^{\textbf{p}'}_{i,i} \varsigma^{\textbf{p}''}_{j,j} }{ |\mathcal{G}^{\textbf{n}}\boldsymbol{\nu}^{\textbf{n}}||\mathcal{G}^{\textbf{n}'}\boldsymbol{\nu}^{\textbf{n}'}||\mathcal{G}^{\textbf{n}''}\boldsymbol{\nu}^{\textbf{n}''}| V^3 } \varsigma^{\textbf{p}''}_{i,j} \mathsf{k}_i m_i'' \hat{I}_{i,j}^{\boldsymbol{\alpha}} \hspace{.5 cm} \text{ , } \hspace{.5 cm} \tilde{J}_{i}^{\boldsymbol{\alpha}} = \frac{ \eta^{\textbf{m}}\eta^{\textbf{m}'}\eta^{\textbf{m}''} \mathcal{G}_{i,i}^{\textbf{n}'} \varsigma^{\textbf{p}'}_{i,i} }{ |\mathcal{G}^{\textbf{n}'}\boldsymbol{\nu}^{\textbf{n}'}| V^3 } \varsigma^{\textbf{p}''}_{i,3} \mathsf{k}_i m_i'' \hat{I}_{i,3}^{\boldsymbol{\alpha}}   \text{ , } \end{split}  \end{equation}
					in which the terms $\hat{I}_{i,j}^{\boldsymbol{\alpha}}$ are integrals of sinusoidal functions.  In order to evaluate the terms $\hat{I}_{i,j}^{\boldsymbol{\alpha}}$, for a wave number triad $\boldsymbol{\mu} = (m,m',m'')$ and phase triad $\boldsymbol{\phi} = (p,p',p'')$ we define
					\[ E^{(\boldsymbol{\mu},\boldsymbol{\phi})} := \int \nolimits_0^{2\pi} s^{m,p}(y) s^{m',p'}(y) s^{m'',p''}(y) dy \text{ . } \]
					For each phase triad $\boldsymbol{\phi}$ the integrals $E^{(\boldsymbol{\mu},\boldsymbol{\phi})}$ are easily evaluated, since the are given in terms of elementary sinusoidal functions.  Furthermore there are only eight cases to evaluate, since $\boldsymbol{\phi} \in \{ 0,1\}^3$.  The difficulty in evaluating the terms $\hat{I}_{i,j}^{\boldsymbol{\alpha}}$ lies in determining how to express these in terms of $E^{(\boldsymbol{\mu},\boldsymbol{\phi})}$.  However, using the phase maps defined in \eqref{PhaseMaps}, the substitutions $y = \ShaOne m_1, y = \ShaTwo m_2, y = 2\ShaThree m_3 $ respectively, and the fact $\ShaOne \ShaTwo \ShaThree = 1$ this is done as follows:
					\begin{equation} \label{NonlinDeriv_SinIntegrals} \hat{I}_{i,j}^{\boldsymbol{\alpha}} = \frac{1}{2} E^{(\boldsymbol{\mu}^1,\pi^{i,j,1}(\boldsymbol{\phi}^1))} E^{(\boldsymbol{\mu}^2,\pi^{i,j,2}(\boldsymbol{\phi}^2))} E^{(\frac{\boldsymbol{\mu}^3}{2},\pi^{i,j,3}(\boldsymbol{\phi}^3))} \text{ . } \end{equation}
					We therefore see that we must evaluate the eight integrals $E^{(\boldsymbol{\mu},\boldsymbol{\phi})}$ for $\boldsymbol{\phi} \in \{0,1\}^3$, where $\boldsymbol{\mu}$ consists of integers, or certain special cases $\boldsymbol{\mu}$ can consist of half-integers.  For any $\boldsymbol{\mu} \in \mathbb{Z}^3_{\geq 0}$ one has
					\[ E^{(\boldsymbol{\mu},(0,0,1))} = E^{(\boldsymbol{\mu},(0,1,0))} = E^{(\boldsymbol{\mu},(1,0,0))} = E^{(\boldsymbol{\mu},(1,1,1))} = 0 \text{ . } \]
					since these are integrals of odd functions over the domain $[-\pi,\pi]$ (after using a substitution).  This then gives the phase condition \eqref{PhaseCompatibilityCond} on $p_k,p_k',p_k''$.  One does not need to consider the half integer case here, since the phase maps $\pi^{i,j,k}$ map the subgroup $\{ \boldsymbol{\xi}^{\tilde{j}} \}_{\tilde{j}\leq 4}$ into itself, and $\boldsymbol{\phi}^3 = \boldsymbol{\xi}^1$.  So one need only consider the remaining four integrals, where $\boldsymbol{\mu}$ can have either integer or half integer components.  In order to evaluate these, one can repeatedly apply the angle addition formulas:
					
					\vspace{-.2 cm}
					\small
					
					\begin{equation} \label{NonlinDeriv_TrigAddition} \begin{split}  \cos (m' y) \cos (m'' y) = \frac{\cos ( (m' + m'')y) + \cos ((m' - m'')y )}{2} & \text{ , } \cos (m' y) \sin (m'' y) = \frac{\sin ((m'+m'')y) - \sin ( (m' -m'')y )}{2} \text{ , } \\ \sin (m'y) \sin (m''y) = \frac{\cos ((m'-m'')y ) -\cos ((m'+m'')y)}{2} & \text{ , } \sin (m'y) \cos (m''y) = \frac{\sin ((m'+m'')y) + \sin ((m'-m'')y )}{2}. \end{split} \end{equation}
					
					\normalsize
					
					\noindent To see how the evaluation procedure works, consider the specific example 
					\[ E^{(\boldsymbol{\mu},(0,1,1))} = \int \nolimits_0^{2\pi} \cos(my) \sin (m'y) \sin( m''y) dy . \]
					In this case, $E^{(\boldsymbol{\mu},(0,1,1))} = 0$ if $m' = 0$ or $m'' = 0$ since $\sin(0) = 0$, so $E^{(\boldsymbol{\mu},(0,1,1))}$ should be proportional to the amplitude factor $A^{(\boldsymbol{\mu},0,1,1)}:=\iota^{m'} \iota^{m''}$.  Expanding via \eqref{NonlinDeriv_TrigAddition}, one obtains 
					\[  E^{(\boldsymbol{\mu},0,1,1)} = \frac{1}{4} \int_0^{2\pi} \big [ \cos \big ( (m + m'-m'') y \big ) + \cos \big ( (m - m' + m'') y \big ) - \cos \big ( (m - m'-m'') y \big ) - \cos \big ( (m+ m'+m'') y \big ) \big ] dy .  \]
					If no convolution condition is satisfied (i.e. $m \neq |m'\pm m''|$) then in the case where $m,m',m''$ are integers one obtains $w^{-1} \sin(wx)$ for each of the integrands (for some $w \neq 0$), which then evaluates to zero due to periodicity.  If at least one of $m,m',m''$ is a half integer, then a convolution-type condition must still be satisfied, since the function $w^{-1} \sin(wx)$ is zero at both boundaries if $w \neq 0$ is a half integer.  On the other hand, if $m',m''> 0$ and a convolution-type condition is satisfied, then the integral is non-zero.  If $m = m' + m''$ then it is negative, otherwise it is positive.  The absolute value is equal to $\frac{\pi}{2}$ if $m,m',m''$ are all non-zero, or $\pi$ if $m = 0$.  This same argument applies for each of the integrals $E^{(\boldsymbol{\mu},\boldsymbol{\xi}^j)}$, $j = 2,3,4$, hence one arrives at the following general statements:
					\begin{enumerate}
						\item $E^{(\boldsymbol{\mu},\boldsymbol{\phi})} = 0$ if the triad $\boldsymbol{\mu}$ doesn't satisfy a convolution-type condition of the form $m = | m' \pm m''|$.  
						\item If $\boldsymbol{\mu} \neq \textbf{0}$ satisfies a convolution condition and $\boldsymbol{\phi} \in \{\boldsymbol{\xi}^i \}_{i \leq 4}$ then 
						\begin{equation} \label{NonlinDeriv_EvalResult} E^{(\boldsymbol{\mu},\boldsymbol{\phi})} = \frac{4\pi}{(\eta^{m}\eta^{m'}\eta^{m''})^2} \sigma^{(\boldsymbol{\mu},\boldsymbol{\phi})}  A^{(\boldsymbol{\mu},\boldsymbol{\phi})} , \end{equation}
						for the sign coefficients $\sigma^{(\boldsymbol{\mu},\boldsymbol{\phi})}$ given in \eqref{def_SignCoefs}, and amplitude factors $A^{(\boldsymbol{\mu},\boldsymbol{\phi})} $.
					\end{enumerate}
					For $E^{(\boldsymbol{\mu},\boldsymbol{\xi}^1)}$, the most of the same argument holds except that in this note that the integral is non-zero even when $\boldsymbol{\mu} = \textbf{0}$.  In this case the amplitude of the integral should be $2\pi$, hence the amplitude factor $A^{(\boldsymbol{\mu},\boldsymbol{\xi}^1)}$ is as defined in \eqref{def_SignCoefs}. 
					
					\subsection{Explicit construction of low-dimensional truncations} 
					
					\label{app:ConsProced}
					
					For illustration, we provide explicit examples of the construction procedure for very simple, low-dimensional truncated models.  This procedure follows three steps: mode selection, evaluation of the linear coupling structure and evaluation of the nonlinear coupling structure.  
					
					\paragraph{The Lorenz-Stenflo model} 
					
					\begin{enumerate}
						\item (Mode selection): The classical Lorenz '63 model can be found by fixing a phase $\textbf{p} = (0,0)$ and including only the largest scale mode $\textbf{m}^1 = (1,0,1)$ for $\textbf{u}$ and $\theta$ which allows for a nontrivial buoyancy force together with an additional mode $\textbf{m}^2 = (0,0,2)$ for $\theta$ which allows for nonlinear interaction.  The Lorenz-Stenflo model is the simplest extension of the classical Lorenz '63 model which allows for rotation, hence one includes also an additional phase $\textbf{p}^2 = (1,1)$.  Since in both $\textbf{m}^1,\textbf{m}^2$ at least one wave number is zero, the only possible value of $c$ is $c =1$.  Denoting 
						$\textbf{n}^1 :=  (\textbf{m}^1, \textbf{p}^1,c)$, $\textbf{n}^2 :=  (\textbf{m}^2, \textbf{p}^1,c)$, $\textbf{n}^{3} = (\textbf{m}^1,\textbf{p}^2,c)$, the resulting truncation has index sets
						\[ \mathscr{N}^{LS}_{\textbf{u}} = \{ \textbf{n}^1, \textbf{n}^3 \} \hspace{.25 cm} \text{ , } \hspace{.25 cm} \mathscr{N}^{LS}_{\theta} = \{ \textbf{n}^1, \textbf{n}^2 \} \text{ . } \]
						We then look for the explicit form of the ordinary differential equation for $\textbf{X}^{LS} := (u^{LS,\textbf{n}^1},u^{LS,\textbf{n}^3},\theta^{LS,\textbf{n}^1},\theta^{LS,\textbf{n}^2})^T$
						\[ \frac{d}{dt} \textbf{X}^{LS} = \mathcal{L}^{LS} \textbf{X}^{LS} + \textbf{N}^{LS}(\textbf{X}^{LS}) \text{ . }  \]
						
						\item (Linear couplings): To evaluate the linear coupling, first note from \eqref{GeneralBoussinesqODE_Vel}, \eqref{GeneralBoussinesqODE_Temp} that only variables with the same wave number $\textbf{m}$ are coupled linearly, hence one immediately has a block diagonal structure.  The only variable with wave number $\textbf{m}^2$ is $\theta^{\textbf{n}^2}$, hence the only linear term here is the diffusive term.  For the variables $u^{\textbf{n}^1} , u^{\textbf{n}^3} , \theta^{\textbf{n}^1}$, note that all three have a dissipative self-interaction term, $u^{\textbf{n}^1} , \theta^{\textbf{n}^1}$ are coupled via the buoyancy force, $u^{\textbf{n}^1} , u^{\textbf{n}^3}$ are coupled via the Coriolis force, and $u^{\textbf{n}^3} , \theta^{\textbf{n}^1}$ have no linear coupling due to the difference in phase.  Thus one finds
						\begin{equation}\label{e:Lorenz63}
							\mathcal{L}^{LS} = 
							\begin{pmatrix}
								-\Pra|\mathcal{K}\textbf{m}^1|^2 & \Pra \Rot g_2 & \Pra \Ray g_1 & 0\\  
								- \Pra \Rot g_2 & -\Pra|\mathcal{K}\textbf{m}^1|^2 & 0 & 0\\ 
								g_1   & 0 &  -|\mathcal{K}\textbf{m}^1|^2 & 0\\
								0 & 0 & 0 & -|\mathcal{K}\textbf{m}^2|^2
							\end{pmatrix} \text{ , }
						\end{equation}
						in which $|\mathcal{K}\textbf{m}^1|^2 = \ShaOne^2 + \ShaThree^2$, $|\mathcal{K}\textbf{m}^2|^2 = 4\ShaThree^2$, $g_1 = \frac{\mathcal{G}_{3,3}^{\textbf{n}^1} \nu_3^{\textbf{n}^1}}{|\mathcal{G}^{\textbf{n}^1} \boldsymbol{\nu}^{\textbf{n}^1|}} = \frac{-\ShaOne }{|\mathcal{K}\textbf{m}^1|}$, $g_2 = \frac{\mathcal{G}_{2,2}^{\textbf{n}^3} \mathcal{G}_{1,1}^{\textbf{n}^1} \nu_1^{\textbf{n}^1} - \mathcal{G}_{1,1}^{\textbf{n}^3} \mathcal{G}_{2,2}^{\textbf{n}^1} \nu_2^{\textbf{n}^1} }{|\mathcal{G}^{\textbf{n}^3} \boldsymbol{\nu}^{\textbf{n}^3}||\mathcal{G}^{\textbf{n}^1} \boldsymbol{\nu}^{\textbf{n}^1}| } = \frac{\ShaThree}{|\mathcal{K}\textbf{m}^1|} $.
						
						\item (Nonlinear couplings): To evaluate the nonlinear coupling, first note that the convolution and phase compatibility conditions in \eqref{WaveCompatibilityCond}, \eqref{PhaseCompatibilityCond} must be satisfied.  Since both velocity variables $\textbf{u}^{\textbf{n}^1}, \textbf{u}^{\textbf{n}^3}$ have the same wave vector $\textbf{m}^1$, no convolution condition can be satisfied, hence the equations for these variables have no nonlinear terms.  On the other hand, the convolution conditions $m_1^2 = m_1^1 - m_1^1$, $m_3^2 = m_3^1 + m_3^1$ and phase conditions $\boldsymbol{\phi}^{1} = \boldsymbol{\phi}^{2} = \boldsymbol{\xi}^1$ are satisfied, hence one has 
						\[ \textbf{N}^{LS}(\textbf{X}^{LS}) = (0,0,I^{\boldsymbol{\alpha}^{1,2}}_\theta u^{LS,\textbf{n}^1} \theta^{LS,\textbf{n}^2} , I^{\boldsymbol{\alpha}^{2,1}}_\theta u^{LS,\textbf{n}^1} \theta^{LS,\textbf{n}^1} )^T \text{ , } \]
						where $\boldsymbol{\alpha}^{i,j} = (\textbf{n}^i,\textbf{n}^1,\textbf{n}^j)$.  One must then determine the coefficients $I^{\boldsymbol{\alpha}^1}_\theta, I^{\boldsymbol{\alpha}^2}_\theta$ from \eqref{def:NonlinearCoefs} - \eqref{def_SignCoefs}, although one has
						\[  I^{\boldsymbol{\alpha}^{1,2}}_\theta = \langle f^{\textbf{n}^1} [ \textbf{v}^{\textbf{n}^1} \cdot \nabla f^{\textbf{n}^2} ] \rangle = - \langle f^{\textbf{n}^2} [ \textbf{v}^{\textbf{n}^1} \cdot \nabla f^{\textbf{n}^1} ] \rangle = - I^{\boldsymbol{\alpha}^{2,1}}_\theta\text{ , } \]
						hence one only need determine $I^{\boldsymbol{\alpha}^{1,2}}_\theta$.  Since $m_1'' = m_2'' = 0$ only $i=3$ is non-zero in \eqref{def:NonlinearCoefs} and one has $\varsigma_{3,3}^{\textbf{p}''} = 1$, hence
						\[ I^{\boldsymbol{\alpha}^{1,2}}_{\theta} = \frac{ 8\mathcal{G}_{3,3}^{\textbf{n}'} \mathsf{k}_3 2 I_{3,3}^{\boldsymbol{\alpha}^{1,2}}}{(\sqrt{2})^5 |\mathcal{G}^{\textbf{n}'}\boldsymbol{\nu}^{\textbf{n}'}| V} = \frac{ 2\sqrt{2} \ShaOne \mathsf{k}_3 I_{3,3}^{\boldsymbol{\alpha}^{1,2}}}{ \sqrt{ \ShaOne^2 + \ShaThree^2 } V}  \text{ . } \]
						Finally, the factor $I_{3,3}^{\boldsymbol{\alpha}^{1,2}}$ must be evaluated via \eqref{def:NonlinearCoefs2}, \eqref{def_SignCoefs}.  One finds
						\[ \begin{split} A^{(\boldsymbol{\mu}^{1},\pi^{3,3,1}(\boldsymbol{\phi}^1))} = A^{((1,1,0),\boldsymbol{\xi}^1)} = 1 \hspace{.5 cm} & \text{ , } \hspace{.5 cm} \sigma^{(\boldsymbol{\mu}^{1},\pi^{3,3,1}(\boldsymbol{\phi}^1))} = \sigma^{((1,1,0),\boldsymbol{\xi}^1)} = 1 \text{ , } \\ A^{(\boldsymbol{\mu}^{2},\pi^{3,3,2}(\boldsymbol{\phi}^2))} = A^{((0,0,0),\boldsymbol{\xi}^1)} = \frac{1}{2} \hspace{.5 cm} & \text{ , } \hspace{.5 cm} \sigma^{(\boldsymbol{\mu}^{2},\pi^{3,3,2}(\boldsymbol{\phi}^2))} = \sigma^{((0,0,0),\boldsymbol{\xi}^1)} = 1 \text{ , } \\ A^{(\boldsymbol{\mu}^{3},\pi^{3,3,3}(\boldsymbol{\phi}^3))} = A^{((1,1,2),\boldsymbol{\xi}^4)} = 1 \hspace{.5 cm} & \text{ , } \hspace{.5 cm} \sigma^{(\boldsymbol{\mu}^{3},\pi^{3,3,3}(\boldsymbol{\phi}^3))} = \sigma^{((1,1,2),\boldsymbol{\xi}^4)} = -1 \text{ , } \end{split} \]
						and so $I^{\boldsymbol{\alpha}^{1,2}}_{\theta} = \frac{-  \ShaOne \mathsf{k}_3 }{ \sqrt{ 2(\ShaOne^2 + \ShaThree^2) } V} $.  Combining the above results, one has the explicit ODE model \eqref{e:Lorenz63}.
					\end{enumerate}
					
					\paragraph{Linear 3D flow model}
					
					While the above Lorenz-Stenflo model illustrates much of the model construction process, it is an essentially 2D fluid flow model and does not illustrate the full complexity of the linear coupling structure for 3D flows.  To this end we construct an appropriate 3d flow model via the process above.  For simplicity we choose only a single wave vector $\textbf{m}$, so that no convolution conditions in \eqref{WaveCompatibilityCond} can be satisfied and the model is strictly linear.
					
					\begin{enumerate}
						\item (Mode selection): For this model we choose only one wave vector with strictly positive entries $\textbf{m}^1 = (1,1,1)$, and two of the four available phases, so as to capture the Coriolis coupling.  Hence let $\textbf{p}^{1} = (0,0)$, $\textbf{p}^{2} = (1,1)$.  Since all entries of $\textbf{m}^1$ are positive, both $c=1$ and $c=2$ are available for $\textbf{u}$.  Hence we define
						\[ \textbf{n}^1 :=  (\textbf{m}^1, \textbf{p}^1,1) \hspace{.25 cm} \text{ , } \hspace{.25 cm} \textbf{n}^2 :=  (\textbf{m}^1, \textbf{p}^2,1) \hspace{.25 cm} \text{ , } \hspace{.25 cm} \textbf{n}^3 :=  (\textbf{m}^1, \textbf{p}^1,2) \hspace{.25 cm} \text{ , } \hspace{.25 cm} \textbf{n}^4 :=  (\textbf{m}^1, \textbf{p}^2,2) \text{ . } \] 
						We choose the following index sets
						\[ \mathscr{N}^{L3d}_{\textbf{u}} = \{ \textbf{n}^1, \textbf{n}^2, \textbf{n}^3, \textbf{n}^4 \} \hspace{.25 cm} \text{ , } \hspace{.25 cm} \mathscr{N}^{L3d}_{\theta} = \{ \textbf{n}^1, \textbf{n}^2 \} \text{ , } \]
						and look for an explicit ODE for $\textbf{X}^{L3d} := (u^{\textbf{n}^1,L3d},u^{\textbf{n}^2,L3d},u^{\textbf{n}^3,L3d},u^{\textbf{n}^4,L3d},\theta^{\textbf{n}^1,L3d},\theta^{\textbf{n}^2,L3d})^T$.  Note that since only one wave vector has been included, no convolution condition in \eqref{WaveCompatibilityCond} can be satisfied, hence there are no nonlinear couplings in this model.  Thus one has 
						\[ \frac{d}{dt} \textbf{X}^{L3d} = \mathcal{L}^{L3d} \textbf{X}^{L3d} \text{ . }  \]
						
						\item (Linear couplings): In order to determine the linear coupling structure, we use Figure \ref{fig:linear_coupling} below, which shows how in general the linear coupling among components for a give index vector works. 
						
						\definecolor{Color1}{rgb}{1,0,0}
						\definecolor{Color2}{rgb}{0,.8,.1}
						\definecolor{Color3}{rgb}{0,0,1}
						
						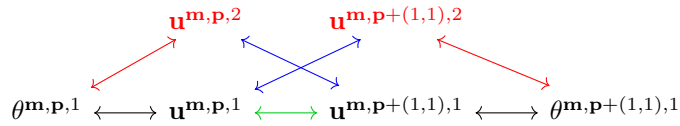
\begin{figure}[H]
							\begin{center}
								
								\[\begin{tikzcd}
									& \textcolor{Color1}{{\textbf{u}^{\textbf{m},\textbf{p},2}}} & \textcolor{Color1}{{\textbf{u}^{\textbf{m},\textbf{p}+(1,1),2}}} \\
									{\theta^{\textbf{m},\textbf{p},1}} & {\textbf{u}^{\textbf{m},\textbf{p},1}} & {{\textbf{u}^{\textbf{m},\textbf{p}+(1,1),1}}} & {{\theta^{\textbf{m},\textbf{p}+(1,1),1}}}
									\arrow[color={Color1}, tail reversed, from=2-1, to=1-2]
									\arrow[tail reversed, from=2-1, to=2-2]
									\arrow[color={Color3}, tail reversed, from=2-2, to=1-3]
									\arrow[color={Color2}, tail reversed, from=2-2, to=2-3]
									\arrow[color={Color3}, tail reversed, from=2-3, to=1-2]
									\arrow[color={Color1}, tail reversed, from=2-4, to=1-3]
									\arrow[ tail reversed, from=2-4, to=2-3]
								\end{tikzcd}\]
								\caption{Illustration of the linear coupling structure for a given index vector $\textbf{n} = (\textbf{m},\textbf{p},c)$. Colors indicate necessary (but not sufficient) condition(s) for the presence of non-trivial coupling terms between terms connected by arrows as follows: 
									Black requires $m_1^{p_1} m_2^{p_2}\neq 0$, 
									\textcolor{Color2}{green requires $\mathsf{S}\neq0$ and exactly one of $m_j =0$ among $j = 1,2,3$}, 
									\textcolor{Color1}{red requires $\textbf{m}>0$}, 
									\textcolor{Color3}{blue requires both $\mathsf{S}\neq0$, $\textbf{m}>0$}.}
								\label{fig:linear_coupling}
							\end{center}
						\end{figure}
						
						Starting with $u^{\textbf{n}^1}$ we note this has the buoyancy coupling to $\theta^{\textbf{n}^1}$, but since $\textbf{m}^1 > 0$ one also has the buoyancy coupling between $\theta^{\textbf{n}^1}$ and $u^{\textbf{n}^3}$.  One also has the Coriolis coupling between $u^{\textbf{n}^1}$ and $u^{\textbf{n}^4}$, but since $\textbf{m}> 0$ the Coriolis coupling between $u^{\textbf{n}^1}$ and $u^{\textbf{n}^2}$ is zero.  Since one includes $u^{\textbf{n}^4}$ there is then a buoyancy coupling between $u^{\textbf{n}^1}$ and $\theta^{\textbf{n}^1}$, and a Coriolis coupling between $u^{\textbf{n}^2}$ and $u^{\textbf{n}^3}$.  Taken together this gives the matrix in \eqref{LinMat_Full3d}.
						
					\end{enumerate}

					\section{Proof of the vorticity balance}
					
					\label{app:VortBal}
					
					\begin{proof}
						For the proof of the vorticity balance, note that the truncated vorticity $\vorticity^{\truncInd} = \nabla \times \textbf{u}^{\truncInd}$ must satisfy
						\begin{equation} \label{TruncatedVorticity} \partial_{t} \vorticity^{\truncInd} = \Pra \Delta \vorticity^{\truncInd} +  \nabla \times \mathcal{P}^{\truncInd}_{\textbf{u}} \big [ \Pra \Ray \theta^{\truncInd} \uVecThree - \Pra \Rot ( \uVecThree \times \textbf{u}^{\truncInd} ) - \textbf{u}^{\truncInd} \cdot \nabla \textbf{u}^{\truncInd}  \big ] , \end{equation}
						Computing the volume integral of \eqref{TruncatedVorticity}, one can eliminate the horizontal derivatives in the viscous term using periodicity, and only $\partial_{x_3}^2 \vorticity$ cannot be eliminated due to the boundary conditions, as for \eqref{Balance_Vort}.  There are then three terms to consider, the buoyancy term, the nonlinear term and the Coriolis term, as follows:
						\[ T_1 =  \langle \nabla \times \mathcal{P}^{\truncInd}_{\textbf{u}} \big [ \theta^{\truncInd} \uVecThree \big ] \rangle \hspace{.5 cm} \text{ , } \hspace{.5 cm} T_2 =  -\langle \nabla \times \mathcal{P}^{\truncInd}_{\textbf{u}} \big [ \uVecThree \times \textbf{u}^{\truncInd} \big ] \rangle \hspace{.5 cm} \text{ , } \hspace{.5 cm} T_3 =  -\langle \nabla \times \mathcal{P}^{\truncInd}_{\textbf{u}} \big [ \textbf{u}^{\truncInd} \cdot \nabla \textbf{u}^{\truncInd} \big ] \rangle \text{ . } \]
						In order to transform these terms into those in \eqref{Balance_Vort}, we consider their expression in Fourier space and must determine the effect of the curl and projection operators.  For a general vector valued function $\textbf{F}$ one can use the definition of the projection operator $\mathcal{P}_{\textbf{u}}^{\truncInd}$, apply the curl operator and compute the spatial integral, and then eliminate all of the terms involving derivatives in $x_1,x_2$ due to the periodicity, thereby obtaining the following:
						\begin{align} \label{VortConsis_CurlProj} \langle \nabla \times \mathcal{P}^{\truncInd}_{\textbf{u}} \big [ \textbf{F} \big ] \rangle = \sum_{ \textbf{n} \in \mathscr{N}_{\textbf{u}}^{\truncInd}} \big \langle \textbf{F} \cdot \textbf{v}^{\textbf{n}} \big \rangle \big \langle \nabla \times \textbf{v}^{\textbf{n}} \big \rangle = \sum_{ \textbf{n} \in \mathscr{N}_{\textbf{u}}^{\truncInd}} \big \langle \textbf{F} \cdot \textbf{v}^{\textbf{n}} \big \rangle \big ( \langle  \partial_{x_3} v_1^{\textbf{n}} \rangle \uVecTwo - \langle \partial_{x_3} v_2^{\textbf{n}} \rangle \uVecOne \big ) . \end{align} 
						Furthermore note that only the terms for which $m_1 = m_2 = 0$ are non-zero, since $\partial_{x_3} v_1^{\textbf{n}}, \partial_{x_3} v_2^{\textbf{n}}$ otherwise have zero mean.  Since $m_1 = m_2 = 0$, it follows from \eqref{PhaseIndexSets} that $\textbf{p} = (0,1)$ or $(1,0)$, and from the definition \eqref{VectorFieldDef_FreeSlip} one has 
						\begin{equation} \label{VortBal_VertStratModes} \textbf{v}^{(0,0,m_3,1,0,1)} = -\frac{\sqrt{2}}{V} \uVecOne \cos (\ShaThree m_3 x_3 ) \hspace{.5 cm} \text{ , } \hspace{.5 cm} \textbf{v}^{(0,0,m_3,0,1,1)} = \frac{\sqrt{2}}{V} \uVecTwo \cos (\ShaThree m_3 x_3 ) \text{ . } \end{equation}
						One can therefore evaluate the integrals in \eqref{VortConsis_CurlProj}, and one finds that the only non-zero terms in the sum must have $m_3$ odd.  Thus the curl of the projection gives the following sum: 
						\begin{align} \label{VortConsis_Sum} \langle \nabla \times \mathcal{P}^{\truncInd}_{\textbf{u}} \big [ \textbf{F} \big ] \rangle = \sum_{\substack{\textbf{n} \in \mathscr{N}_{\textbf{u}}^{\truncInd} \cap \mathscr{N}_{\textbf{u}}^{*} \\  m_3 \text{ odd }}} 4 \sqrt{2\pi} \ShaThree  \big \langle \textbf{F} \cdot \textbf{v}^{\textbf{n}} \big \rangle \big ( \delta^{\textbf{p},(1,0)} \uVecTwo + \delta^{\textbf{p},(0,1)} \uVecOne \big ) . \notag \end{align}
						Each of the terms $T_1,T_2,T_3$ are thus sums over such restricted indices $\textbf{n}$ obtained by inserting different expressions for $\textbf{F}$.
						
						First we consider the terms $T_1$, $T_2$.  The buoyancy term $T_1$ can immediately be eliminated since from \eqref{BasisRelations} one has
						\[ \langle  \theta^{\truncInd} \uVecThree \cdot \textbf{v}^{\textbf{n}} \rangle = \sum_{\tilde{\textbf{n}} \in \mathscr{N}_{\theta}^{\truncInd}} \delta^{\textbf{m},\tilde{\textbf{m}}} \delta^{\textbf{p},\tilde{\textbf{p}}} \frac{\mathcal{G}_{3,3}^{\textbf{n}}\nu_3^{\textbf{n}}}{|\mathcal{G}^{\textbf{n}}\boldsymbol{\nu}^{\textbf{n}}|} \theta^{\truncInd,\tilde{\textbf{n}}} \text{ , }  \] 
						and by definition one has $\mathcal{G}_{3,3}^{\textbf{n}}=0$ since $m_1 = m_2 = 0$.  The Coriolis term $T_2$ only appears for $\Rot \neq 0$, hence the condition in Criterion \ref{Crit:VortCrit} (ii).  From the above one has
						\[ T_2 = -4 \sqrt{2\pi} \ShaThree \sum_{\substack{\textbf{n} \in \mathscr{N}_{\textbf{u}}^{\truncInd}\cap \mathscr{N}_{\textbf{u}}^{*} \\ m_3 \text{ odd }}} \sum_{ \tilde{\textbf{n}} \in \mathscr{N}_{\textbf{u}}^{\truncInd} } u^{\truncInd,\tilde{\textbf{n}}} \big \langle (\uVecThree \times \textbf{v}^{\tilde{\textbf{n}}} ) \cdot \textbf{v}^{\textbf{n}} \big \rangle \big ( \delta^{\textbf{p},(1,0)} \uVecTwo + \delta^{\textbf{p},(0,1)} \uVecOne \big ) \text{ . }  \]
						Inserting $m_1 = m_2 = 0$ into the definition of $\mathcal{G}^{\textbf{n}}$ one finds the coefficient from \eqref{BasisRelations} is given by $(-1)^{p_2}$, hence one has:
						\begin{equation} \label{VortBal_T2Bal_LHS} T_2 = 4 \sqrt{2\pi} \ShaThree \sum_{\substack{\textbf{n} \in \mathscr{N}_{\textbf{u}}^{\truncInd}\cap \mathscr{N}_{\textbf{u}}^{*} \\ m_3 \text{ odd }}} \sum_{ \tilde{\textbf{n}} = (\textbf{m},p_1+1,p_2+1,1) \in \mathscr{N}_{\textbf{u}}^{\truncInd} } u^{\truncInd,\tilde{\textbf{n}}} \big ( \delta^{\textbf{p},(0,1)} \uVecOne - \delta^{\textbf{p},(1,0)} \uVecTwo \big ) \text{ . }  \end{equation}
						In particular, note that the term $\textbf{u}^{(\textbf{m},p_1+1,p_2+1,1)}$ appears in the above sum iff both $(\textbf{m},p_1,p_2,1)$ and $(\textbf{m},p_1+1,p_2+1,1)$ are included in $\mathscr{N}_{\textbf{u}}^{\truncInd}$.  On the other hand for the vorticity balance to be satisfied one must have
						\[ T_2 = \langle \partial_{x_3} \textbf{u} \rangle \text{ , } \]
						so expanding the right hand side one obtains the following, where all terms with $\textbf{m}_h := (m_1,m_2) \neq \textbf{0}$ are already eliminated due to periodicity and the term $\partial_{x_3} u_3$ is eliminated due to the fundamental theorem of calculus and the non-penetration boundary condition:
						\[ \langle \partial_{x_3} \textbf{u}^{\truncInd} \rangle = \sum_{ \textbf{n} \in \mathscr{N}_{\textbf{u}}^{\truncInd} \cap \mathscr{N}_{\textbf{u}}^* }  u^{\truncInd,\textbf{n}} \big ( \uVecOne \langle \partial_{x_3} v_1^{\textbf{n}} \rangle +   \uVecTwo \langle \partial_{x_3} v_1^{\textbf{n}} \rangle \big ) \text{ . } \]
						One can easily evaluate the integrals and to obtain 
						\begin{equation} \label{VortBal_T2Bal_RHS} \langle \partial_{x_3} \textbf{u}^{\truncInd} \rangle = 4 \sqrt{2\pi} \ShaThree \sum_{ \textbf{n} \in \mathscr{N}_{\textbf{u}}^{\truncInd} \cap \mathscr{N}_{\textbf{u}}^* }  u^{\truncInd,\textbf{n}} \big ( \uVecOne \delta^{\textbf{p},(1,0)} -  \uVecTwo \delta^{\textbf{p},(0,1)} \big ) \text{ . } \end{equation}
						Comparing the above expressions, one sees that the terms in \eqref{VortBal_T2Bal_RHS} exactly agree with those in the sum \eqref{VortBal_T2Bal_LHS} if and only if Criterion \ref{Crit:VortCrit} (ii) is satisfied.
						
						The final term to consider is $T_3$.  By expanding the nonlinear term, integrating by parts and using $m_1 = m_2 = 0$, one finds 
						\begin{equation} \label{VortConsis_Nonlin} \langle \big [ ( \textbf{u}^{\truncInd} \cdot \nabla ) \textbf{u}^{\truncInd} \big ] \cdot \textbf{v}^{\textbf{n}} \rangle =  - \sum_{\textbf{n}' \in \mathscr{N}_{\textbf{u}}^{\truncInd} } \sum_{\textbf{n}'' \in \mathscr{N}_{\textbf{u}}^{\truncInd} } u^{\truncInd,\textbf{n}'} u^{\truncInd,\textbf{n}''} \big \langle v^{\textbf{n}'}_3 \textbf{v}^{\textbf{n}''} \cdot \partial_{x_3} \textbf{v}^{\textbf{n}} \big  \rangle .  \end{equation}
						Due to the orthogonality of the sinusoids, the only terms which remain in this sum must satisfy $\textbf{m}_h' := (m_1',m_2') = (m_1'',m_2'') =: \textbf{m}_h''$ and $m_3 = |m_3' \pm m_3''|$.  Furthermore,  $v_3^{\textbf{n}'}$ must be non-zero, which occurs iff $\nu_3^{\textbf{n}'} \neq 0$ and $\textbf{m}_h' \neq \textbf{0}$.  It is also clear that some phase condition must be satisfied by $\textbf{p},\textbf{p}',\textbf{p}''$, which we now consider.  Since $m_1 = m_2 = 0$ one must have either $\textbf{p} = (0,1)$ or $(1,0)$, and from \eqref{VortBal_VertStratModes} one sees that the dot product in \eqref{VortConsis_Nonlin} involves only one component of $\textbf{v}^{\textbf{n}''}$, namely $v^{\textbf{n}''}_1$ if $\textbf{p} = (1,0)$ or $v^{\textbf{n}''}_2$ if $\textbf{p} = (0,1)$.  Hence the phase $\textbf{p}'$ of $v_3^{\textbf{n}}$ must match the phase of this component.  One can then insert the definitions of $v_3^{\textbf{n}'}, \textbf{v}^{\textbf{n}''}$ and obtain 
						\[ \begin{split}
							\big \langle v^{\textbf{n}'}_3 \textbf{v}^{\textbf{n}''} \cdot \partial_{x_3} \textbf{v}^{\textbf{n}} \big  \rangle & = \delta^{\textbf{m}_h',\textbf{m}_h''} \frac{\sqrt{2} \ShaThree m_3 \mathcal{G}^{\textbf{n}'}_{3,3} \eta^{\textbf{m}'} \eta^{\textbf{m}''} }{V^3 |\mathcal{G}^{\textbf{n}'}\boldsymbol{\nu}^{\textbf{n}'}| | \mathcal{G}^{\textbf{n}''}\boldsymbol{\nu}^{\textbf{n}''} |  } \big ( \delta^{\textbf{p},(1,0)} \mathcal{G}_{1,1}^{\textbf{n}''} \varsigma_{1,1}^{\textbf{n}''} \big \langle s^{m_1',p_1'}s^{m_2',p_2'}s^{m_3',1} s^{m_1'',p_1''+1}s^{m_2'',p_2''}s^{m_3'',0} s^{m_3,1} \big \rangle \\ & \hspace{4 cm} - \delta^{\textbf{p},(0,1)} \mathcal{G}_{2,2}^{\textbf{n}''} \varsigma_{2,2}^{\textbf{n}''} \big \langle s^{m_1',p_1'}s^{m_2',p_2'}s^{m_3',1} s^{m_1'',p_1''}s^{m_2'',p_2''+1}s^{m_3'',0} s^{m_3,1} \big \rangle \big ) \text{ , } 
						\end{split} \]
						By inspecting the phases of the horizontal sinusoids, one finds that only terms with $\textbf{p} = \textbf{p}' + \textbf{p}''$ are non-zero.  Since it is assumed that $\nu_3^{\textbf{n}''} \neq 0$ and that the phases are matching, the horizontal integrals are simply evaluated to be $\frac{4\pi^2}{\ShaOne \ShaTwo}$, and the vertical integrals can be evaluated as in \eqref{NonlinDeriv_EvalResult}.  Thus one obtains the following:
						\[ \big \langle v^{\textbf{n}'}_3 \textbf{v}^{\textbf{n}''} \cdot \partial_{x_3} \textbf{v}^{\textbf{n}} \big  \rangle = \frac{ \delta^{\textbf{m}_h',\textbf{m}_h''} \ShaThree m_3 \mathcal{G}^{\textbf{n}'}_{3,3} \nu_3^{\textbf{n}'} \sigma^{(\boldsymbol{\mu}^3,\boldsymbol{\xi}^4)} }{V \eta^{m_3''}|\mathcal{G}^{\textbf{n}'}\boldsymbol{\nu}^{\textbf{n}'}| | \mathcal{G}^{\textbf{n}''}\boldsymbol{\nu}^{\textbf{n}''} |  } \big ( \delta^{\textbf{p},(1,0)} \mathcal{G}_{1,1}^{\textbf{n}''} \varsigma_{1,1}^{\textbf{n}''} - \delta^{\textbf{p},(0,1)} \mathcal{G}_{2,2}^{\textbf{n}''} \varsigma_{2,2}^{\textbf{n}''} \big ) \text{ . } \]
						In order to avoid repeated terms in the expression for $T_3$, one can impose $m_3' > m_3''$ by adding the corresponding expression with the symbols $m_3',m_3''$ interchanged.  By collecting the above results, one arrives at the following expression:
						\begin{equation} \label{NonlinDeriv_T3Sum} T_3 = \sum_{\textbf{n}' \in \mathscr{N}_{\textbf{u}}^{\truncInd} } \sum_{\substack{\textbf{n}''  \in \mathscr{N}_{\textbf{u}}^{\truncInd} \\ m_3'+m_3'' \text{ odd } \\ m_3' > m_3'' }}  \frac{2\sqrt{2} \ShaThree^2 \delta^{\textbf{m}_h',\textbf{m}_h''} }{\pi |\mathcal{G}^{\textbf{n}'}\boldsymbol{\nu}^{\textbf{n}'}| | \mathcal{G}^{\textbf{n}''}\boldsymbol{\nu}^{\textbf{n}''} | }  \textbf{c}^{\textbf{n}',\textbf{n}''} u^{\truncInd,\textbf{n}'} u^{\truncInd,\textbf{n}''} \text{ , }  \end{equation}
						in which 
						\begin{equation} \label{VortBalance_LHS}
							\textbf{c}^{\textbf{n}',\textbf{n}''} = \sum_{\substack{\textbf{n} \in \mathscr{N}_{\textbf{u}}^{\truncInd} \cap \mathscr{N}_{\textbf{u}}^{*} \\  m_3 = m_3' \pm m_3''}} m_3 \delta^{\textbf{p}'+\textbf{p}'',\textbf{p}} \begin{pmatrix}
								(-1)^{p_2'} \delta^{\textbf{p},(0,1)} \big ( \frac{\mathcal{G}_{3,3}^{\textbf{n}'} \nu_3^{\textbf{n}'}}{\eta^{m_3''}} \mathcal{G}_{2,2}^{\textbf{n}''} - \frac{\mathcal{G}_{3,3}^{\textbf{n}''} \nu_3^{\textbf{n}''}}{\sqrt{2}}  \sigma^{(m_3,m_3'',m_3',\boldsymbol{\xi}^4)} \mathcal{G}_{2,2}^{\textbf{n}'} \big ) \\ (-1)^{p_1'} \delta^{\textbf{p},(1,0)} \big ( \frac{\mathcal{G}_{3,3}^{\textbf{n}''} \nu_3^{\textbf{n}''}}{\sqrt{2}}  \sigma^{(m_3,m_3'',m_3',\boldsymbol{\xi}^4)} \mathcal{G}_{1,1}^{\textbf{n}'} - \frac{\mathcal{G}_{3,3}^{\textbf{n}'} \nu_3^{\textbf{n}'}}{\eta^{m_3''}} \mathcal{G}_{1,1}^{\textbf{n}''} \big ) \\ 0
							\end{pmatrix} \text{ . }
						\end{equation}
						On the other hand, for the vorticity balance to hold one must have
						\begin{equation} \label{VortBal_T3Bal} T_3 = \langle ( \vorticity^{\truncInd} \cdot \nabla ) \textbf{u}^{\truncInd} \rangle \text{ , } \end{equation}
						hence we consider the expression of the right hand side in Fourier space.  Expanding the right hand side one obtains:
						\[ \begin{split}
							\langle ( \vorticity^{\truncInd} \cdot \nabla ) \textbf{u}^{\truncInd} \rangle = \sum_{\textbf{n}' \in \mathscr{N}_{\textbf{u}}^{\truncInd} } \sum_{\textbf{n}'' \in \mathscr{N}_{\textbf{u}}^{\truncInd} } u^{\truncInd,\textbf{n}'} u^{\truncInd,\textbf{n}''} \big \langle \big ( (\nabla \times \textbf{v}^{\textbf{n}'} )\cdot \nabla \big ) \textbf{v}^{\textbf{n}''} \big \rangle \text{ . }
						\end{split} \]
						For the third component $v_3^{\textbf{n}''}$, one can simply integrate by parts to cancel the various terms from the curl, whereas for the first two components one cannot integrate by parts in the vertical direction without obtaining a boundary term.  Thus one obtains the following:
						\begin{equation} \label{VortBal_T3Bal_CompExp} \begin{split}
								\big \langle \big ( (\nabla \times \textbf{v}^{\textbf{n}'} )\cdot \nabla \big ) \textbf{v}^{\textbf{n}''} \big  \rangle = \sum_{j = 1,2} \big \langle ( -\partial_{x_3} v_2^{\textbf{n}'} \partial_{x_1} + \partial_{x_1} v_2^{\textbf{n}'} \partial_{x_3} + \partial_{x_3} v_1^{\textbf{n}'} \partial_{x_2} - \partial_{x_2} v_1^{\textbf{n}'} \partial_{x_3} ) v_j^{\textbf{n}''} \big \rangle  \uVecJ =: \sum_{j=1,2} b_j \uVecJ
						\end{split} \end{equation}
						Inserting the definitions of $\textbf{v}^{\textbf{n}'},\textbf{v}^{\textbf{n}''}$ gives
						\begin{equation} \label{VortBal_T3Bal_CompExp2} \begin{split}
								b_1 = \frac{\eta^{\textbf{m}'}\eta^{\textbf{m}''}}{|\mathcal{G}^{\textbf{n}'}\boldsymbol{\nu}^{\textbf{n}'}| | \mathcal{G}^{\textbf{n}''}\boldsymbol{\nu}^{\textbf{n}''} |V^2} \mathcal{G}^{\textbf{n}''}_{1,1} \varsigma_{1,1}^{\textbf{n}''} \Big [ & \mathcal{G}^{\textbf{n}'}_{2,2} \varsigma_{2,2}^{\textbf{n}'} \ShaOne \ShaThree \big ( (-1)^{p_1''} m_3'm_1'' \big \langle s^{m_1',p_1'}s^{m_2',p_2'+1}s^{m_3',1} s^{m_1'',p_1''}s^{m_2'',p_2''}s^{m_3'',0} \big \rangle \\ & \hspace{1.5 cm} + (-1)^{p_1'} m_1' m_3'' \big \langle s^{m_1',p_1'+1}s^{m_2',p_2'+1}s^{m_3',0} s^{m_1'',p_1''+1}s^{m_2'',p_2''}s^{m_3'',1} \big \rangle \big ) \\ & + \mathcal{G}^{\textbf{n}'}_{1,1} \varsigma_{1,1}^{\textbf{n}'} \ShaTwo \ShaThree \big ( (-1)^{p_2''} m_3' m_2'' \big \langle s^{m_1',p_1'+1}s^{m_2',p_2'}s^{m_3',1} s^{m_1'',p_1''+1}s^{m_2'',p_2''+1}s^{m_3'',0} \big \rangle \\ & \hspace{1 cm} + (-1)^{p_2'+1} m_2' m_3'' \big \langle s^{m_1',p_1'+1}s^{m_2',p_2'+1}s^{m_3',0} s^{m_1'',p_1''+1}s^{m_2'',p_2''}s^{m_3'',1} \big \rangle \big ) \Big ] \text{ . }
						\end{split} \end{equation}
						From \eqref{VortBal_T3Bal_CompExp} $b_1 = 0$ if $v_1^{\textbf{n}''} = 0$, so the non-trivial terms satisfy $\nu_1^{\textbf{n}''} \neq 0$.  By examining each integral in \eqref{VortBal_T3Bal_CompExp2}, one sees from the orthogonality of the sinusoids that all of the above integrals are zero unless $\textbf{m}_h' = \textbf{m}_h''$ and $\textbf{p}' + \textbf{p}'' = (0,1)$.  Furthermore since every term is multiplied by one of $m_1',m_2',m_1'',m_2''$ the non-trivial terms satisfy $\textbf{m}_h' \neq \textbf{0}$.  If these hold then one can compute the integrals, giving the following:
						\[ b_1 = (-1)^{p_2'} \frac{\eta^{m'_3}\eta^{m''_3}\ShaThree \mathcal{G}^{\textbf{n}''}_{1,1} \nu_1^{\textbf{n}''}}{|\mathcal{G}^{\textbf{n}'}\boldsymbol{\nu}^{\textbf{n}'}| | \mathcal{G}^{\textbf{n}''}\boldsymbol{\nu}^{\textbf{n}''} | \pi}  \Big ( \mathcal{G}^{\textbf{n}'}_{2,2} \ShaOne m_1' - \mathcal{G}^{\textbf{n}'}_{1,1} \ShaTwo m_2' \Big ) ( 1 - (-1)^{m_3'+m_3''}) \text{ , } \]
						from which it is clear that $m_3'+m_3''$ is odd for the non-trivial terms.  By similar arguments one finds $b_2$ is given as follows when $\textbf{p}'+\textbf{p}'' = (1,0)$:
						\[ b_2 = (-1)^{p_1'} \frac{\eta^{m'_3}\eta^{m''_3}\ShaThree \mathcal{G}^{\textbf{n}''}_{2,2} \nu_2^{\textbf{n}''}}{|\mathcal{G}^{\textbf{n}'}\boldsymbol{\nu}^{\textbf{n}'}| | \mathcal{G}^{\textbf{n}''}\boldsymbol{\nu}^{\textbf{n}''} | \pi}  \Big ( \mathcal{G}^{\textbf{n}'}_{2,2} \ShaOne m_1' - \mathcal{G}^{\textbf{n}'}_{1,1} \ShaTwo m_2' \Big ) ( 1 - (-1)^{m_3'+m_3''}) \text{ . } \]
						In order to avoid duplicate terms we enforce $m_3' > m_3''$, hence one has the following:
						\[ \begin{split}
							\langle ( \vorticity^{\truncInd} \cdot \nabla ) \textbf{u}^{\truncInd} \rangle = \sum_{\textbf{n}' \in \mathscr{N}_{\textbf{u}}^{\truncInd} } \sum_{\substack{\textbf{n}'' \in \mathscr{N}_{\textbf{u}}^{\truncInd} \\ m_3' + m_3'' \text{ odd } \\ m_3' > m_3''}}  \frac{2\sqrt{2}\ShaThree \delta^{\textbf{m}_h',\textbf{m}_h''} }{\pi |\mathcal{G}^{\textbf{n}'}\boldsymbol{\nu}^{\textbf{n}'}| | \mathcal{G}^{\textbf{n}''}\boldsymbol{\nu}^{\textbf{n}''} | } \eta^{m''_3} \tilde{\textbf{c}}^{\textbf{n}',\textbf{n}''} u^{\truncInd,\textbf{n}'} u^{\truncInd,\textbf{n}''} \text{ . }
						\end{split} \]
						in which
						\begin{align} 
							\tilde{c}^{\textbf{n}',\textbf{n}''}_1 & = (-1)^{p_2'} \delta^{\textbf{p}'+\textbf{p}'',(0,1)} \Big ( \mathcal{G}^{\textbf{n}''}_{1,1} \nu_1^{\textbf{n}''}  \big ( \mathcal{G}^{\textbf{n}'}_{2,2} \ShaOne m_1' - \mathcal{G}^{\textbf{n}'}_{1,1} \ShaTwo m_2' \big ) - \mathcal{G}^{\textbf{n}'}_{1,1} \nu_1^{\textbf{n}'} \big ( \mathcal{G}^{\textbf{n}''}_{2,2} \ShaOne m_1' - \mathcal{G}^{\textbf{n}''}_{1,1} \ShaTwo m_2' \big ) \Big ) \text{ , } \notag \\ \tilde{c}^{\textbf{n}',\textbf{n}''}_2 & = (-1)^{p_1'} \delta^{\textbf{p}'+\textbf{p}'',(1,0)} \Big ( \mathcal{G}^{\textbf{n}''}_{2,2} \nu_2^{\textbf{n}''}  \big ( \mathcal{G}^{\textbf{n}'}_{2,2} \ShaOne m_1' - \mathcal{G}^{\textbf{n}'}_{1,1} \ShaTwo m_2' \big ) - \mathcal{G}^{\textbf{n}'}_{2,2} \nu_2^{\textbf{n}'}  \big ( \mathcal{G}^{\textbf{n}''}_{2,2} \ShaOne m_1' - \mathcal{G}^{\textbf{n}''}_{1,1} \ShaTwo m_2' \big ) \Big )  \text{ , } \notag \\ \label{VortBalance_RHS}  \tilde{c}^{\textbf{n}',\textbf{n}''}_3 & = 0 \text{ . }
						\end{align}
						Thus \eqref{VortBal_T3Bal} holds iff $\ShaThree \textbf{c}^{\textbf{n}',\textbf{n}''}= \eta^{m''_3} \tilde{\textbf{c}}^{\textbf{n}',\textbf{n}''} $ for all $\textbf{n}',\textbf{n}'' \in \mathscr{N}_{\textbf{u}}^{\truncInd}$ with $\textbf{m}_h' = \textbf{m}_h''$, $m_3'+m_3''$ odd, $m_3' > m_3''$ and $\textbf{p}'+\textbf{p}'' = \textbf{p}$, where either $\textbf{p} = (0,1)$ or $(1,0)$.  In particular it is clear that the expression \eqref{VortBalance_LHS} depends on which vertically stratified modes are included in the model, and hence one only has equality if certain such modes are included.  Conceptually the remainder of the proof just consists of checking which conditions guarantee equality, but the challenge is that the expressions in \eqref{VortBalance_LHS} and \eqref{VortBalance_RHS} are heavily case dependent.  Note that in the case $\textbf{p} = (0,1)$ one need only check the first component, and hence equality holds iff
						\begin{align} \label{VortBal_FirstComp}
							\ShaThree  \sum_{\substack{\textbf{n} \in \mathscr{N}_{\textbf{u}}^{\truncInd} \cap \mathscr{N}_{\textbf{u}}^{*} \\  \textbf{p} = (0,1) \\ m_3 = m_3' \pm m_3''}} & m_3 \Big ( \frac{\mathcal{G}_{3,3}^{\textbf{n}'} \nu_3^{\textbf{n}'}}{\eta^{m_3''}} \mathcal{G}_{2}^{\textbf{n}''} - \frac{\mathcal{G}_{3,3}^{\textbf{n}''} \nu_3^{\textbf{n}''}}{\sqrt{2}}  \sigma^{(m_3,m_3'',m_3',\boldsymbol{\xi}^4)} \mathcal{G}_{2,2}^{\textbf{n}'} \Big ) \\ & = \eta^{m''_3} \Big ( \mathcal{G}^{\textbf{n}''}_{1,1} \nu_1^{\textbf{n}''} \big ( \mathcal{G}^{\textbf{n}'}_{2,2} \ShaOne m_1' -  \mathcal{G}^{\textbf{n}'}_{1,1} \ShaTwo m_2' \big ) - \mathcal{G}^{\textbf{n}'}_{1,1} \nu_1^{\textbf{n}'}  \big ( \mathcal{G}^{\textbf{n}''}_{2,2} \ShaOne m_1' - \mathcal{G}^{\textbf{n}''}_{1,1} \ShaTwo m_2' \big ) \Big ) \text{ , } \notag
						\end{align}
						Similarly in the case $\textbf{p} = (1,0)$ one need only check the second component, and hence equality holds iff
						\begin{align} \label{VortBal_SecondComp}
							\ShaThree  \sum_{\substack{\textbf{n} \in \mathscr{N}_{\textbf{u}}^{\truncInd} \cap \mathscr{N}_{\textbf{u}}^{*} \\  \textbf{p} = (0,1) \\ m_3 = m_3' \pm m_3''}} & m_3 \Big ( -\frac{\mathcal{G}_{3,3}^{\textbf{n}'} \nu_3^{\textbf{n}'}}{\eta^{m_3''}} \mathcal{G}_{1,1}^{\textbf{n}''} + \frac{\mathcal{G}_{3,3}^{\textbf{n}''} \nu_3^{\textbf{n}''}}{\sqrt{2}}  \sigma^{(m_3,m_3'',m_3',\boldsymbol{\xi}^4)} \mathcal{G}_{1,1}^{\textbf{n}'} \Big ) \\ & = \eta^{m''_3} \Big ( \mathcal{G}^{\textbf{n}''}_{2,2} \nu_2^{\textbf{n}''} \big ( \mathcal{G}^{\textbf{n}'}_{2,2} \ShaOne m_1' -  \mathcal{G}^{\textbf{n}'}_{1,1} \ShaTwo m_2' \big ) - \mathcal{G}^{\textbf{n}'}_{2,2} \nu_2^{\textbf{n}'}  \big ( \mathcal{G}^{\textbf{n}''}_{2,2} \ShaOne m_1' - \mathcal{G}^{\textbf{n}''}_{1,1} \ShaTwo m_2' \big ) \Big ) \text{ , } \notag
						\end{align}
						From here, the various cases must be considered individually.  The following summarizes the results of a case by case comparison of \eqref{VortBal_FirstComp} and \eqref{VortBal_SecondComp} supported by Mathematica, where it is found that Criterion \ref{Crit:VortCrit} (i) is necessary and sufficient to ensure \eqref{VortBal_T3Bal}.
						\begin{enumerate}
							\item \underline{Case $m_1',m_2',m_3'' > 0$:}  In this case one has $\nu_3^{\textbf{n}'} = \nu_3^{\textbf{n}''} = \nu_1^{\textbf{n}'} = \nu_1^{\textbf{n}''} =1 $ and $\eta^{m_3''} = \sqrt{2}$, so for $\textbf{p} = (0,1)$ one needs to check 
							\[ \begin{split}
								\sum_{\substack{\textbf{n} \in \mathscr{N}_{\textbf{u}}^{\truncInd} \cap \mathscr{N}_{\textbf{u}}^{*} \\  \textbf{p} = (0,1) \\ m_3 = m_3' \pm m_3''}} & \frac{\ShaThree m_3}{2} \Big ( \mathcal{G}_{3,3}^{\textbf{n}'}  \mathcal{G}_{2,2}^{\textbf{n}''} -  \sigma^{(m_3,m_3'',m_3',\boldsymbol{\xi}^4)}  \mathcal{G}_{3,3}^{\textbf{n}''} \mathcal{G}_{2,2}^{\textbf{n}'} \Big ) =  \mathcal{G}^{\textbf{n}''}_{1,1} \big ( \mathcal{G}^{\textbf{n}'}_{2,2} \ShaOne m_1' -  \mathcal{G}^{\textbf{n}'}_{1,1} \ShaTwo m_2' \big ) - \mathcal{G}^{\textbf{n}'}_{1,1} \big ( \mathcal{G}^{\textbf{n}''}_{2,2} \ShaOne m_1' - \mathcal{G}^{\textbf{n}''}_{1,1} \ShaTwo m_2' \big ) \text{ , } 
							\end{split} \]
							for the cases $(c',c'') = (1,1),(1,2),(2,1),(2,2)$.  The right hand side is non-zero for $\ShaOne m_1' \neq \ShaTwo m_2'$ or $(c',c'') \neq (1,1)$, and hence one finds that \eqref{VortBal_FirstComp} holds if and only if both $(0,0,m_3' + m_3'',\textbf{p},1)$ and $(0,0,m_3' - m_3'',\textbf{p},1)$ are included in $\mathscr{N}_{\textbf{u}}^{\truncInd}$.  For $\textbf{p} = (1,0)$ one must check a similar expression, and one finds that the analogous result holds.
							
							\item \underline{Case $m_1',m_2'> 0,$ $m_3'' = 0$:} In this case one has $\nu_3^{\textbf{n}'} = \nu_1^{\textbf{n}'} = \nu_1^{\textbf{n}''} =1 $, but $\nu_3^{\textbf{n}''}  = 0$ and $\eta^{m_3''} = 1$.  For $\textbf{p} = (0,1)$ one therefore needs to check 
							\begin{align} 
								\ShaThree m_3' \mathcal{G}_{3,3}^{\textbf{n}'} \mathcal{G}_{2,2}^{\textbf{n}''} = \mathcal{G}^{\textbf{n}''}_{1,1} \big ( \mathcal{G}^{\textbf{n}'}_{2,2} \ShaOne m_1' -  \mathcal{G}^{\textbf{n}'}_{1,1} \ShaTwo m_2' \big ) - \mathcal{G}^{\textbf{n}'}_{1,1} \big ( \mathcal{G}^{\textbf{n}''}_{2,2} \ShaOne m_1' - \mathcal{G}^{\textbf{n}''}_{1,1} \ShaTwo m_2' \big ) \text{ , } \notag
							\end{align}
							for the cases $(c',c'') = (1,1),(1,2)$.  Here again the right hand side is non-zero for $\ShaOne m_1' \neq \ShaTwo m_2'$ or $(c',c'') \neq (1,1)$, and hence one finds that \eqref{VortBal_FirstComp} holds if and only if $(0,0,m_3',\textbf{p},1)$ is included in $\mathscr{N}_{\textbf{u}}^{\truncInd}$.  For $\textbf{p} = (1,0)$ one must check a similar expression for each of the same values of $(c',c'')$, and here one again finds \eqref{VortBal_SecondComp} holds if and only if both $(0,0,m_3' + m_3'',\textbf{p},1)$ and $(0,0,m_3' - m_3'',\textbf{p},1)$ are included in $\mathscr{N}_{\textbf{u}}^{\truncInd}$.
							
							\item \underline{Case $m_1',m_3''> 0,$ $m_2' = 0$:} In this case one has $c' = c'' = 1$ and $\eta^{m_3''} = \sqrt{2}$.  Since $m_1' > 0$ the value of $p_1'$ is irrelevant, but since $m_2' = 0$ the value of $p_2'$ must be treated case by case.  In the case $\textbf{p}= (0,1)$, the first component is given as follows when $(p_2',p_2'') = (0,1)$
							\[ \sum_{\substack{\textbf{n} \in \mathscr{N}_{\textbf{u}}^{\truncInd} \cap \mathscr{N}_{\textbf{u}}^{*} \\  \textbf{p} = (0,1) \\ m_3 = m_3' \pm m_3''}} \frac{\ShaThree m_3}{2} \mathcal{G}_{3,3}^{\textbf{n}'} \mathcal{G}_{2,2}^{\textbf{n}''}  =  - \mathcal{G}^{\textbf{n}'}_{1,1} \mathcal{G}^{\textbf{n}''}_{2,2} \ShaOne m_1'  \text{ , }  \]
							When $(p_2',p_2'') = (1,0)$ the first component is given by
							\[ -\sum_{\substack{\textbf{n} \in \mathscr{N}_{\textbf{u}}^{\truncInd} \cap \mathscr{N}_{\textbf{u}}^{*} \\  \textbf{p} = (0,1) \\ m_3 = m_3' \pm m_3''}} \frac{\ShaThree m_3}{2} \sigma^{(m_3,m_3'',m_3',\boldsymbol{\xi}^4)}\mathcal{G}_{3,3}^{\textbf{n}''} \mathcal{G}_{2,2}^{\textbf{n}'}  = \mathcal{G}^{\textbf{n}''}_{1,1} \mathcal{G}^{\textbf{n}'}_{2,2} \ShaOne m_1' \text{ . } \]
							On the other hand, in the case $\textbf{p}= (1,0)$, the second component is given as follows when $(p_2',p_2'') = (0,0)$
							\[ \sum_{\substack{\textbf{n} \in \mathscr{N}_{\textbf{u}}^{\truncInd} \cap \mathscr{N}_{\textbf{u}}^{*} \\  \textbf{p} = (1,0) \\ m_3 = m_3' \pm m_3''}} \frac{\ShaThree m_3}{2} \big ( \mathcal{G}_{3,3}^{\textbf{n}'} \mathcal{G}_{1,1}^{\textbf{n}''} - \sigma^{(m_3,m_3'',m_3',\boldsymbol{\xi}^4)}\mathcal{G}_{3,3}^{\textbf{n}''} \mathcal{G}_{1,1}^{\textbf{n}'} \big ) = 0 \text{ , } \]
							but is simply given by $0 = 0 $ when $(p_2',p_2'') = (1,1)$.  One sees that when $p_2'+p_2'' = 1$ the right hand side is non-zero, and hence equality holds iff $(0,0,m_3' + m_3'',\textbf{p},1)$ and $(0,0,m_3' - m_3'',\textbf{p},1)$ are included in $\mathscr{N}_{\textbf{u}}^{\truncInd}$.  On the other hand, for $(p_2',p_2'') = (0,0)$ one must either include both $(0,0,m_3' + m_3'',\textbf{p},1)$ and $(0,0,m_3' - m_3'',\textbf{p},1)$ in $\mathscr{N}_{\textbf{u}}^{\truncInd}$, or neither.  Finally, when $(p_2',p_2'') = (1,1)$ the inclusion or exclusion of $(0,0,m_3' + m_3'',\textbf{p},1)$ and $(0,0,m_3' - m_3'',\textbf{p},1)$ in $\mathscr{N}_{\textbf{u}}^{\truncInd}$ has no effect on the vorticity balance. 
							\item \underline{Case $m_2',m_3''> 0,$ $m_1' = 0$:} This is analogous to the case above, with the roles of $m_1',m_2'$ and $p_1',p_2'$ reversed.
							\item \underline{Case $m_1'> 0,$ $m_2' = m_3'' = 0$:} In this case one has $c' = c'' = 1$, $\eta^{m_3''} = 1$ and $\nu_3^{\textbf{n}''} = 0$.  Again the value of $p_1'$ is irrelevant, and $p_2'$ must be treated case by case.  However, due to \eqref{PhaseIndexSets} one must have $p_2'' = 1$.  For the case $\textbf{p}= (0,1)$, the first component is given as follows when $(p_2',p_2'') = (0,1)$
							\[ \sum_{\substack{\textbf{n} \in \mathscr{N}_{\textbf{u}}^{\truncInd} \cap \mathscr{N}_{\textbf{u}}^{*} \\  \textbf{p} = (0,1) \\ m_3 = m_3' }} \ShaThree m_3 \mathcal{G}_{3,3}^{\textbf{n}'} \mathcal{G}_{2,2}^{\textbf{n}''}  =  - \mathcal{G}^{\textbf{n}'}_{1,1} \mathcal{G}^{\textbf{n}''}_{2,2} \ShaOne m_1'  \text{ , }  \]
							Here one sees that right hand side is non-zero, and hence equality holds iff $(0,0,m_3',\textbf{p},1)$ is included in $\mathscr{N}_{\textbf{u}}^{\truncInd}$.  On the other hand, in the case $\textbf{p}= (1,0)$, the second component is simply given by $0 = 0$ when $(p_2',p_2'') = (1,1)$, hence in this case the inclusion or exclusion of $(0,0,m_3',\textbf{p},1)$ in $\mathscr{N}_{\textbf{u}}^{\truncInd}$ has no effect on the vorticity balance. 
							
							\item \underline{Case $m_2'> 0,$ $m_1' = m_3'' = 0$:} This is analogous to the case above, with the roles of $m_1',m_2'$ and $p_1',p_2'$ reversed.
						\end{enumerate}
						
					\end{proof}
					
					\section*{Acknowledgements}
					
					This paper is a contribution to the project M2 of the Collaborative Research Centre TRR 181 "Energy Transfers in Atmosphere and Ocean" funded by the Deutsche Forschungsgemeinschaft (DFG, German Research Foundation) - Projektnummer 274762653.  The authors thank Dr. Maxim Kirsebom for participation during the early stages of this paper and Dr. Fabian Bleitner for advice regarding Paraview.
					
					
					\section*{Data Availability}
					
					The code generated to create the figures in this work has been made available on GitHub at \url{https://github.com/rkwelter/RRBC_SpectralCodeRepo}, while the data can be made available upon request to the author.
					
					
					
					\printbibliography

@article{Grmela97,
  title = {Dynamics and thermodynamics of complex fluids.  I. Development of a general formalism},
  author = {Grmela, Miroslav and \"Ottinger, Hans Christian},
  journal = {Phys. Rev. E},
  volume = {56},
  issue = {6},
  pages = {6620--6632},
  numpages = {0},
  year = {1997},
  month = {Dec},
  publisher = {American Physical Society},
  doi = {10.1103/PhysRevE.56.6620},
  url = {https://link.aps.org/doi/10.1103/PhysRevE.56.6620}
}

@article{WIMMER2020109016,
title = {Energy conserving upwinded compatible finite element schemes for the rotating shallow water equations},
journal = {Journal of Computational Physics},
volume = {401},
pages = {109016},
year = {2020},
issn = {0021-9991},
doi = {https://doi.org/10.1016/j.jcp.2019.109016},
url = {https://www.sciencedirect.com/science/article/pii/S0021999119307223},
author = {Golo A. Wimmer and Colin J. Cotter and Werner Bauer}
}

@article{SANDERSE2025106473,
title = {Energy-consistent discretization of viscous dissipation with application to natural convection flow},
journal = {Computers and Fluids},
volume = {286},
pages = {106473},
year = {2025},
issn = {0045-7930},
doi = {https://doi.org/10.1016/j.compfluid.2024.106473},
url = {https://www.sciencedirect.com/science/article/pii/S0045793024003049},
author = {B. Sanderse and F.X. Trias}
}

@article{Prugger02112022,
author = {A. Prugger and J. D. M. Rademacher and J. Yang and},
title = {Geophysical fluid models with simple energy backscatter: explicit flows and unbounded exponential growth},
journal = {Geophysical \& Astrophysical Fluid Dynamics},
volume = {116},
number = {5-6},
pages = {374--410},
year = {2022},
publisher = {Taylor \& Francis},
doi = {10.1080/03091929.2021.2011269},


URL = { 
    
        https://doi.org/10.1080/03091929.2021.2011269
    
    

},
eprint = { 
    
        https://doi.org/10.1080/03091929.2021.2011269
    
    

}

}

@Inbook{Holst2024,
author="Holst, Paul
and Rademacher, Jens D.  M.
and Yang, Jichen",
editor="Henry, David",
title="Rotating Convection and Flows with Horizontal Kinetic Energy Backscatter",
bookTitle="Nonlinear Dispersive Waves: Based on the 2023 Workshop at University College Cork, Ireland",
year="2024",
publisher="Springer Nature Switzerland",
address="Cham",
pages="133--171",
isbn="978-3-031-63512-0",
doi="10.1007/978-3-031-63512-0_7",
url="https://doi.org/10.1007/978-3-031-63512-0_7"
}

@article{Andrews_2024_Enforcing,
author = {B. Andrews and P. Farrell},
title = {Enforcing conservation laws and dissipation inequalities numerically via auxiliary variables},
year = {2024},
URL = {https://arxiv.org/abs/2407.11904}
}

@article{ConstantinDoering_1996,
	title = {Variational bounds on energy dissipation in incompressible flows. III. Convection},
	author = {Doering, Charles R. and Constantin, Peter},
	journal = {Phys. Rev. E},
	volume = {53},
	issue = {6},
	pages = {5957--5981},
	numpages = {0},
	year = {1996},
	month = {June},
	publisher = {American Physical Society},
	doi = {10.1103/PhysRevE.53.5957},
	url = {https://link.aps.org/doi/10.1103/PhysRevE.53.5957}
}

@article{GluhovskyTongAgee_2002,
      author = "Alexander Gluhovsky and Christopher Tong and Ernest Agee",
      title = "Selection of Modes in Convective Low-Order Models",
      journal = "Journal of the Atmospheric Sciences",
      year = "2002",
      publisher = "American Meteorological Society",
      address = "Boston MA, USA",
      volume = "59",
      number = "8",
      doi = "https://doi.org/10.1175/1520-0469(2002)059<1383:SOMICL>2.0.CO;2",
      pages=      "1383 - 1393",
      url = "https://journals.ametsoc.org/view/journals/atsc/59/8/1520-0469_2002_059_1383_somicl_2.0.co_2.xml"
}

@article{HermizGuzdarFinn_1995,
  title = {Improved low-order model for shear flow driven by Rayleigh-B\'enard convection},
  author = {Hermiz, K. B. and Guzdar, P. N. and Finn, J. M.},
  journal = {Phys. Rev. E},
  volume = {51},
  issue = {1},
  pages = {325--331},
  numpages = {0},
  year = {1995},
  month = {January},
  publisher = {American Physical Society},
  doi = {10.1103/PhysRevE.51.325},
  url = {https://link.aps.org/doi/10.1103/PhysRevE.51.325}
}

@article{HowardKrishnamurti_1986, 
    title={Large-scale flow in turbulent convection: a mathematical model}, 
    volume={170}, 
    DOI={10.1017/S0022112086000940}, 
    journal={Journal of Fluid Mechanics}, 
    publisher={Cambridge University Press}, 
    author={Howard, L. N. and Krishnamurti, R.}, year={1986}, 
    pages={385–410}}

@article{iyer2020classical,
author = {Iyer, Kartik and Scheel, Janet and Schumacher, Joerg and Sreenivasan, Katepalli},
year = {2020},
month = {03},
pages = {201922794},
title = {Classical 1/3 scaling of convection holds up to Ra = 10 15},
volume = {117},
journal = {Proceedings of the National Academy of Sciences},
doi = {10.1073/pnas.1922794117}
}

@misc{OlsonDoering2022,
	doi = {10.48550/ARXIV.2203.02067},
	url = {https://arxiv.org/abs/2203.02067},
	author = {Olson, Matthew L. and Doering, Charles R.},
	title = {Heat transport in a hierarchy of reduced-order convection models},
	publisher = {arXiv},
	year = {2022},
	copyright = {arXiv.org perpetual, non-exclusive license}
}

@article{Welter2025Rotating,
	author = {Welter, Roland },
	title = {Rotating Rayleigh-Benard convection: Attractors, bifurcations and heat transport via a Galerkin hierarchy},
	journal = {SIAM Journal on Applied Dynamical Systems},
	year = {2025}
}

@article{WenDingChiniKerswell2022,
	author = {Wen, Baole  and Ding, Zijing  and Chini, Gregory P.  and Kerswell, Rich R. },
	title = {Heat transport in Rayleigh–Bénard convection with linear marginality},
	journal = {Philosophical Transactions of the Royal Society A: Mathematical, Physical and Engineering Sciences},
	volume = {380},
	number = {2225},
	pages = {20210039},
	year = {2022},
	doi = {10.1098/rsta.2021.0039},
	
	URL = {https://royalsocietypublishing.org/doi/abs/10.1098/rsta.2021.0039},
	eprint = {https://royalsocietypublishing.org/doi/pdf/10.1098/rsta.2021.0039}
}

@article{wen_goluskin_doering_2022, title={Steady Rayleigh–Bénard convection between no-slip boundaries}, volume={933}, DOI={10.1017/jfm.2021.1042}, journal={Journal of Fluid Mechanics}, publisher={Cambridge University Press}, author={Wen, Baole and Goluskin, David and Doering, Charles R.}, year={2022}, pages={R4}}

@article{LStenflo_1996,
	doi = {10.1088/0031-8949/53/1/015},
	url = {https://dx.doi.org/10.1088/0031-8949/53/1/015},
	year = {1996},
	month = {01},
	publisher = {},
	volume = {53},
	number = {1},
	pages = {83},
	author = {L Stenflo},
	title = {Generalized Lorenz equations for acoustic-gravity waves in the atmosphere},
	journal = {Physica Scripta}
}

@article{Geisselmann_2024_Energy,
author = {J. Giesselmann and A. Karsai and T. Tscherpel},
title = {Energy-consistent Petrov-Galerkin time discretization of port-Hamiltonian systems},
year = {2024},
URL = {https://arxiv.org/abs/2404.12480}
}

@article{Miquel2021Coral, doi = {10.21105/joss.02978}, url = {https://doi.org/10.21105/joss.02978}, year = {2021}, publisher = {The Open Journal}, volume = {6}, number = {65}, pages = {2978}, author = {Benjamin Miquel}, title = {Coral: a parallel spectral solver for fluid dynamics and partial differential equations}, journal = {Journal of Open Source Software} }

@article{StevensClercxLohse_2010,
	title = {Optimal Prandtl number for heat transfer in rotating Rayleigh–Bénard convection},
	journal = {New Journal of Physics},
	volume = {12},
	year = {2010},
	doi = {10.1088/1367-2630/12/7/075005},
	url = {https://iopscience.iop.org/article/10.1088/1367-2630/12/7/075005},
	author = {Richard J. A. M. Stevens and Herman J. H. Clercx and Detlef Lohse}
}

@article{ThiffeaultHorton_1996,
	title = {Energy‐conserving truncations for convection with shear flow},
	volume = {8},
	issn = {1070-6631},
	url = {https://doi.org/10.1063/1.868956},
	doi = {10.1063/1.868956},
	number = {7},
	journal = {Physics of Fluids},
	author = {Thiffeault, Jean‐Luc and Horton, Wendell},
	month = jul,
	year = {1996},
	pages = {1715--1719},
}

@article{WhiteheadDoering_2011,
	title = {Ultimate State of Two-Dimensional Rayleigh-B\'enard Convection between Free-Slip Fixed-Temperature Boundaries},
	author = {Whitehead, Jared P. and Doering, Charles R.},
	journal = {Phys. Rev. Lett.},
	volume = {106},
	issue = {24},
	pages = {244501},
	numpages = {4},
	year = {2011},
	month = {June},
	publisher = {American Physical Society},
	doi = {10.1103/PhysRevLett.106.244501},
	url = {https://link.aps.org/doi/10.1103/PhysRevLett.106.244501}
}

@article{WhiteheadDoering_2012,
author = {Whitehead, Jared and Doering, Charles},
year = {2012},
month = {09},
pages = {241-259},
title = {Rigid bounds on heat transport by a fluid between slippery boundaries},
volume = {707},
journal = {Journal of Fluid Mechanics},
doi = {10.1017/jfm.2012.274}
}

@book{eden2019energy,
  title={Energy Transfers in Atmosphere and Ocean},
  author={Eden, Carsten and Iske, Armin and Franzke, Christian L. E. and Oliver, Marcel and Rademacher, and Jens D. M. and Badin, Gualtiero and von Storch, Jin-Song and Olbers, Dirk and Pollmann, Friederike},
  isbn={978-3-030-05703-9},
  doi={https://doi.org/10.1007/978-3-030-05704-6},
  series={Mathematics of Planet Earth},
  url={https://link.springer.com/book/10.1007/978-3-030-05704-6},
  year={2019},
  publisher={Springer}
}
					
				\end{document}